\documentclass[reprint,amssymb, amsmath, aps, superscriptaddress, showpacs, footinbib,  prb]{revtex4-1}
\pdfoutput=1
\usepackage{graphicx}
\usepackage{color}

\newcommand{\eff}{\text{eff}}
\newcommand{\AFM}{\text{AFM}}

\newcommand{\sub}{\text{sub}}
\newcommand{\iso}{\text{iso}}
\newcommand{\imp}{\text{imp}}
\newcommand{\Cels}{$^\circ$C}

\begin{document}

\title{Short-range order of Br and three-dimensional magnetism in (CuBr)LaNb$_2$O$_7$}
\author{Alexander A. Tsirlin}
\email{altsirlin@gmail.com}
\affiliation{Max Planck Institute for Chemical Physics of Solids, N\"{o}thnitzer
Str. 40, 01187 Dresden, Germany}

\author{Artem M. Abakumov}
\email{Artem.Abakumov@ua.ac.be}
\affiliation{EMAT, University of Antwerp, Groenenborgerlaan 171, B-2020 Antwerp, Belgium}

\author{Clemens~Ritter}
\affiliation{Institut Laue-Langevin, BP 156, F-38042 Grenoble, France}

\author{Paul F. Henry}
\affiliation{MI-1, Helmholtz Center Berlin for Materials and Energy, Hahn-Meitner Platz 1, 14109 Berlin, Germany}
\affiliation{European Spallation Source, ESS AB, PO Box 176, 221 00 Lund, Sweden}

\author{Oleg Janson}
\author{Helge Rosner}
\affiliation{Max Planck Institute for Chemical Physics of Solids, N\"{o}thnitzer
Str. 40, 01187 Dresden, Germany}

\begin{abstract}
We present a comprehensive study of the crystal structure, magnetic structure, and microscopic magnetic model of (CuBr)LaNb$_2$O$_7$, the Br analog of the spin-gap quantum magnet (CuCl)LaNb$_2$O$_7$. Despite similar crystal structures and spin lattices, the magnetic behavior and even peculiarities of the atomic arrangement in the Cl and Br compounds are very different. The high-resolution x-ray and neutron data reveal a split position of Br atoms in (CuBr)LaNb$_2$O$_7$. This splitting originates from two possible configurations developed by [CuBr] zigzag ribbons. While the Br atoms are locally ordered in the $ab$ plane, their arrangement along the $c$ direction remains partially disordered. The predominant and energetically more favorable configuration features an additional doubling of the $c$ lattice parameter that was not observed in (CuCl)LaNb$_2$O$_7$. (CuBr)LaNb$_2$O$_7$ undergoes long-range antiferromagnetic ordering at $T_N=32$~K, which is nearly 70~\% of the leading exchange coupling $J_4\simeq 48$~K. The Br compound does not show any experimental signatures of low-dimensional magnetism, because the underlying spin lattice is three-dimensional. The coupling along the $c$ direction is comparable to the couplings in the $ab$ plane, even though the shortest \mbox{Cu--Cu} distance along $c$ (11.69~\r A) is three times larger than nearest-neighbor distances in the $ab$ plane (3.55~\r A). The stripe antiferromagnetic long-range order featuring columns of parallel spins in the $ab$ plane and antiparallel spins along $c$ is verified experimentally and confirmed by the microscopic analysis.
\end{abstract}

\pacs{61.66.Fn, 75.30.Cr, 75.30.Et, 61.72.Ff}
\maketitle

\section{Introduction}
\label{sec:intro}
Transition-metal halides are one of the best playgrounds for studying diverse magnetism of low-dimensional spin systems. Considering four stable halogen elements (F, Cl, Br, and I), chlorine and bromine are most appealing because of the similar chemistry, yet different ionic radii that strongly influence superexchange pathways and relevant magnetic interactions. The change in the coupling regime between a chloride and an isostructural bromide is a fairly common scenario.\cite{manaka1997a,*manaka1997b,valenti2003,*zaharko2006,ono2005,*foyevtsova2011} The differences between chlorides and bromides can be generally understood in terms of interatomic distances and angles, depending on the size of the ligand. While these geometrical parameters determine individual exchange couplings, the influence on the spin lattice may be more involved, especially in mixed Cl/Br systems with a variable crystallographic symmetry\cite{[{For example: }][{}]krueger2010} and effects of bond randomness.\cite{[{For example: }][{}]manaka2001,*manaka2002,*goto2008} In the following, we will present an even more striking example, where Br atoms not only change the magnetic ground state but also trigger the incomplete structural order in the compound. We will consider (CuBr)LaNb$_2$O$_7$, a long-range ordered antiferromagnet,\cite{oba2006} which is remarkably different from its Cl analog (CuCl)LaNb$_2$O$_7$ showing a gapped singlet ground state with zero ordered magnetic moment on the Cu sites.\cite{kageyama2005a,*kageyama2005b} 

\begin{figure}
\includegraphics{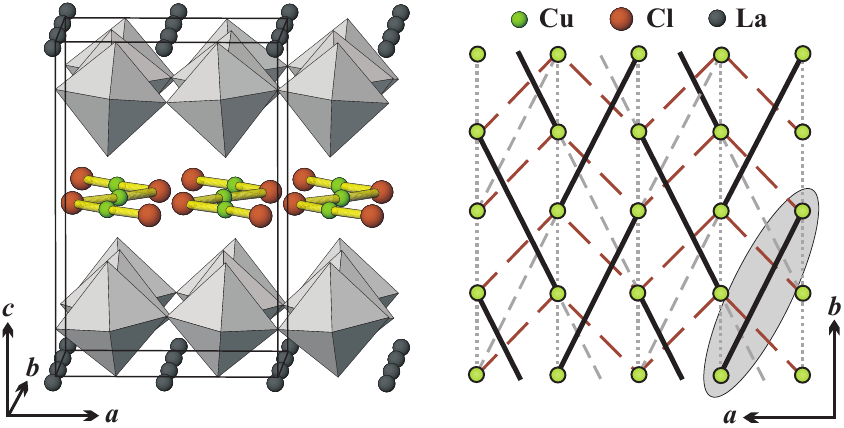}
\caption{\label{fig:structure-Cl}
(Color online) Left panel: low-temperature crystal structure of (CuCl)LaNb$_2$O$_7$ featuring the [CuCl] zigzag ribbons and tilted NbO$_6$ octahedra.\cite{tsirlin2010a,tassel2010,hernandez2011} Right panel: spin lattice in the $ab$ plane. Although the spatial arrangement of Cu atoms resembles the square lattice, numerous inequivalent interactions are present.\cite{tsirlin2010b} Note the spin dimers formed on the fourth-neighbor Cu atoms, as shown by a shaded oval. For notation of individual couplings, see Fig.~\ref{fig:lattice}.
}
\end{figure}
Both (CuX)LaNb$_2$O$_7$ compounds (X = Cl, Br) feature flat [Cu$^{+2}$X] magnetic layers separated by non-magnetic [LaNb$_2^{+5}$O$_7$] slabs with two layers of corner-sharing NbO$_6$ octahedra (Fig.~\ref{fig:structure-Cl}). The original structural model described by Kodenkandath \mbox{\textit{et al.}\cite{koden1999}} is deceptively simple, with a regular square-lattice arrangement of Cu and halogen atoms, and the ensuing tetragonal symmetry. Recent studies\cite{tsirlin2010a,tassel2010,hernandez2011} questioned the completeness of this structural model, and identified an orthorhombic superstructure related to the doubling of both the $a$ and $b$ lattice parameters. In (CuCl)LaNb$_2$O$_7$, main features of this superstructure are: i) cooperative tilts of the NbO$_6$ octahedra; and ii) displacements of Cu and Cl atoms in the $ab$ plane. The Jahn-Teller distortion inherent to Cu$^{+2}$ results in shorter and longer Cu--Cl bonds, so that Cu and Cl atoms form zigzag ribbons running along the $b$ direction (Fig.~\ref{fig:structure-Cl}).\cite{tsirlin2009,tsirlin2010a,tassel2010,ren2010} The orthorhombic structural model was derived from x-ray\cite{tsirlin2010a} and neutron\cite{tassel2010} powder data, and later confirmed in a single-crystal x-ray experiment.\cite{hernandez2011} This model satisfactorily accounts for the spin-gap magnetic behavior of (CuCl)LaNb$_2$O$_7$ (note spin dimers in the right panel of Fig.~\ref{fig:structure-Cl}), although details of the spin lattice remain controversial.\cite{tassel2010,tsirlin2010b}

In contrast to the Cl compound, its Br analogue (CuBr)LaNb$_2$O$_7$ still lacks a detailed structural and microscopic investigation. Experimental data evidence magnetic couplings of $J\simeq 30-40$~K, both ferromagnetic (FM) and antiferromagnetic (AFM),\footnote{More precise estimates of the magnetic couplings can be found in Sec.~\ref{sec:simul}} and a stripe (columnar) AFM long-range order that sets in below $T_N\simeq 32$~K.\cite{oba2006} While these observations would conform to the square-lattice arrangement of Cu$^{+2}$ atoms with the dominating next-nearest-neighbor AFM coupling,\cite{[{For example: }][{}]shannon2004} an attempt to quantify the square-lattice model essentially failed and predicted a strongly frustrated regime of (CuBr)LaNb$_2$O$_7$, in agreement with naive computational results\cite{ren2010} and in apparent contradiction with the high N\'eel temperature of the compound ($T_N/J\simeq 1$).\cite{oba2006} 

Yoshida \mbox{\textit{et al.}\cite{yoshida2008}} used nuclear magnetic resonance (NMR) to demonstrate sizable deviations from the tetragonal symmetry at the Cu and Br sites, thus invalidating the available structural model as well as the square-lattice magnetic model. They also proposed a low-symmetry structure common for (CuCl)LaNb$_2$O$_7$ and (CuBr)LaNb$_2$O$_7$,\cite{yoshida2008,yoshida2007} but recent diffraction studies of the Cl compound refuted this conjecture.\cite{tsirlin2010a,tassel2010,hernandez2011} Ren and Chen\cite{ren2010} put forward an alternative orthorhombic structure later confirmed for (CuCl)LaNb$_2$O$_7$ and -- as we will show below -- roughly matching the experimental structure of (CuBr)LaNb$_2$O$_7$, which however features a disordered Br position (compare Figs.~\ref{fig:structure-Cl} and~\ref{fig:structure}). No microscopic description for the magnetism of (CuBr)LaNb$_2$O$_7$ has been given so far.

In this paper, we first establish the crystal structure of (CuBr)LaNb$_2$O$_7$ and further consider the microscopic magnetic model, thus providing a comprehensive description of this compound in terms of the precise atomic positions and ensuing exchange couplings. We show that (CuBr)LaNb$_2$O$_7$ is weakly frustrated and, moreover, reveals only weak quantum fluctuations because of the enhanced connectivity of the spin lattice. Our results are naturally separated into the crystal-structure (Sec.~\ref{sec:structure}) and microscopic-model parts (Sec.~\ref{sec:model}) followed by a discussion and summary in Sec.~\ref{sec:discussion}, where we compare (CuBr)LaNb$_2$O$_7$ to the isostructural compounds and consider broader implications for other Cu$^{+2}$ bromides.

\section{Methods}
\label{sec:methods}
Powder samples of (CuBr)LaNb$_2$O$_7$ were prepared by a two-step procedure following Ref.~\onlinecite{koden1999}. First, we synthesized RbLaNb$_2$O$_7$ by mixing stoichiometric amounts of La$_2$O$_3$ and Nb$_2$O$_5$ with a 25\% excess of Rb$_2$CO$_3$, followed by an annealing at 850~\Cels\ (1123~K, 8 hours) and 1050~\Cels\ (1323~K, 30 hours) with one intermediate regrinding. The La$_2$O$_3$ and Nb$_2$O$_5$ powders were dried at 800~\Cels\ prior to the experiment. The RbLaNb$_2$O$_7$ powder was washed with water to remove excess rubidium oxide. 

On the second step, RbLaNb$_2$O$_7$ was mixed with a twofold amount of anhydrous CuBr$_2$, pressed into a pellet, sealed into an evacuated quartz tube, and heated at 350~\Cels\ (623~K) for 48 hours. The resulting sample was again washed with water to eliminate the excess of CuBr$_2$ as well as RbBr formed during the reaction. The quality of starting materials, the RbLaNb$_2$O$_7$ precursor, and the final (CuBr)LaNb$_2$O$_7$ product was carefully checked with powder x-ray diffraction (XRD) measured using Huber G670 Guinier camera (CuK$_{\alpha1}$ radiation, ImagePlate detector, $2\theta=3-100$~deg. angular range). The successful refinement of the high-resolution x-ray and neutron data (Sec.~\ref{sec:structure}) further confirms the high purity of our powder samples. 

High-resolution XRD data for structure refinement were collected at the ID31 beamline of European Synchrotron Radiation Facility (ESRF, Grenoble, France) using a constant wavelength of $\lambda\simeq 0.4$~\r A and eight scintillation detectors, each preceded by a Si (111) analyzer crystal, in the angular range $2\theta=1-40$~deg. The powder sample was contained in a thin-walled borosilicate glass capillary that was spun during the experiment. The temperature of the sample was controlled by a He-flow cryostat (temperature range $10-200$~K), a liquid-nitrogen cryostream ($200-350$~K), and hot-air blower ($350-750$~K). 

\begin{figure*}
\includegraphics{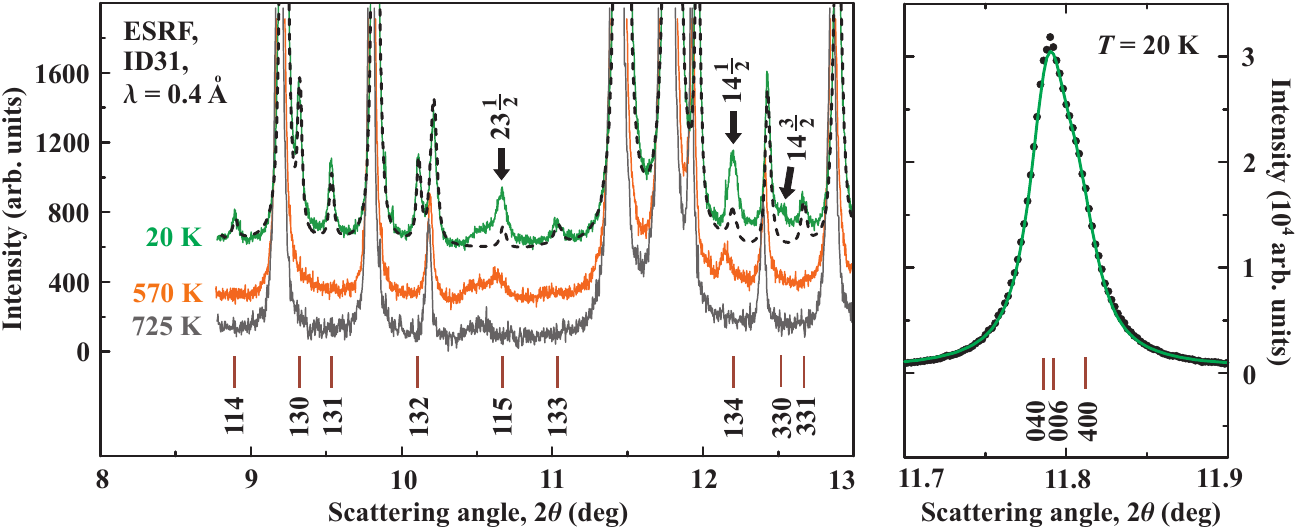}
\caption{\label{fig:xrd}
(Color online) Left panel: high-resolution XRD patterns of (CuBr)LaNb$_2$O$_7$ measured at 20~K, 570~K, and 725~K for the $\alpha$-, $\beta$-, and $\gamma$-polymorphs, respectively. Only the superstructure reflections are indexed. The dashed line is the Rietveld refinement of the 20~K pattern. Note the diffuse scattering that is missing in the refinement. The diffuse features can be ascribed to a unit cell with the doubled $c$ parameter. This diffuse scattering is denoted by arrows and persists at 570~K in the $\beta$-polymorph, whereas the sharp superstructure reflections with even $h+k$ disappear. Right panel: a part of the 20~K pattern showing the splitting of the 400 reflection.
}
\end{figure*}
Neutron diffraction data were collected at the E9 high-resolution diffractometer ($\lambda\simeq 1.797$~\r A, $Q=1.2-6.4$~\r A$^{-1}$, $T=20$~K and 300~K) installed at the Hahn-Meitner Institute (Helmholtz Center Berlin for Materials and Energy, Germany). Unfortunately, these data were somewhat limited in the $q$ range and insufficient to detect weak magnetic reflections. Therefore, we performed further experiments at the high-resolution instrument D2B ($\lambda\simeq 1.595$~\r A, $Q=0.9-7.6$~\r A$^{-1}$, $T=10$~K) and the high-flux instrument D20 ($\lambda\simeq 2.417$~\r A, $Q=0.35-4.9$~\r A$^{-1}$, $T=1.5$~K and 40~K), both installed at Institute Laue-Langevin (ILL, Grenoble, France). A 6~g powder sample used for all neutron measurements was loaded into a cylindrical vanadium container and cooled down with the standard Orange He-flow cryostat. Crystal and magnetic structures of (CuBr)LaNb$_2$O$_7$ were refined with \texttt{JANA2006}\cite{jana2006} and \texttt{FullProf}\cite{fullprof} programs, respectively.

The specimen for the electron diffraction (ED) study was prepared by crushing the sample under ethanol and depositing a drop of suspension on a holey carbon grid. ED patterns were taken at room temperature with a Tecnai G2 transmission electron microscope operated at 200 kV. Owing to the instability of the compound under intense electron beam, high-resolution imaging was not possible.

Thermogravimetric analysis (TGA) was performed with the STA409 Netzsch thermal balance in the $300-1050$~K temperature range. For differential scanning calorimetry (DSC), we used a Perkin Elmer 8500 instrument in the temperature range $100-670$~K. In both measurements, a heating rate of 10~K/min, corundum crucibles, and Ar atmosphere were used.

The magnetic susceptibility of (CuBr)LaNb$_2$O$_7$ was measured with Quantum Design MPMS SQUID magnetometer in the temperature range $2-380$~K in applied fields between 0.1~T and 5~T. 

To evaluate lattice energies, individual exchange couplings, and the microscopic magnetic model, we performed full-potential scalar-relativistic density-functional-theory (DFT) band-structure calculations using the \texttt{FPLO9.01-35} code.\cite{fplo} We used both the local density approximation (LDA)\cite{pw92} and the generalized gradient approximation (GGA)\cite{pbe} for the exchange-correlation potential. The symmetry-irreducible part of the first Brillouin zone was sampled by a fine mesh of up to 570~points for the 48-atom crystallographic unit cell, 48~points for 96-atom (doubled) supercells, and 8~points for 144-atom (tripled) supercells. Convergence with respect to the $k$ mesh was carefully checked. Correlation effects in the Cu $3d$ shell were treated either on the model level (Hubbard model constructed on top of LDA band structure) or in the framework of the mean-field DFT+$U$ approach. Structure relaxations for large supercells were also performed in the \texttt{VASP} code\cite{vasp1,*vasp2} with the basis set of projected augmented waves\cite{paw1,*paw2} and the energy cutoff of 400~eV. Further details of the computational procedures are given in the respective sections.

The microscopic magnetic model was further refined against the experimental data using quantum Monte Carlo (QMC) simulations of the magnetic susceptibility, the magnetization isotherms, the N\'eel temperature, and of the ordered magnetic moment. These simulations were performed with the \texttt{loop}\cite{loop} and \verb|dirloop_sse|\cite{dirloop} algorithms of the \texttt{ALPS} simulation package,\cite{alps} as further described in Sec.~\ref{sec:simul}.
\section{Crystal structure}
\label{sec:structure}

\subsection{Low-temperature structure}
\label{sec:lowT}

\begin{figure}
\includegraphics{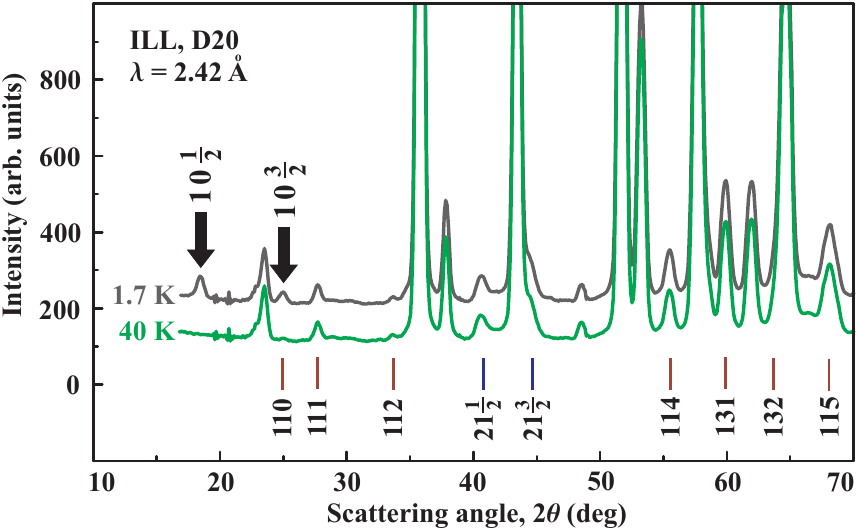}
\caption{\label{fig:neutron}
(Color online) Neutron diffraction patterns of (CuBr)LaNb$_2$O$_7$ measured at 1.7~K and 40~K, below and above the magnetic transition at $T_N\simeq 32$~K, respectively. Only the superstructure reflections are indexed. Note two weak nuclear reflections that can be indexed with the doubled $c$ parameter, only. Magnetic reflections are labeled with arrows. The patterns are offset for clarity.
}
\end{figure}

\begin{figure}
\includegraphics{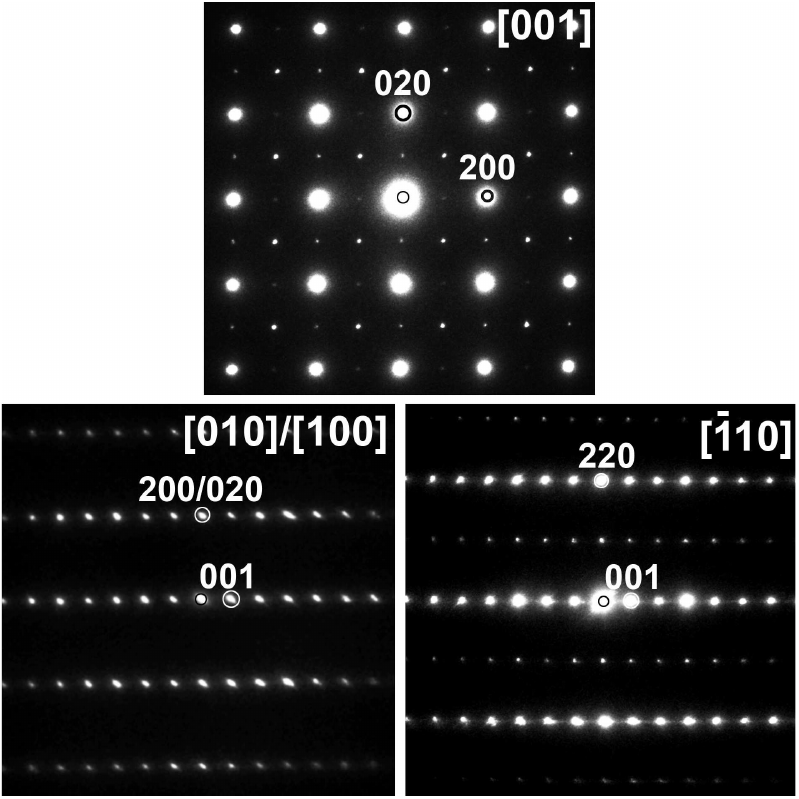}
\caption{\label{fig:ED}
Room-temperature electron diffraction patterns of (CuBr)LaNb$_2$O$_7$. Bright dots are the subcell reflections. Less bright dots with odd $h$ and $k$ are visible in the $[001]$ and $[\bar 110]$ patterns, and evidence the superstructure with the $C$-centered $2a_{\sub}\times 2a_{\sub}\times c_{\sub}$ unit cell. Faint dots with odd $h+k$ violate the $C$-centered unit cell and originate from the short-range order of Br atoms (see text for details).
}
\end{figure}

Low-temperature x-ray and neutron diffraction patterns of (CuBr)LaNb$_2$O$_7$ (Figs.~\ref{fig:xrd} and~\ref{fig:neutron}), as well as electron diffraction patterns measured at room temperature (Fig.~\ref{fig:ED}),\footnote{Although the electron diffraction study was done at RT, its results are relevant to the low-temperature structure, because (CuBr)LaNb$_2$O$_7$ does not demonstrate phase transitions upon cooling below room temperature (Sec.~\ref{sec:evolution}).} 
showed weak reflections violating the tetragonal unit cell proposed in earlier studies.\cite{koden1999} Additionally, the high-resolution x-ray data revealed a weak orthorhombic distortion, evidenced by the splitting of the 400 reflection shown in the right panel of Fig.~\ref{fig:xrd}. Most of the superstructure reflections could be indexed in a $C$-centered pseudotetragonal $2a_{\sub}\times 2a_{\sub}\times c_{\sub}$ unit cell, where $a_{\sub}\simeq 3.9$~\r A and $c_{\sub}\simeq 11.7$~\r A are parameters of the tetragonal subcell reported in Ref.~\onlinecite{koden1999}. The data also showed several diffuse features (Figs.~\ref{fig:xrd} and~\ref{fig:neutron}) along with faint electron-diffraction spots violating the $C$-centering. These features originate from the short-range order of Br atoms and will be discussed separately in Sec.~\ref{sec:short-range}.

Since no reflection conditions other than the general requirement $h+k=2n$ for the $C$-centering were observed, the $Cmmm$ space group or its subgroups could be used for the structure refinement. In the $Cmmm$ space group, the origin matches the inversion center that can be placed either on Cu atoms, on Br atoms, or in between the two neighboring Cu atoms. Only the latter setting allowed the successful refinement. In this case, the NbO$_6$ octahedra follow the $a^0b^-c^0$ tilting distortion,\footnote{%
We refer to the conventional Glazer's notation, where $a$, $b$, and $c$ show different rotations around the respective crystallographic directions, $+/-$ denote the in-phase/out-of-phase tilt, and 0 indicates the absence of rotations.} whereas the Cu and Br atoms are constrained to the $z=0$ plane by the respective mirror symmetry. Additionally, the $m_y$ and $m_x$ planes allow the displacements of the Cu and Br atoms along the $a$ and $b$ directions, respectively. While the $a$ and $b$ axes are interchangeable, we use the above setting with Cu on the $m_y$ plane and Br on the $m_x$ plane for the sake of consistency with the previously published data on (CuCl)LaNb$_2$O$_7$.\cite{tsirlin2010a}

\begin{table}
\caption{\label{tab:cell}
Lattice parameters (in~\r A) and space groups for structure refinements of (CuBr)LaNb$_2$O$_7$ at different temperatures $T$ (in~K), as given in Tables~\ref{tab:str-1} and~\ref{tab:str-3}. The error bars are based on the Rietveld refinement.
}\smallskip
\begin{ruledtabular}
\begin{tabular}{ccccc}
   $T$        & $x/a$     & $y/b$     & $z/c$      & Space group \\
   20         & 7.7856(2) & 7.7983(2) & 11.6938(2) & $Cmmm$      \\
   300        & 7.7926(2) & 7.8009(2) & 11.7017(2) & $Cmmm$      \\
   720        & 3.9077(1) & 3.9077(1) & 11.7192(1) & $P4/mmm$    \\
\end{tabular}
\end{ruledtabular}
\end{table}
The refinement of the high-resolution neutron data was somewhat unstable, because the weak orthorhombic splitting caused large fluctuations of the $a$ and $b$ lattice parameters. To improve the refinement, we additionally considered the high-resolution x-ray data, where the orthorhombic splitting is better resolved (Fig.~\ref{fig:xrd}). We have also used the neutron data from the high-flux D20 instrument that is most sensitive to weak superstructure reflections. Altogether, the low-temperature structure of (CuBr)LaNb$_2$O$_7$ was simultaneously refined against three datasets.\cite{supplement} Diffuse scattering and weak magnetic reflections observed in the D20 experiment were excluded from the refinement and analyzed separately (Secs.~\ref{sec:short-range} and~\ref{sec:magstructure}). 

\begin{figure}
\includegraphics{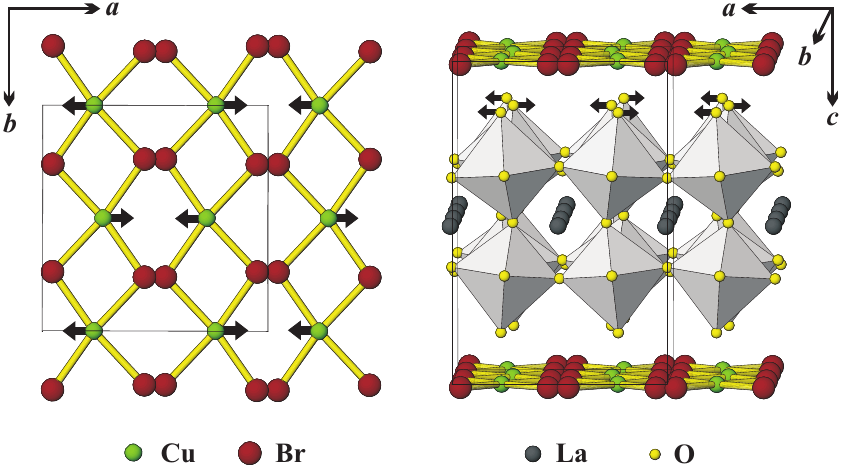}
\caption{\label{fig:structure}
(Color online) Low-temperature crystal structure of (CuBr)LaNb$_2$O$_7$ with the split position of Br atoms (the displacements along the $c$ direction are too small to be visualized on this scale). Arrows show the displacements of oxygen atoms upon the $a^0b^-c^0$ tilting distortion (right panel) and the ensuing displacements of Cu atoms in the $ab$ plane (left panel). The lines denote short Cu--Br bonds in the $ab$ plane.
}
\end{figure}

The refinement of atomic positions and atomic displacement parameters (ADPs) produced an unusually large ADP for Br, $U_{\iso}\simeq 0.050$~\r A$^2$ indicating a disorder of the Br atoms. A subsequent refinement of the anisotropic ADP showed a drastic elongation of the ellipsoid along the $a$ direction. Therefore, a split position with the Br atom shifted from the $m_x$ plane was introduced (Fig.~\ref{fig:structure}). This improved the refinement and reduced the ADP of Br down to $1.0\times 10^{-2}$~\r A$^2$. A displacement out of the $m_z$ plane allowed for a further reduction in the ADP and resulted in a non-zero $z$ coordinate. Therefore, the Br atoms occupy the split 16-fold position in the averaged $C$-centered orthorhombic crystal structure.

Final atomic positions are summarized in Table~\ref{tab:str-1}.\cite{supplement,foot1} Figure~\ref{fig:structure} depicts the low-temperature structure of (CuBr)LaNb$_2$O$_7$, which is further referred to as the $\alpha$-polymorph. This structure bears apparent similarities to the low-temperature $\alpha$-(CuCl)LaNb$_2$O$_7$ structure that also has the orthorhombic symmetry. Both Cl and Br compounds reveal the $a^0b^-c^0$ tilting distortion with the out-of-phase rotation of the NbO$_6$ octahedra about the $b$ axis. The tilting angle measured as the tilt of the Nb--O3 bond with respect to the $c$ axis is 4.8~deg in $\alpha$-(CuBr)LaNb$_2$O$_7$ compared to 5.7~deg in the Cl compound (Ref.~\onlinecite{hernandez2011}). Following this rotation, terminal oxygen atoms displace along the $a$ direction and promote similar displacements of Cu atoms in the $ab$ plane (Fig.~\ref{fig:structure}). 

\begin{table}
\caption{\label{tab:str-1}\label{tab:str-2}
Atomic positions and atomic displacement parameters ($U_{\iso}$, in $10^{-2}$~\r A$^2$) for $\alpha$-(CuBr)LaNb$_2$O$_7$ at 20~K (first line) and 300~K (second line), as refined from the synchrotron XRD and neutron data. The space group is $Cmmm$. The $U_{\iso}$ of oxygen atoms were refined as a single parameter. The error bars are based on the Rietveld refinement. See text and Supplementary information for details.
}
\begin{ruledtabular}
\begin{tabular}{cccccc}
  Atom & Position                & $x/a$     & $y/b$     & $z/c$     & $U_{\iso}$  \\
  Cu   &  $4g$                   & 0.7296(2) & $\frac12$ & 0         & 0.36(3)     \\
       &                         & 0.7265(5) & $\frac12$ & 0         & 0.97(7)     \\
  Br\footnotemark[1] & $16r$     & 0.4536(4) & 0.2385(3) & 0.0068(9) & 0.54(13)    \\
\footnotetext[1]{Fixed occupancy $g_{\text{Br}}=\frac14$.}
       &                         & 0.4568(8) & 0.2424(5) & 0.012(1)  & 1.2(3)      \\
  La   &  $4j$                   & 0         & 0.2588(2) & $\frac12$  & 0.02(2)    \\
       &                         & 0         & 0.2568(3) & $\frac12$  & 0.51(3)     \\
  Nb   &  $8o$                   & 0.7467(2) & $\frac12$ & 0.31005(4) & 0.00(2)    \\
       &                         & 0.7476(2) & $\frac12$ & 0.30929(8) & 0.51(3)     \\
  O1   &  $4l$                   & 0         & $\frac12$ & 0.3265(4)  & 0.20(1)    \\
       &                         & 0         & $\frac12$ & 0.3280(11) & 1.47(5)     \\
  O2   &  $8m$                   & $\frac14$ & $\frac34$ & 0.3456(3)  & 0.20(1)    \\
       &                         & $\frac14$ & $\frac34$ & 0.3428(9)  & 1.47(5)     \\
  O3   &  $8o$                   & 0.2276(2) & 0         & 0.15958(7) & 0.20(1)    \\
       &                         & 0.2352(7) & 0         & 0.1583(2)  & 1.47(5)     \\
  O4   &  $4h$                   & 0.7831(3) & $\frac12$ & $\frac12$  & 0.20(1)    \\
       &                         & 0.7676(10)& $\frac12$ & $\frac12$  & 1.47(5)     \\
  O5   &  $4k$                   & 0         & 0         & 0.3621(4)  & 0.20(1)    \\
       &                         & 0         & 0         & 0.3634(10) & 1.47(5)     \\
\end{tabular}
\end{ruledtabular}
\end{table}

An important difference between $\alpha$-(CuCl)LaNb$_2$O$_7$ and $\alpha$-(CuBr)LaNb$_2$O$_7$ is the arrangement of halogen atoms. While the Cl atoms are fully ordered and lie in the $ab$ plane, the Br atoms are disordered in the $ab$ plane and additionally show marginal out-of-plane displacements. This increases the symmetry from $Pbam$ in the Cl compound to $Cmmm$ in the Br compound. Both structures reveal shortened Cu--X distances of about 2.39~\r A (X = Cl)\cite{tassel2010} and $2.49-2.55$~\r A (X = Br), compared to 2.76~\r A in the tetragonal structures reported earlier.\cite{koden1999} The elongated Cu--Br distances of $2.95-3.10$~\r A are also present. The shortened and elongated bonds reflect the Jahn-Teller distortion driven by the $d^9$ electronic configuration of Cu$^{+2}$ (Ref.~\onlinecite{tsirlin2009}). 

Each Cu atom forms two short bonds to the O3 atoms of the NbO$_6$ octahedra (see Table~\ref{tab:dist}). Therefore, only two short bonds to Br may be formed to avoid overbonding. In $\alpha$-(CuCl)LaNb$_2$O$_7$, the \textit{trans}-arrangement of the short Cu--Cl bonds (Cl--Cu--Cl angle close to $180^{\circ}$) is well defined, whereas the split Br position could allow for different local environments of Cu. However, electronic effects strongly disfavor a random arrangement of Br and lead to the local ordering that largely resembles the $\alpha$-(CuCl)LaNb$_2$O$_7$ structure. Details of the short-range order in $\alpha$-(CuBr)LaNb$_2$O$_7$ are discussed below in Sec.~\ref{sec:short-range} followed by microscopic considerations in Sec.~\ref{sec:energetics}.

\subsection{Short-range order}
\label{sec:short-range}
\begin{figure}
\includegraphics{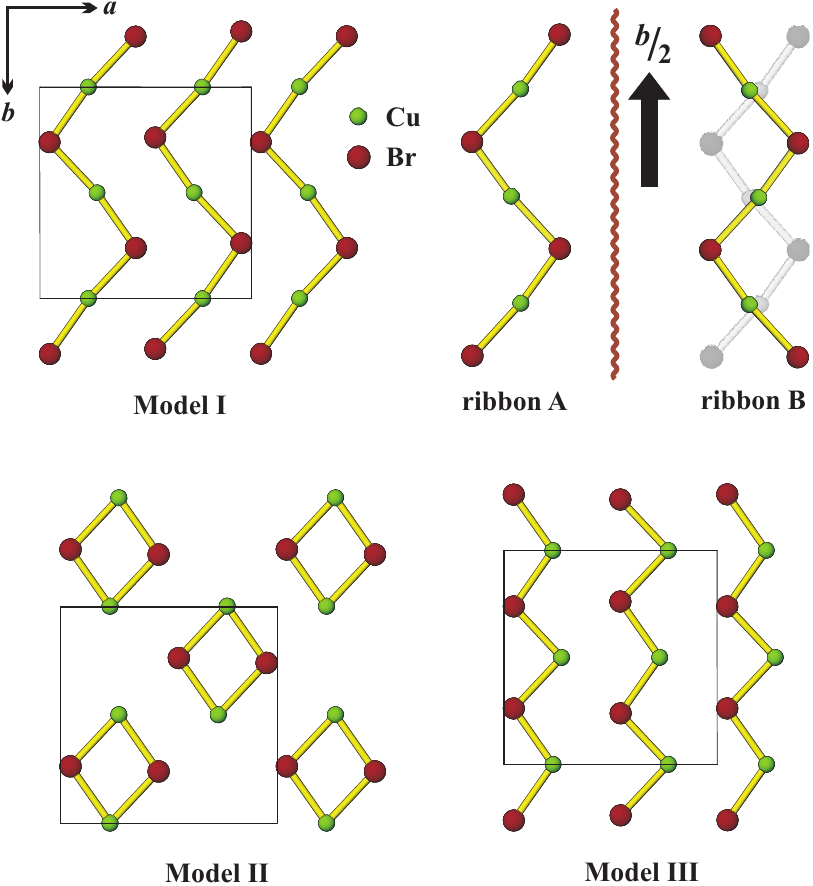}
\caption{\label{fig:antiphase}
(Color online) Possible local order of Br atoms in $\alpha$-(CuBr)LaNb$_2$O$_7$. The top panel shows the only stable configuration (\textit{trans}-arrangement of short Cu--Br bonds) and the principal scheme of disorder by random shifts of zigzag [CuBr] ribbons for one half of the $b$ lattice parameter (compare to the refined structure shown in Fig.~\ref{fig:structure}). The bottom panel shows the unstable configurations (see Sec.~\ref{sec:short-range}).
}
\end{figure}

To explore the short-range structural order in $\alpha$-(CuBr)LaNb$_2$O$_7$, we consider the NMR data reported by Yoshida~\textit{et al.},\cite{yoshida2008} who evaluated electric field gradients (EFGs) on Cu and Br sites. Their results evidence single crystallographic positions for both Cu and Br within the experimental resolution. The refined crystal structure allows for several local configurations (Fig.~\ref{fig:antiphase}): i) \textit{trans}-arrangement of short Cu--Br bonds forming zigzag ribbons along the $b$ direction; ii) \textit{cis}-arrangement of short bonds forming dimers (similar to the earlier proposal in Ref.~\onlinecite{yoshida2008}); iii) \textit{cis}-arrangement of short bonds forming chains along $b$. The EFGs calculated for these configurations along with the experimental results are listed in Table~\ref{tab:efg}. In DFT calculations, we use the LSDA+$U$ method with the on-site Coulomb repulsion $U_d=5$~eV, the on-site Hund's exchange $J_d=1$~eV, and the around-mean-field (AMF) double-counting correction. The variation of these parameters in a reasonable range has little influence on the computed EFG values. Further details of the computational method are given in Sec.~\ref{sec:dft}.

\begin{table}
\caption{\label{tab:efg}
Calculated and experimental parameters of the EFG tensor for the Cu and Br sites in $\alpha$-(CuBr)LaNb$_2$O$_7$. $V_{zz}$ is the leading component (in~$10^{21}$~V/m$^2$) and $\eta=(V_{yy}-V_{xx})/V_{zz}$ is the asymmetry. The models of local order are shown in Fig.~\ref{fig:antiphase}. Model III features two inequivalent Cu sites with slightly different Cu--Br bond lengths. Experimental values are taken from Ref.~\onlinecite{yoshida2008}. Note that the sign of $V_{zz}$ can not be determined experimentally.
}
\begin{ruledtabular}
\begin{tabular}{rrcrc}
     & \multicolumn{2}{c}{Cu} & \multicolumn{2}{c}{Br} \\
                      & $V_{zz}$  & $\eta$    & $V_{zz}$ & $\eta$ \\
  Model I (trans)     & $-10.3$   & 0.27      & $-45.0$  &  0.28  \\
  Model II (cis)      &  12.4     & 0.34      & $-42.3$  &  0.26  \\
  Model III (cis)     & 12.4/12.0 & 0.06/0.38 & $-42.3$  &  0.04  \\
  Experiment          & $\pm 10.6$ & --       & $\pm 30.0$ &  0.33  \\
\end{tabular}
\end{ruledtabular}
\end{table}

The EFG tensor is quantified by its leading component $V_{zz}$ and the asymmetry parameter $\eta=(V_{yy}-V_{xx})/V_{zz}$. Considering experimental accuracy and resolution, computational estimates show rather small differences between three possible configurations. Although $V_{zz}$ of Cu changes sign depending on the \textit{cis}- or \textit{trans}-arrangement, this effect can not be captured experimentally, because NMR does not resolve the sign of $V_{zz}$ unambiguously. Nevertheless, experiments safely establish that only one local arrangement is present in the crystal structure, because the respective $V_{zz}$ values on Cu sites differ for about 20~\% and should be easily resolvable. According to the Cu $V_{zz}$, the \textit{trans}-arrangement of the Cu--Br bonds is more likely, as further confirmed by energetic considerations in Sec.~\ref{sec:energetics}. All three configurations systematically overestimate $V_{zz}$ for the Br site. Since a similar 30~\% overestimate has been reported for the $V_{zz}$ of Cl in $\alpha$-(CuCl)LaNb$_2$O$_7$ (see Table~V in Ref.~\onlinecite{tsirlin2010a}), the discrepancy is likely related to drawbacks of the computational method and not to the inaccuracies of our structural model.

The \textit{trans}-arrangement of short Cu--Br bonds results in the formation of planar CuO$_2$Br$_2$ plaquettes and [CuBr] zigzag ribbons, similar to the [CuCl] ribbons in the structure of $\alpha$-(CuCl)LaNb$_2$O$_7$ (see Fig.~\ref{fig:structure-Cl}). The disorder in the Br position implies that such ribbons randomly displace for one half of the $b$ lattice parameter to form another ribbon with a different arrangement of Br atoms (see the top panel of Fig.~\ref{fig:antiphase}). However, the ribbons within the $ab$ plane should be ordered, because the shift of the ribbon for $b/2$ creates Cu atoms with the \textit{cis}-configuration of short bonds at the interface (see Fig.~\ref{fig:antiphase-2}). Since the \textit{cis}-arrangement of short bonds is energetically highly unfavorable (see Sec.~\ref{sec:energetics}) and is not observed by NMR, the disorder within the $ab$ plane can be ruled out. Thus, the only plausible scenario is the random arrangement of the [CuBr] ribbons in neighboring planes. 

Labeling one possible arrangement of the [CuBr] ribbon as A and the ribbon displaced for $b/2$ as B (Fig.~\ref{fig:antiphase}), one can construct structures with different orderings along the $c$ direction. The simplest examples are AAAA (or the equivalent BBBB) and ABAB shown in Fig.~\ref{fig:models}. The former is equivalent to the $\alpha$-(CuCl)LaNb$_2$O$_7$ structure and has the $2a_{\sub}\times 2a_{\sub}\times c_{\sub}$ unit cell with the $Pbam$ symmetry. The latter features the doubled translation along $c$ ($2a_{\sub}\times 2a_{\sub}\times 2c_{\sub}$ cell). Ribbon A is transformed into ribbon B by a $[\frac12,\frac12,0]$ translation supplied with the $[0,0,\frac12]$ translation between the neighboring layers. The structure is, therefore, body-centered and retains all symmetry elements of the AAAA configuration, thus resulting in the $Ibam$ space group. While the AAAA structure should manifest itself by additional superstructure reflections violating the $C$-centering (i.e., $hkl$ with odd $h+k$ and integer $l$), the ABAB structure gives rise to the reflections with odd $h+k$ and half-integer $l$. A closer inspection of the diffraction data shows that the reflections of the latter type are indeed present.

The low-temperature neutron diffraction patterns (Fig.~\ref{fig:neutron}) feature two weak and somewhat diffuse reflections that can be indexed with odd $h+k$ and half-integer $l$. The missing intensity in the refinement of the low-temperature XRD data (Fig.~\ref{fig:xrd}) can be explained in a similar way. Finally, the faint spots at $hk0$ ($h+k=\text{odd}$) in the $[001]$ ED pattern (Fig.~\ref{fig:ED}) are traces of the diffuse intensity lines intersecting the [001] reciprocal lattice plane. The x-ray data additional show very broad features at the positions of $hkl$ reflections with odd $h+k$ and integer $l$ (e.g., the broad hump around $2\theta\simeq 10.5$~deg in Fig.~\ref{fig:xrd}). This intensity may signal traces of the AAAA type ($Pbam$) ordering, which however is not seen in the neutron data. Since neutrons probe the bulk of the sample, we conclude that in $\alpha$-(CuBr)LaNb$_2$O$_7$ the ABAB type of the local order prevails, although sporadic interruptions of this configuration give rise to the disordered position of Br in the averaged crystal structure with the $Cmmm$ symmetry. Small regions of the AAAA type order that manifests itself in the x-ray and electron diffraction data may also be present in the sample, although their concentration is low. Further studies on single crystals should better resolve the diffuse intensity and provide additional insight into the short-range order of Br atoms in $\alpha$-(CuBr)LaNb$_2$O$_7$.

A refinement of the data in the $Ibam$ space group accounts well for the intensities of all superstructure reflections. However, it also shows the sizable broadening of the reflections with odd $h+k$, and arrives at a relatively large ADP of Br ($U_{\text{iso}}\simeq 0.02$~\r A$^2$). This confirms that the Br atoms are only locally ordered, and the averaged structure should be described by the $Cmmm$ space group. The disorder in the positions of Br atoms along the $c$ direction and the preference for the ABAB-type order are further confirmed by microscopic considerations in Sec.~\ref{sec:energetics}.
\subsection{Temperature evolution}
\label{sec:evolution}
In (CuCl)LaNb$_2$O$_7$, the Cl-containing counterpart of (CuBr)LaNb$_2$O$_7$, the study of the temperature evolution provided a valuable insight into the formation of the superstructure, and clarified the interplay of displacements observed for different atomic species.\cite{tsirlin2010a} Upon heating, the tilting distortion is eliminated above $T_1\simeq 500$~K ($\alpha\rightarrow\beta$ transition). Since the terminal oxygen atoms of the NbO$_6$ octahedra are no longer displaced, the ordered arrangement of Cu atoms is also lost (see Fig.~\ref{fig:structure}). However, the ordered positions of Cl atoms are preserved up to the second $\beta\rightarrow\gamma$ structural transition at $T_2\simeq 640$~K. Above $T_2$, (CuCl)LaNb$_2$O$_7$ displays the tetragonal crystal structure ($a_{\sub}\times a_{\sub}\times c_{\sub}$ unit cell, $\gamma$-phase) with random displacements of both Cu and Cl atoms within the [CuCl] planes. Note, however, that neither Cu nor Cl occupy the high-symmetry positions on the four-fold axes and rather remain displaced, in order to retain the short Cu--Cl bonds (Jahn-Teller distortion) at least locally.\cite{tsirlin2010a}

On the experimental side, the two-step formation of the $\alpha$-(CuCl)LaNb$_2$O$_7$ superstructure implies the appearance of the $hkl$ superstructure reflections in two stages. The superstructure reflections with odd $h+k$ are observed in the $\beta$-polymorph below $T_2$ (ordering of Cl atoms), while the reflections with even $h+k$ appear in the $\alpha$-polymorph below $T_1$ only (tilting distortion and ordering of Cu atoms). 

\begin{table}
\caption{\label{tab:str-3}
Atomic positions and atomic displacement parameters ($U_{\iso}$, in $10^{-2}$~\r A$^2$) for $\gamma$-(CuBr)LaNb$_2$O$_7$ refined using the synchrotron XRD data at 720~K. The space group is $P4/mmm$. The $U_{\iso}$ of oxygen atoms were refined as a single parameter. The error bars are based on the Rietveld refinement. See text and Supplementary information for details.
}
\begin{ruledtabular}
\begin{tabular}{cccccc}
  Atom & Position & $x$       & $y$       & $z$        & $U_{\iso}$  \\
  Cu\footnotemark[1] &  $4n$    & 0.559(1)  & $\frac12$ & 0          & 1.2(1)      \\
  Br\footnotemark[1] &  $8s$    & 0.068(2)  & 0         & 0.0189(5)  & 3.1(3)      \\\footnotetext[1]{Fixed occupancies: $g_{\text{Cu}}=\frac14$, $g_{\text{Br}}=\frac18$.}
  La   &  $1b$    & 0         & 0         & $\frac12$  & 1.08(2)     \\
  Nb   &  $2h$    & $\frac12$ & $\frac12$ & 0.30821(5) & 0.79(2)     \\
  O1   &  $4i$    & 0         & $\frac12$ & 0.3432(2)  & 1.62(6)     \\
  O2   &  $2h$    & $\frac12$ & $\frac12$ & 0.1637(4)  & 1.62(6)     \\
  O3   &  $1d$    & $\frac12$ & $\frac12$ & $\frac12$  & 1.62(6)     \\
\end{tabular}
\end{ruledtabular}
\end{table}
The temperature evolution of (CuBr)LaNb$_2$O$_7$ follows the same line. The superstructure reflections with even $h+k$ disappear above $T_1\simeq 500$~K, whereas the diffuse scattering centered at $hkl$ with odd $h+k$ and non-integer $l$ persists up to $T_2\simeq 620$~K (see Fig.~\ref{fig:xrd}). Although the Br atoms are locally ordered, their arrangement is not influenced by the onset of the tilting distortion in the [LaNb$_2$O$_7$] perovskite slabs. 

\begin{figure}
\includegraphics{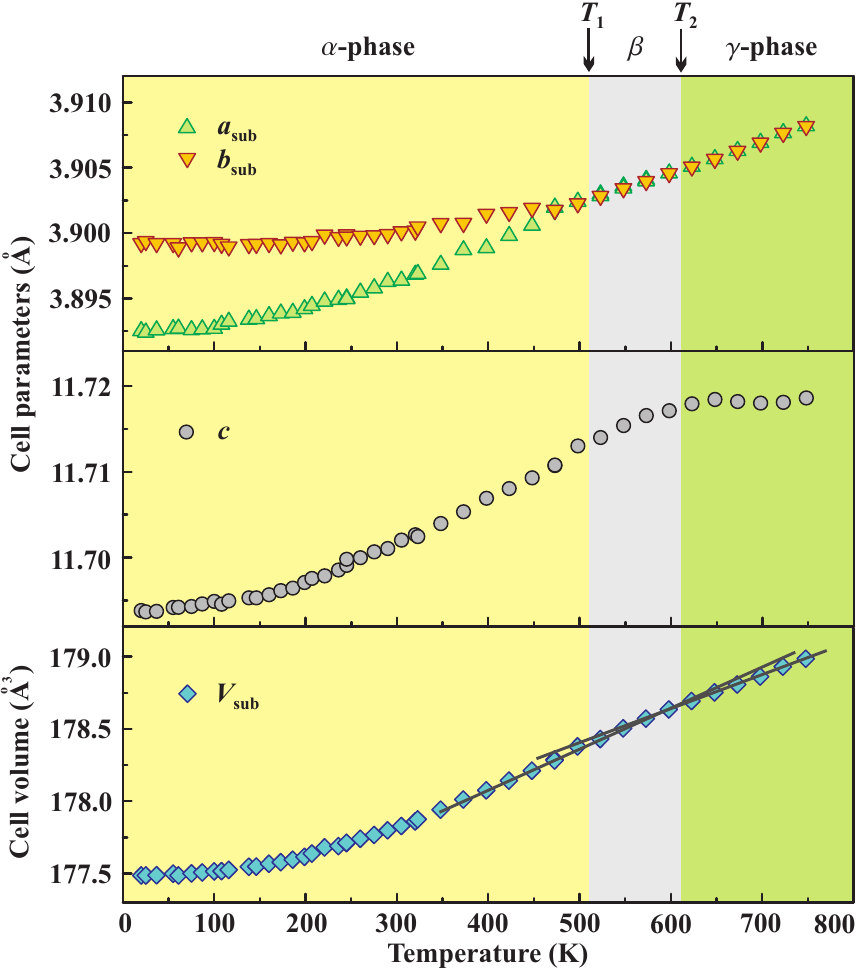}
\caption{\label{fig:evolution}
(Color online) Temperature evolution of subcell parameters and subcell volume for (CuBr)LaNb$_2$O$_7$. Lines are guide for the eye.
}
\end{figure}

Upon heating, the weak orthorhombic distortion present in (CuBr)LaNb$_2$O$_7$ is gradually reduced and eventually disappears around $T_1$ (Fig.~\ref{fig:evolution}). Unfortunately, this effect can not be tracked precisely because of the sizable reflection broadening that prevented us from the independent refinement of the $a$ and $b$ parameters above 450~K. Above $T_2$, the $c$ parameter becomes temperature-independent, whereas the temperature evolution of the cell volume shows a slight change in the slope. DSC does not reveal any anomalies at $T_1$ and $T_2$, because the structural changes are weak and associated with only marginal changes in the entropy. TGA confirms the stability of (CuBr)LaNb$_2$O$_7$ within the studied temperature range and shows the onset of decomposition (weight loss) above 750~K, only.\cite{supplement}

Above $T_2$, the crystal structure of (CuBr)LaNb$_2$O$_7$ can be refined in a tetragonal $a_{\sub}\times a_{\sub}\times c_{\sub}$ unit cell ($\gamma$-polymorph).\footnote{Although the very weak diffuse scattering at $hkl$ with odd $h+k$ and integer $l$ is still present above $T_2$ (e.g., the broad feature at $2\theta\simeq 10.5$~deg in Fig.~\ref{fig:xrd}), it can not be taken into account in the conventional Rietveld refinement and should be considered in future neutron and/or single-crystal studies.} The regular [LaNb$_2$O$_7$] slabs have the four-fold symmetry, while the Cu atoms are disordered over four equivalent positions in the $z=0$ plane, and the Br atoms are disordered over eight equivalent positions lying above and below this plane (Table~\ref{tab:str-3}).\cite{supplement} The displacements of Cu and Br atoms ensure the formation of short Cu--Br bonds that preserve local Jahn-Teller distortions for Cu atoms. The out-of-plane displacement of Br [$\Delta z_{\text{Br}}=z_{\text{Br}}/c=0.0189(5)$] is notably increased compared to $\Delta z_{\text{Br}}\simeq 0.007(1)$ at low temperatures. This feature is clearly different from $\gamma$-(CuCl)LaNb$_2$O$_7$, where Cl atoms remain in the mirror plane and have a larger ADP ($U_{\text{iso}}=0.047(2)$~\r A$^2$) with a stronger tendency to the in-plane displacements.\footnote{The refinement of the XRD data presented in Ref.~\onlinecite{tsirlin2010a} results in $U_{11}=0.048(6)$~\r A$^2$, $U_{22}=0.024(5)$~\r A$^2$, and $U_{33}=0.03(1)$~\r A$^2$ for Cl atoms at 660~K.} 

\begin{figure}
\includegraphics{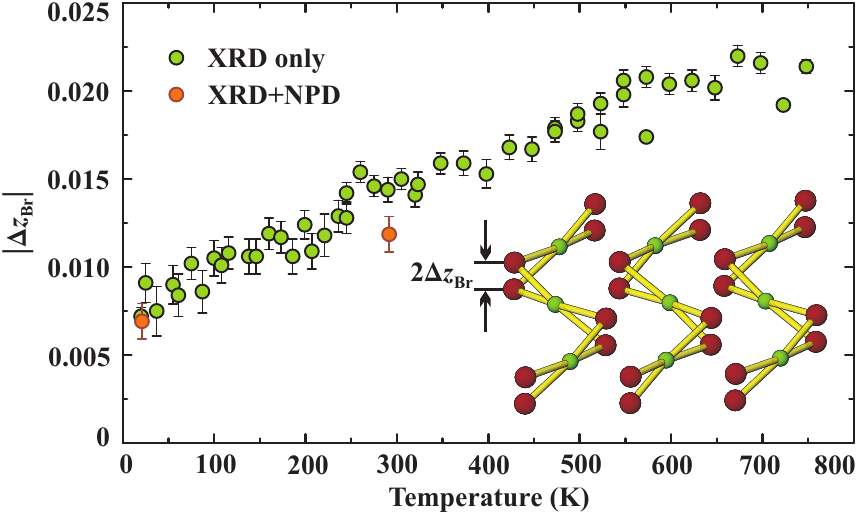}
\caption{\label{fig:Br}
(Color online) Temperature-dependent out-of-plane displacement of the Br atoms ($\Delta z_{\text{Br}}$), as obtained from the refinement of the high-resolution XRD data and the joint refinement of x-ray and neutron data (XRD+NPD). The inset shows the [CuBr] ribbons with the split position of Br above and below the mirror plane at $z=0$.
}
\end{figure}

\begin{figure}
\includegraphics{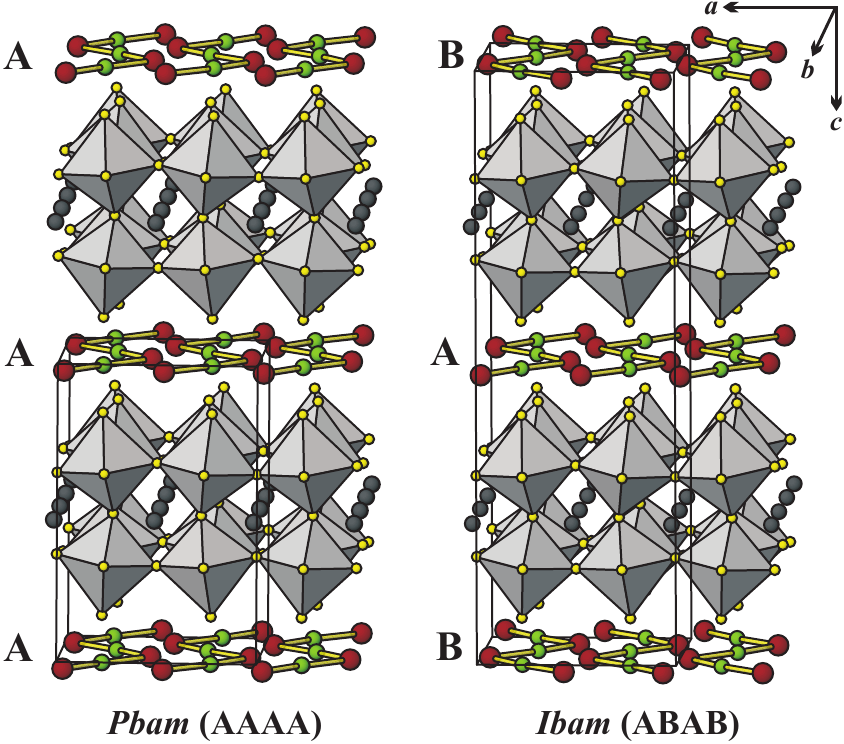}
\caption{\label{fig:models}
(Color online) Structural models representing the short-range order of Br atoms in (CuBr)LaNb$_2$O$_7$. The prevailing $Ibam$ model (right panel) results in the doubling of the $c$ lattice parameter and manifests itself by diffuse scattering in the x-ray, neutron, and electron diffraction data.
}
\end{figure}

The out-of-plane Br displacements systematically increase with temperature (Fig.~\ref{fig:Br}), as confirmed by Rietveld refinements of the XRD data and the joint refinement of the x-ray and neutron data at room temperature (Table~\ref{tab:str-2}). The three-fold increase in $\Delta z_{\text{Br}}$ upon heating shows that the displacements are largely driven by dynamic effects, such as soft phonon modes. However, at low temperatures the sizable displacements are still present and manifest one of the differences between (CuBr)LaNb$_2$O$_7$ and its Cl-containing counterpart.

The smooth temperature evolution of lattice parameters and cell volume (Fig.~\ref{sec:evolution}) suggests that the phase transitions at $T_1$ and $T_2$ are of second order, similar to (CuCl)LaNb$_2$O$_7$ (Ref.~\onlinecite{tsirlin2010a}).

\subsection{Energetics}
\label{sec:energetics}
We will now consider the crystal structure of (CuBr)LaNb$_2$O$_7$ from a microscopic viewpoint, and compare total energies for different arrangements of Br atoms. First, we investigate two ordered models (Fig.~\ref{fig:models}) that are consistent with the experimental diffraction data reported in Sections~\ref{sec:lowT} and~\ref{sec:short-range}. One model represents the $Pbam$ structure of $\alpha$-(CuCl)LaNb$_2$O$_7$ type with the $c_{\sub}$ lattice parameter and the equivalent arrangement of [CuBr] ribbons along the $c$ direction (AAAA). The other model features the $Ibam$ symmetry and the doubled lattice parameter along $c$ because of the alternating arrangement of the [CuBr] ribbons (ABAB). Both models were optimized\footnote{%
Following Ref.~\onlinecite{tsirlin2010a}, we performed structure optimization using GGA+$U$. Although the LDA+$U$ functional delivers similar structural models, it underestimates the Cu--Br distances and leads to less accurate structural data. Relaxed structures are nearly independent of the $U_d$ and $J_d$ parameters as well as the double-counting-correction scheme.} 
until residual forces dropped below 0.01~eV/\r A. The ensuing interatomic distances and angles are listed in Table~\ref{tab:dist} along with the experimental results for the low-temperature crystal structure from Table~\ref{tab:str-1}.

The optimized crystal structures show remarkable agreement with the experimental data, thus suggesting an excellent potential of DFT+$U$ for structure prediction. The different arrangement of the [CuBr] ribbons in the $Pbam$ and $Ibam$ models has no appreciable influence on individual interatomic distances and angles. Indeed, both structures have similar energy, with a slight preference for the $Ibam$ model ($\sim 2$~meV/f.u.) in agreement with the experimental data (Sec.~\ref{sec:short-range}). The marginal energy difference between the $Pbam$ and $Ibam$ structures is a natural explanation for the incomplete structural order in (CuBr)LaNb$_2$O$_7$. 

\begin{figure}
\includegraphics{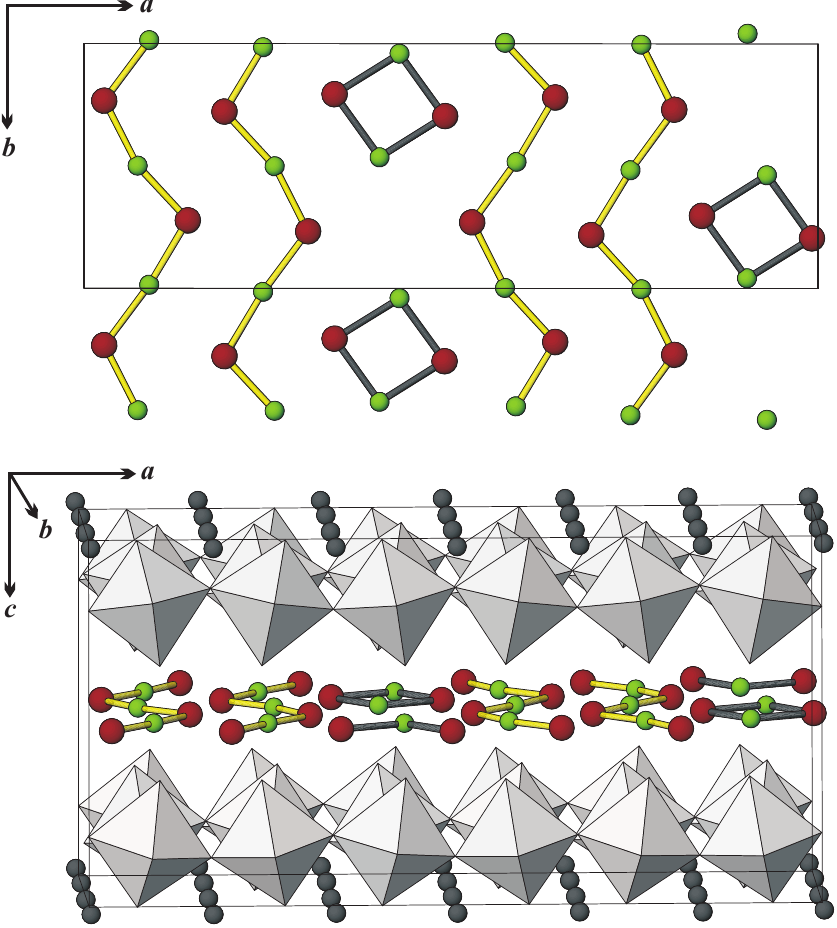}
\caption{\label{fig:antiphase-2}
(Color online) Sketch of the possible disorder in the $ab$ plane: the A- and B-type [CuBr] ribbons coexist within the same layer of the (CuBr)LaNb$_2$O$_7$ structure. Light and dark lines show the short Cu--Br bonds in the \textit{trans}- and \textit{cis}-configurations, respectively. Note that the change in the configuration of the ribbons does not disrupt the tilting distortion in the [LaNb$_2$O$_7$] slabs (bottom panel).
}
\end{figure}

\begin{table}
\caption{\label{tab:dist}
Experimental (at $T=20$~K) and relaxed interatomic distances (in~\r A) and angles (in~deg) in the (CuBr)LaNb$_2$O$_7$ structure. The structures were relaxed using GGA+$U$ with $U_d=5$~eV. Note that the experimental structure features a split position of Br, hence the number of Cu--Br distances is doubled.}
\begin{ruledtabular}
\begin{tabular}{lrrr}
              &  Experiment &  Relaxation & Relaxation \\
  Space group &   $Cmmm$    &    $Pbam$   &  $Ibam$    \\
  Cu--O3      &  1.866(1)   &  1.848      &  1.848     \\
  Cu--Br      &  $2\times 2.490(3)$   &  2.517      &  2.516     \\
  Cu--Br      &  $2\times 2.550(3)$   &  2.545      &  2.543     \\
  Nb--O1      &  1.981(2)   &  1.968      &  1.968     \\
  Nb--O2      &  1.994(1)   &  2.000      &  2.005     \\
  Nb--O2      &  1.994(1)   &  2.005      &  2.001     \\
  Nb--O3      &  1.766(1)   &  1.788      &  1.788     \\
  Nb--O4      &  2.239(1)   &  2.271      &  2.271     \\
  Nb--O5      &  2.015(2)   &  2.046      &  2.046     \\
  Cu--Br--Cu  &  101.8(2)   &  101.6      &  101.7     \\
\end{tabular}
\end{ruledtabular}
\end{table}
For the sake of completeness, we also estimated the energies for other possible configurations of Br atoms. The structures with the \textit{cis}-arrangement of short bonds, as depicted in Fig.~\ref{fig:antiphase}, appeared to be highly unstable and converged to the structure with the \textit{trans}-arrangement. However, the \textit{cis}-arrangement of short bonds can be stabilized as a defect in the regular structure with Cu atoms featuring a more stable \textit{trans}-arrangement. To introduce such a defect, we used the unit cell doubled or tripled along the $a$ direction, and created inequivalent [CuBr] ribbons. Fig.~\ref{fig:antiphase-2} depicts the fully relaxed structure with the \textit{cis}-arrangement of Cu--Br bonds on the interface. Although such structures are metastable and form local minima on the potential energy surface, they are highly unfavorable with respect to the $Pbam$ and $Ibam$ structures shown in Fig.~\ref{fig:models}. The relaxations performed for the doubled and tripled unit cells show excess energies of 0.158~eV/f.u. and 0.125~eV/f.u., respectively, thus yielding the energy cost of $8\times 0.158/4$~eV and $12\times 0.125/4$~eV or $0.30-0.35$~eV for each Cu atom with the \textit{cis}-arrangement of short bonds. This energy cost is significantly higher than the thermal energy of 0.055~eV available at the preparation temperature of $600-650$~K. Therefore, the formation of Cu atoms with the \textit{cis}-configuration of short bonds is extremely unlikely.

Our structure relaxations show that different arrangements of Br atoms have little influence on the tilting distortion developed by the [LaNb$_2$O$_7$] perovskite slabs. The formation of defects in the [CuBr] layers (Fig.~\ref{fig:antiphase-2}) does not disrupt the uniform $a^0b^-c^0$ tilting pattern. This illustrates the remarkable independence of the Br displacements and the tilting distortion that set in upon two different phase transitions and remain essentially decoupled.

\begin{figure}
\includegraphics{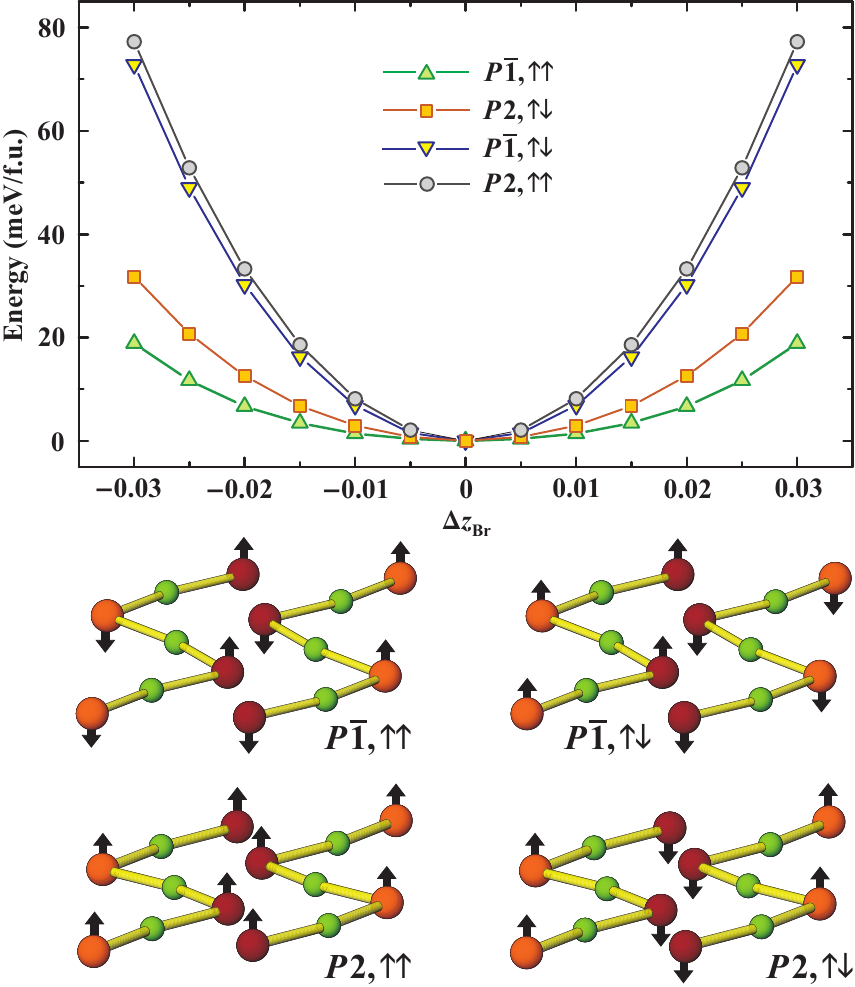}
\caption{\label{fig:Br-energy}
(Color online) Top panel: relative energies for different out-of-plane displacements of Br atoms, with all other atomic positions fully optimized. Bottom panel: sketch of the Br displacements, see text for notation details. Different colors show the Br$_a$ and Br$_b$ sites, as explained in the text.
}
\end{figure}
Finally, we examined the possible displacements of Br atoms out of the $z=0$ plane. Structures with the out-of-plane displacements lack the $m_z$ mirror plane and have either $P2$ ($P112$) or $P\bar 1$ symmetries. Both space groups feature two inequivalent Br positions, Br$_a$ and Br$_b$, and allow for two different out-of-plane displacements each. The structures with Br$_a$ and Br$_b$ displaced in the same/opposite directions with respect to the $z=0$ plane are denoted $\uparrow\uparrow$ and $\uparrow\downarrow$, respectively. Altogether, four different starting configurations are possible (see the bottom panel of Fig.~\ref{fig:Br-energy}), but all of them relax to the same $Pbam$ (or $Ibam$, depending on the chosen ordering type) structure with $z_{\text{Br}}=0$ and the mirror-plane symmetry restored. The lowest energy of the $Pbam$ and $Ibam$ structures can also be seen from the upper panel of Fig.~\ref{fig:Br-energy}, where total energies for fully relaxed structures with fixed Br displacements are presented.\footnote{In the $P2$ space group, we also fixed the $z$ coordinate for one of the Cu sites. Otherwise, the origin can be arbitrarily shifted along the $c$ direction.} 

The out-of-plane Br displacements are not equivalent in terms of the energy (Fig.~\ref{fig:Br-energy}). The displacements require lower energies in the $P\bar1$ $\uparrow\uparrow$ and $P2$ $\uparrow\downarrow$ structures, where the two Br atoms of the same CuO$_2$Br$_2$ plaquette are shifted in the opposite directions. This way, the whole plaquette is slightly tilted, while its overall shape is preserved. The displacements in the $P\bar1$ $\uparrow\downarrow$ and $P2$ $\uparrow\uparrow$ structures require much higher energies, because the shape of the plaquette is changed. Finally, the preference of the $P\bar1$ $\uparrow\uparrow$ structure compared to the $P2$ $\uparrow\downarrow$ structure can be explained by the increase in the Br--Br distances between the negatively charged Br ions of neighboring [CuBr] ribbons. 

The out-of-plane displacements require relatively low energies even for the unfavorable configurations $P\bar 1$ $\uparrow\downarrow$ and $P2$ $\uparrow\uparrow$. For example, the thermal energy of about 40~meV at 500~K is sufficient to induce random displacements with $\Delta z_{\text{Br}}=\pm0.02$. This estimate compares well to the experimental displacement of $\Delta z_{\text{Br}}\simeq 0.02$ observed at high temperatures (Fig.~\ref{fig:Br}). Our microscopic study confirms the dynamic nature of the out-of-plane Br displacements. At low temperatures, the displacements should be eliminated to restore the mirror-plane symmetry of the equilibrium crystal structure. The residual out-of-plane displacements observed experimentally at low temperatures are a non-equilibrium feature, presumably related to defects and/or incomplete ordering of Br atoms.

\section{Microscopic magnetic model}
\label{sec:model}

\subsection{Electronic structure}
\label{sec:dft}
To evaluate the microscopic magnetic model, we calculate the band structure and evaluate individual exchange couplings. Our results are based on the fully relaxed crystal structures with $Pbam$ or $Ibam$ symmetry (Fig.~\ref{fig:models}). Since interatomic distances and angles in these structures are nearly indistinguishable (see Table~\ref{tab:dist}), the ensuing exchange couplings are also the same. Reference calculations for ordered structures based on the experimental refinement (neglecting the out-of-plane displacements of Br atoms) produced similar results for individual exchange couplings.

\begin{figure}
\includegraphics{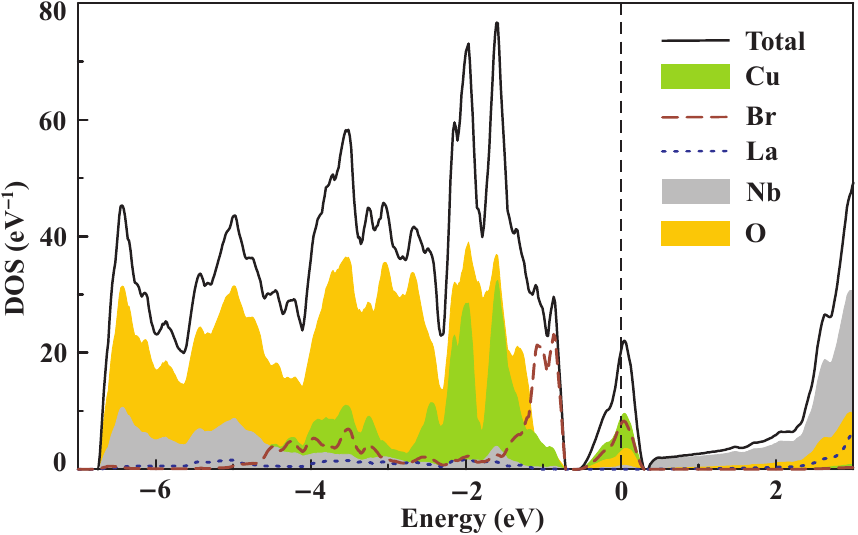}
\caption{\label{fig:dos}
(Color online) LDA density of states for the low-temperature structure of (CuBr)LaNb$_2$O$_7$. The Fermi level is at zero energy.
}
\end{figure}
The LDA band structures of (CuBr)LaNb$_2$O$_7$ and (CuCl)LaNb$_2$O$_7$ are rather similar (compare Fig.~\ref{fig:dos} to Fig.~2 of Ref.~\onlinecite{tsirlin2010b}). Oxygen $2p$ bands span the energy range between $-7$ and $-1$~eV, with Cu $3d$ bands above $-4$~eV and Br $4p$ bands above $-1.5$~eV. The states at the Fermi level are represented by four narrow bands according to four Cu atoms in the unit cell. These bands are formed by Cu $d_{x^2-y^2}$ orbitals with an appreciable admixture of the Br $4p$ and O $2p$ orbitals (here, $x$ and $y$ are directed along the Cu--O and short Cu--Br bonds, respectively). The sizable energy separation between Cu $d_{x^2-y^2}$ and the rest of Cu $3d$ states is driven by the large crystal-field splitting that illustrates the Jahn-Teller distortion underlying the formation of short \mbox{Cu--Br} bonds. According to Ref.~\onlinecite{tsirlin2009}, the simple tetragonal crystal structure from Ref.~\onlinecite{koden1999} entails Cu atoms with four equivalent Cu--Br bonds. The respective orbital state is nearly degenerate, because the two $e_g$ orbitals have similar energies in the effectively octahedral CuO$_2$Br$_4$ coordination. The displacements of Br atoms off the four-fold rotation axis and the ensuing shortening of the two Cu--Br bonds stabilize the non-degenerate orbital configuration. In (CuBr)LaNb$_2$O$_7$, the contribution of halogen $p$ states at the Fermi level is slightly larger than in the Cl compound, 36~\% and 33~\%, respectively. This trend follows the higher energy of Br $4p$ compared to the Cl $3p$ orbitals.

\begin{figure}
\includegraphics{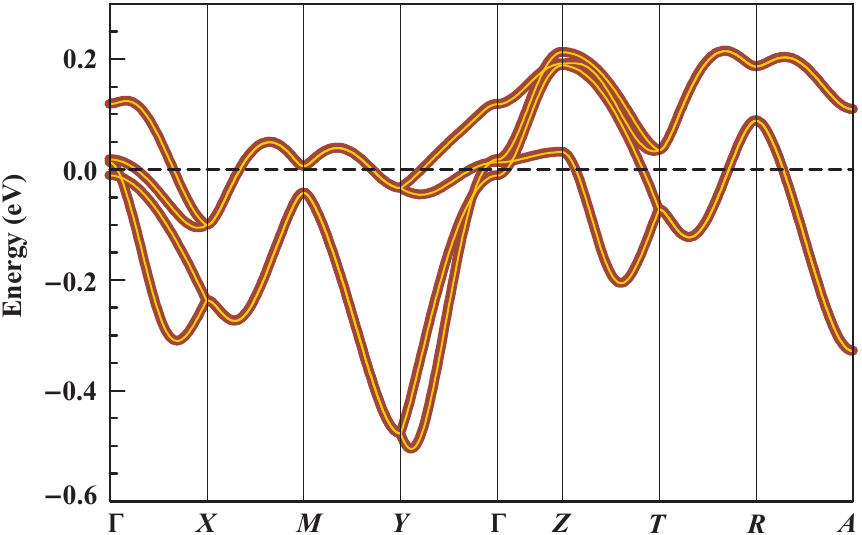}
\caption{\label{fig:bands}
(Color online) LDA band structure in the vicinity of the Fermi level (thin light lines) and the fit with the tight-binding model (thick dark lines). The Fermi level is at zero energy. The notation of $k$ points is as follows: $\Gamma(0,0,0)$, $X(\frac12,0,0)$, $M(\frac12,\frac12,0)$, $Y(0,\frac12,0)$, $Z(0,0,\frac12)$, $T(\frac12,0,\frac12)$, $R(\frac12,\frac12,\frac12)$, $A(0,\frac12,\frac12)$.
}
\end{figure}
The dispersions of the Cu $d_{x^2-y^2}$ bands identify possible electron transfers and magnitudes of AFM interactions in the system (Fig.~\ref{fig:bands}). The electron transfers ($t_i$) are quantified in a tight-binding model based on Wannier functions with the $d_{x^2-y^2}$ orbital character.\cite{wannier} We further supply this tight-binging model with a Hubbard term to account for an effective on-site Coulomb repulsion $U_{\eff}$, and reduce the problem to the strongly correlated limit $t_i\ll U_{\eff}$. This way, one arrives at the Heisenberg model with AFM exchanges $J_i^{\AFM}=4t_i^2/U_{\eff}$ for low-lying excitations. We use $U_{\eff}=4$~eV, according to previous studies of Cu$^{+2}$ halides and oxyhalides.\cite{tsirlin2010b,schmitt2009} 

\begin{figure}
\includegraphics{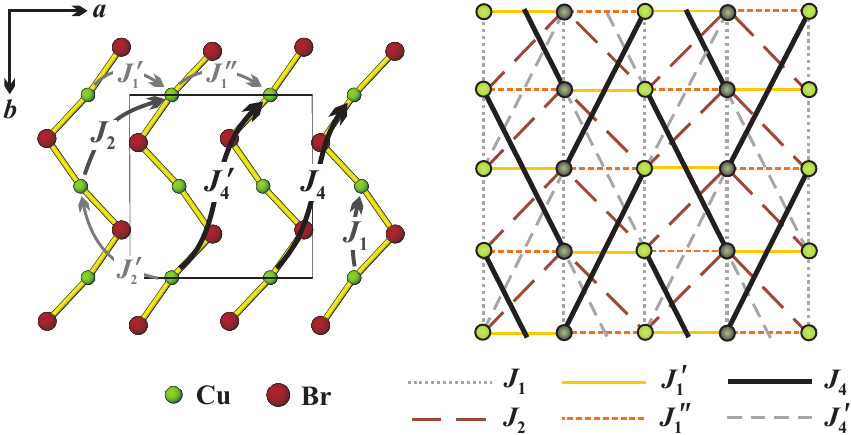}
\caption{\label{fig:lattice}
(Color online) Magnetic couplings in the $ab$ plane (left panel) and the respective spin lattice (right panel). Light and dark spheres denote up and down spins in the ordered antiferromagnetic state, as determined experimentally (Sec.~\ref{sec:magstructure}). The couplings $J_2'$, $J_5$, and $J_5'$ are not shown. The interlayer coupling $J_{\perp}$ runs along the $c$ direction.
}
\end{figure}

\begin{table}
\caption{\label{tab:exchange}
Magnetic couplings in (CuBr)LaNb$_2$O$_7$: Cu--Cu distances, hopping parameters $t_i$ of the tight-binding model together with the ensuing AFM exchange couplings $J_i^{\AFM}=4t_i^2/U_{\eff}$, and full exchange couplings $J_i$ evaluated using DFT+$U$ calculations with the AMF and FLL double-counting correction schemes. See text for details.
}
\begin{ruledtabular}
\begin{tabular}{ccrrrr}
        & Distance & \multicolumn{1}{c}{$t_i$} & $J_i^{\AFM}$ & $J_i$ (AMF) & $J_i$ (FLL) \\
        & (\r A)   & (eV)     &     (K)      &   (K)    &   (K)    \\
  $J_1$ & 3.912    & 0        &      0       &   $-75$  &   $-47$  \\
 $J_1'$ & 3.575    & 0.045    &     24       &      31  &    7     \\
$J_1''$ & 4.211    & 0.034    &     13       &       1  &    4     \\
  $J_2$ & 5.510    & $-0.047$ &     26       &      58  &   12     \\
 $J_2'$ & 5.510    & 0.011    &      1       &    $-4$  &   $-14$  \\
  $J_3$ & 7.786    & $-0.004$ &      0       &      --  &   --     \\
 $J_3'$ & 7.798    & 0.008    &      1       &      --  &   --     \\
  $J_4$ & 8.579    & $-0.097$ &    110       &     144  &   54     \\
 $J_4'$ & 8.862    & $-0.036$ &     15       &      37  &   17     \\
  $J_5$ & 11.020   & 0.011    &      1       &      --  &   --     \\
 $J_5'$ & 11.020   & 0.035    &     14       &   13$^a$ &   1$^a$  \\\footnotetext{Here, only the sum $J_5+J_5'$ was evaluated.}
$J_{\perp}$ & 11.694 & $-0.028$ &    9       &      17  &    6     \\
\end{tabular}
\end{ruledtabular}
\end{table}
The hoppings of the tight-binding model and the resulting AFM exchange couplings are listed in Table~\ref{tab:exchange}. Following Ref.~\onlinecite{tsirlin2010b}, we denote the couplings in the $ab$ plane according to the Cu--Cu distances ($J_1,J_1'$, and $J_1''$ between nearest neighbors, $J_2$ and $J_2'$ between second neighbors, etc.) and use $J_{\perp}$ for the interlayer coupling along $c$ (Fig.~\ref{fig:lattice}). At first glance, the microscopic scenario is somewhat unusual, because the leading coupling $J_4$ runs between fourth neighbors, while the nearest-neighbor and next-nearest-neighbor couplings are relatively weak. This peculiar behavior is rationalized in terms of the magnetic $d_{x^2-y^2}$ orbital lying in the plane of the CuO$_2$Br$_2$ plaquette. The large spatial extent of Br $4p$ orbitals facilitates the long-range hopping between the fourth neighbors, but this effect is highly selective. Only the fourth-neighbor hoppings along the short Cu--Br bonds are enhanced (e.g., along $[210]$ and not along $[2\bar 10]$, or the other way around, depending on the orientation of the plaquette). Moreover, the $t_4$ path is twice more efficient than its $t_4'$ counterpart. This can be explained by the different Cu--Br--Br angles of 156.4~deg and 144.7~deg, respectively. The larger angle renders a more straight and, therefore, more efficient superexchange pathway (see the left panel of Fig.~\ref{fig:lattice}). Note that the interlayer hopping $t_{\perp}$ is comparable to several short-range intralayer interactions, such as $t_1'$, $t_1''$, $t_2$, and $t_4'$. Similar to the Cu--Br--Br--Cu pathway for $J_4$, the long Cu--O--Nb--O--Nb--O--Cu pathway is also rather efficient despite the very long Cu--Cu distance of 11.69~\r A, because oxygen $2p$ orbitals entering the Wannier functions are directed along this pathway, while the low-lying Nb $4d$ states additionally facilitate the superexchange.

Altogether, the LDA-based microscopic model of (CuBr)LaNb$_2$O$_7$ strongly resembles the respective model for the isostructural Cl compound, where the leading coupling between fourth neighbors was first observed experimentally\cite{kageyama2005a} and later understood microscopically.\cite{tsirlin2009,tsirlin2010a,tassel2010} The main difference between the LDA-based models lies in the nearest-neighbor couplings $J_1'$ and $J_1''$ that are sizable in the Br compound, yet rather weak in its Cl counterpart. 

An important deficiency of the LDA-based approach is the neglect of FM contributions that are typical for short superexchange pathways. To evaluate the full exchange integrals ($J_i$), we turn to DFT+$U$ calculations and list representative results in the last two columns of Table~\ref{tab:exchange}, where we used certain values of the on-site Coulomb repulsion parameter $U_d$, depending on the double-counting correction applied in DFT+$U$: $U_d=5$~eV and 12~eV for the around-mean-field (AMF) and fully-localized-limit (FLL) options, respectively. The use of a smaller $U_d$ in AMF and a larger $U_d$ in FLL is a rather general, although empirical, practice to ensure similar results obtained in the two approaches.\cite{tsirlin2010b,tsirlin2010c,tsirlin2010d} The specific choice of $U_d$ is a more subtle issue, because the ratios and, particularly, the energy scale of the couplings are strongly dependent on $U_d$. Following Ref.~\onlinecite{tsirlin2010b}, we adjust the $U_d$ values to experimental quantities, the Curie-Weiss temperature $\theta\simeq 5$~K that is a linear combination of all exchange couplings, and the saturation field $\mu_0H_s\simeq 70-85$~T that measures AFM couplings in the system (see Ref.~\onlinecite{oba2006} and Sec.~\ref{sec:simul}). However, the precise adjustment to both $\theta$ and $H_s$ with the single value of $U_d$ is not possible, and uncertainties in the computed exchange couplings remain. The problem is probably unavoidable, owing to the complex crystal structure, large number of nonequivalent interactions, and their low energy scale.

Combining the LDA-based evaluation of $J_i^{\AFM}$ and the calculation of $J_i$ within DFT+$U$, we arrive at a qualitative microscopic model of (CuBr)LaNb$_2$O$_7$. The main features of this model are: i) leading fourth-neighbor coupling $J_4$ running through the particular, least curved Cu--Br--Br--Cu pathway; ii) large FM coupling between nearest neighbors along $b$: $J_1/J_4\simeq -0.5$; iii) sizable interlayer coupling $J_{\perp}/J_4\simeq 0.3$. Further details of the model remain ambiguous, and probably even obscure. The leading couplings $J_1$ and $J_4$ establish a stripe AFM order with parallel spins along $b$ and antiparallel spins along $a$ (see the right panel of Fig.~\ref{fig:lattice}). While weaker AFM couplings $J_1,J_1'',J_2$, and $J_4'$ support this ordering pattern, the possibly FM $J_2'$ and the small AFM $J_5$ frustrate the spin lattice. The relevance of these couplings is rather difficult to estimate, since their values depend on the computational procedure. For example, the sizable FM $J_2'$ appears in the FLL calculations only, while the AMF calculation results in a negligibly small $J_2'$. Unfortunately, the precise evaluation of such small couplings lies beyond the accuracy of present-day methods. Therefore, we further consider the problem of frustration using phenomenological arguments in Secs.~\ref{sec:simul} and~\ref{sec:discussion}.

\subsection{Magnetic structure}
\label{sec:magstructure}
Oba \textit{et~al.}\cite{oba2006} have claimed the stripe AFM ordering in (CuBr)LaNb$_2$O$_7$, based on the observation of two magnetic reflections in a neutron diffraction experiment. They also reported the ordered magnetic moment of $\mu\simeq 0.60$~$\mu_B$ in surprisingly good agreement with the naive square-lattice model, which is, however, inconsistent with the actual orthorhombic symmetry of $\alpha$-(CuBr)LaNb$_2$O$_7$. We have revisited experimental information on the magnetic structure, because the powder data collected with the high-intensity D20 diffractometer allow to observe a larger number of magnetic reflections. Additionally, the refined value of $\mu$ depends on the scale factor for the nuclear scattering and, therefore, should be reconsidered for the revised structural model.

\begin{figure}
\includegraphics{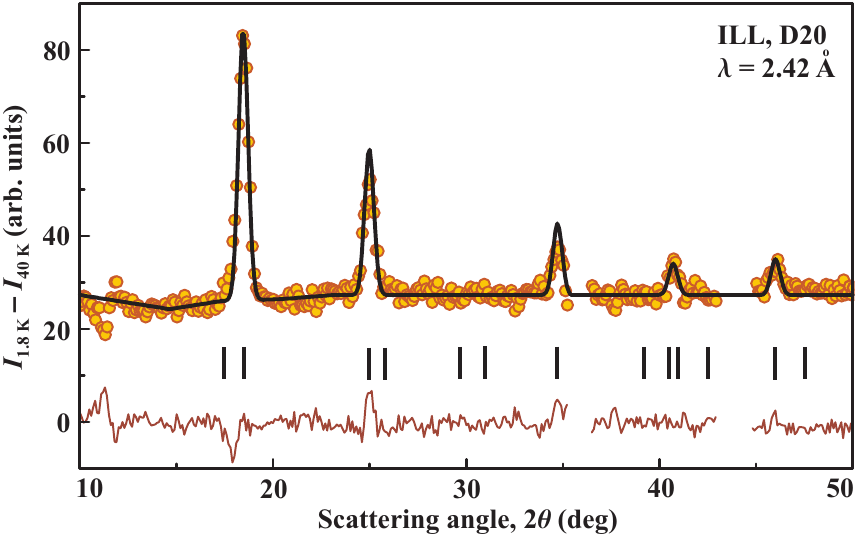}
\caption{\label{fig:magstructure}
(Color online) Refinement of the magnetic structure for the subtracted neutron data ($I_{\text{1.8 K}}-I_{\text{40 K}}$): experimental (circles), calculated (dark line), and difference (light line) patterns. Ticks show positions of the magnetic reflections. The excluded regions around $2\theta=36$~deg and 44~deg contain wiggles due to the incomplete subtraction of the 200/003 and 202 nuclear reflections. Both subtracted and calculated patterns are offset from zero to ensure positive intensities.
}
\end{figure}
Upon cooling below $T_N\simeq 32$~K, at least two weak magnetic reflections were observed (Fig.~\ref{fig:neutron}). To refine the magnetic structure, we subtracted the 40~K pattern measured right above $T_N$ from the low-temperature pattern collected at 1.8~K. Since lattice parameters of (CuBr)LaNb$_2$O$_7$ are nearly unchanged below 40~K (see Fig.~\ref{fig:evolution}), the subtraction effectively eliminates the nuclear scattering and reveals a number of additional, very weak magnetic reflections at higher angles (Fig.~\ref{fig:magstructure}). The subtracted pattern is refined as a purely magnetic phase. 

The magnetic reflections can be indexed by a $\mathbf k=(0,0,\frac12)$ propagation vector that is consistent with $\mathbf k=(0,\frac12,\frac12)$ reported by Oba \textit{et~al.}\cite{oba2006} for the four times smaller $a_{\sub}\times a_{\sub}\times c_{\sub}$ tetragonal subcell. The $4h$ position of Cu and the $\mathbf k=(0,0,\frac12)$ allow for a number of irreducible representations, but only one of them leads to a complete refinement of the difference pattern. This representation corresponds to the stripe AFM structure, with the FM order of spins along either $a$ or $b$. The spins follow the same direction, i.e., the $a$ direction for the FM order along $a$ and the $b$ direction for the FM order along $b$. This magnetic structure is in agreement with the results reported by Oba \textit{et al.}\cite{oba2006} The refined magnetic moment at 1.8~K equals to 0.72(1)~$\mu_B$, which is slightly larger than the refined value of 0.60~$\mu_B$ reported in Ref.~\onlinecite{oba2006}. 

\begin{figure}
\includegraphics{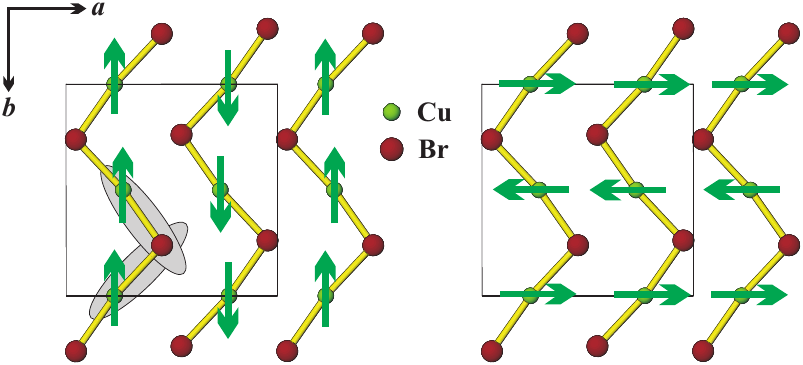}
\caption{\label{fig:hyperfine}
(Color online) Possible magnetic structures with stripes of parallel spins directed along $b$ (left panel) and $a$ (right panel). Only the left structure is consistent with the NMR data, because Br atoms are strongly bonded to two Cu atoms with same direction of spin, thus featuring a large hyperfine field observed experimentally.
}
\end{figure}

The stripe pattern (Fig.~\ref{fig:magstructure}) features antiparallel spins along one of the intralayer direction ($a$ or $b$) and along the interlayer direction $c$. The tiny difference between the $a$ and $b$ lattice parameters, as well as the limited resolution of the experiment do not allow us to determine the direction of stripes experimentally (Fig.~\ref{fig:hyperfine}). However, the microscopic magnetic model shows clearly that the coupling along the [CuBr] zigzag ribbons ($b$ direction) is FM, while the coupling along $a$ is AFM (Table~\ref{tab:exchange}). Therefore, we conclude that stripes of parallel spins run along the $b$ direction. 

This conclusion is further confirmed by NMR results showing a large hyperfine field of 16.4~T at the Br site.\cite{yoshida2008} According to the structural models shown in Fig.~\ref{fig:models}, the Br atoms are surrounded by four Cu atoms, with shorter Cu--Br distances of about 2.5~\r A to the Cu atoms of the same [CuBr] ribbon, and longer distances of more than 2.9~\r A to the Cu atoms of the neighboring ribbon. If stripes of parallel spins run along $a$, each Br atom will be strongly coupled to the spin-up and spin-down Cu atoms of the same ribbon, so that the hyperfine field on the Br site is strongly reduced or even canceled (Fig.~\ref{fig:hyperfine}, right). By contrast, stripes of parallel spins running along the $b$ direction (left panel of Fig.~\ref{fig:hyperfine}) ensure the strong hyperfine coupling of Br to the two closest Cu atoms with the same spin, and the ensuing large hyperfine field observed experimentally.
\subsection{Model simulations}
\label{sec:simul}
We will now consider experimental information on the magnetic behavior of (CuBr)LaNb$_2$O$_7$ and discuss the dimensionality of the underlying spin lattice as well as its possible frustration. The DFT results (Table~\ref{tab:exchange}) show clearly that the spin lattice is three-dimensional (3D), because sizable couplings are present in the $ab$ plane and along the $c$ direction ($J_{\perp}$). The 3D nature of the spin lattice severely restricts the set of simulation techniques that could be applied to this system. Particularly, a finite lattice of sufficient size can not be treated with exact diagonalization to perform a realistic simulation of a frustrated 3D magnet. The method of choice is QMC that allows to handle large finite lattices, yet fails for frustrated systems at low temperatures because of the sign problem. To perform QMC simulations, we neglect the potentially frustrating couplings $J_2'$, $J_5$, and $J_5'$, and consider the remaining interactions $J_1,J_1',J_1'',J_2,J_4,J_4'$, and $J_{\perp}$. The reduction to the non-frustrated spin lattice is justified phenomenologically by the high N\'eel temperature ($T_N/J_4\simeq 0.67$, see below) and the sizable ordered moment of 0.72~$\mu_B$, which is approaching the classical value of 1~$\mu_B$ for spin-$\frac12$ (see the discussion in the end of this section). \textit{A posteriori}, a remarkable agreement between the experimental data and simulations for the non-frustrated spin lattice also disfavors frustration in (CuBr)LaNb$_2$O$_7$.

\begin{figure}
\includegraphics{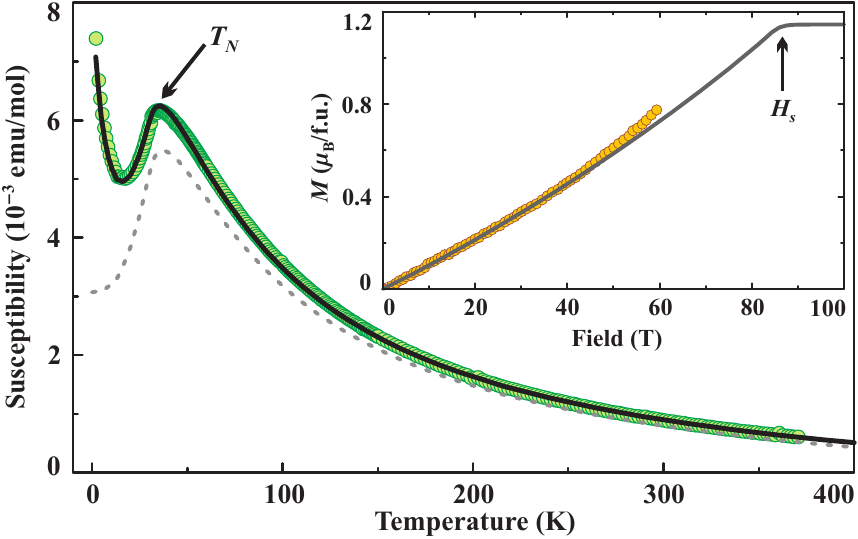}
\caption{\label{fig:fits}
(Color online) Magnetic susceptibility of (CuBr)LaNb$_2$O$_7$ measured in the applied field of 0.5~T (circles) and the fit with $J_4=48$~K, $J_1/J_4=-0.6$, $J_2/J_4=0.25$, and $J_1'=J_1''=J_2=J_4'=J_{\perp}$ (solid line), see text for details. The dashed line shows the intrinsic susceptibility of (CuBr)LaNb$_2$O$_7$ without the Curie-Weiss impurity contribution $C_{\imp}/(T+\theta_{\imp})$. The inset depicts an experimental magnetization isotherm measured at 1.3~K (Ref.~\onlinecite{oba2006}, circles) and the fit with the same model parameters (solid line).
}
\end{figure}
The spin lattice of (CuBr)LaNb$_2$O$_7$ features seven inequivalent couplings that render fits to the experimental data rather ambiguous. To reduce this ambiguity, we introduce the constraint $J_1'=J_1''=J_2=J_4'=J_{\perp}$ since all these couplings are of the same order (see Table~\ref{tab:exchange}), and thus arrive at three variable parameters, $J_4$, $J_1$, and $J_2$, as well as the $g$-value. Additionally, we use: i) the temperature-independent term $\chi_0$ to account for core diamagnetism and van Vleck paramagnetism contributing to the magnetic susceptibility; and ii) the Curie-Weiss term $C_{\imp}/(T+\theta_{\imp})$ to describe the low-temperature susceptibility upturn\cite{imai2008} that presumably originates from weakly coupled impurity spins.%
\footnote{Impurity spins typically arise from local defects and/or surface effects that are inevitable in powder samples, especially those prepared at low temperatures. The contribution of impurity spins is usually approximated by the Curie-Weiss law with a small Weiss constant $\theta_{\text{imp}}$, as shown, e.g., for the impurity spins in herbertsmithite (Ref.~\onlinecite{imai2008}).}
 The temperature dependence of the susceptibility and the magnetization isotherm are both fitted with $J_4=48(1)$~K, $J_1/J_4=-0.6(1)$, $J_2/J_4=0.25(5)$, $g=2.29(2)$, $\chi_0=-7.4(3)\times 10^{-4}$~emu/mol, $C_{\imp}=0.032(1)$~emu~K/mol, and $\theta_{\imp}=6(1)$~K (Fig.~\ref{fig:fits}).\footnote{%
We performed simulations for the finite lattice with $12\times 12\times 6$ sites and periodic boundary conditions. This lattice size is sufficient to avoid finite-size effects in the relevant temperature range. The magnetization curve was simulated at $T/J_4=0.03$ corresponding to 1.45~K vs. the experimental temperature of 1.3~K (Ref.~\onlinecite{oba2006}).} Note that this set of exchange couplings perfectly reproduces the magnetic ordering temperature $T_N\simeq 32$~K, and the $g$-value is in the typical range for Cu$^{+2}$ systems.\cite{banks2009,*janson2011,*povarov2011} Because the $g$-value exceeds 2.0 and the saturated magnetization is $M_s=gS\mu_B\simeq 1.15$~$\mu_B$/f.u. for $S=\frac12$, our estimate for the saturation field $\mu_0H_s\simeq 85$~T is somewhat larger than $\mu_0H_s\simeq 70$~T reported in Ref.~\onlinecite{oba2006}.

The fitted value of $C_{\imp}$ corresponds to about 8~wt.\% of an impurity that shows non-negligible AFM interactions with the energy scale of $\theta_{\imp}=6$~K. Since our XRD and neutron data do not reveal any crystalline impurities, the large Curie-Weiss term should be attributed to defects that are possibly related to random out-of-plane displacements of Br atoms. Presently, we have not studied the effect of such displacements microscopically, because even the model based on the equilibrium crystal structure is rather complex. The role of the out-of-plane displacements should be addressed in future studies.

Despite the large number of variable parameters in the fitting, their estimates arise from different features of the experimental data, and appear to be robust under the constraint $J_1'=J_1''=J_2=J_4'=J_{\perp}$ introduced for individual exchange couplings. For example, $C_{\imp}$ and $\theta_{\imp}$ are determined by the low-temperature part of the susceptibility that has a nearly temperature-independent intrinsic contribution of (CuBr)LaNb$_2$O$_7$ (see Fig.~\ref{fig:fits}). The values of $\chi_0$ and $g$ rest upon the high-temperature data, where the susceptibility follows the Curie-Weiss law. Finally, the overall energy scale given by a linear combination of AFM couplings is fixed by the saturation field (slope of the magnetization curve), and the shape of the susceptibility above $T_N$ determines $J_1/J_4$ and $J_2/J_4$. 

Once the constraint on the exchange couplings is released, the ambiguity appears. Similar to (CuCl)LaNb$_2$O$_7$, the couplings can be largely redistributed between the weaker AFM exchanges $J_1', J_1'', J_2$, and $J_4'$. Unfortunately, at the present stage this ambiguity is unavoidable, given the large number of inequivalent exchange couplings and the limited experimental data collected on powder samples. A further refinement of individual parameters would require inelastic neutron scattering measurements on single crystals that provide access to the whole spectrum of magnetic excitations.

We also evaluated the ordered magnetic moment and compared it to the experimental estimate (Sec.~\ref{sec:magstructure}). Following Ref.~\onlinecite{sandvik1997}, we calculated the staggered magnetization ($m_s$) for finite lattices of different size up to $24\times 24\times 12$ and performed finite-size scaling.\footnote{%
The inverse temperature was set to $\beta=(T/J_4)^{-1}=4L$ for the $2L\times 2L\times L$ finite lattices.} The $m_s$ values were obtained from the static structure factor taken at the propagation vector of the magnetic structure, and from the spin-spin correlation taken at the largest separation on the finite lattice. The two approaches yield the staggered magnetization of 0.879~$\mu_B$ and 0.880~$\mu_B$ and confirm the 3D nature of the spin system (compare to $m_s=0.60$~$\mu_B$ and 0.83~$\mu_B$ for the spin-$\frac12$ square lattice\cite{sandvik1997} and cubic lattice,\cite{[{QMC result. See also: }][{}]schmidt2002} respectively). To compare this result with the experimental $\mu$ determined by neutron diffraction, one has to scale $m_s$ with the $g$-value ($\mu=gSm_s$), and account for the Cu--Br hybridization that spreads the spin polarization to the ligand site, thus reducing the observed magnetic moment on Cu. Using $g=2.29$ and the DFT+$U$ results showing the $20-35$~\% reduction in the moment due to the hybridization,\footnote{%
More specifically, the magnetic moment on the Cu site is 0.64~$\mu_B$ in the AMF calculation with $U_d=5$~eV and 0.80~$\mu_B$ in the FLL calculation with $U_d=12$~eV, compared to 1~$\mu_B$ in a system without hybridization. Note that these numbers are obtained from DFT+$U$ calculations that miss both quantum effects and spin-orbit coupling. Therefore, neither the deviation of $g$ from the free-electron value, nor quantum fluctuations are taken into account.} we estimate the ordered moment $\mu=0.65-0.80$~$\mu_B$ in remarkable agreement with the experimental result of 0.72(1)~$\mu_B$.

\section{Discussion}
\label{sec:discussion}
Our study provides detailed information on the crystal structure and magnetic behavior of (CuBr)LaNb$_2$O$_7$. We will now compare (CuBr)LaNb$_2$O$_7$ to (CuCl)LaNb$_2$O$_7$, and discuss their structural differences as well as the origin of different magnetic ground states. In agreement with earlier expectations,\cite{yoshida2007,ren2010} we find that the Cl and Br compounds have similar crystal structures featuring a tilting distortion within the [LaNb$_2$O$_7$] perovskite slabs and ordered displacements of Cu and halogen atoms in the $ab$ plane. However, a salient feature of (CuBr)LaNb$_2$O$_7$ is the split Br position in the averaged crystal structure and an additional short-range order of Br resulting in the doubling of the $c$ lattice parameter. The Br atoms are also prone to out-of-plane displacements that become sizable at elevated temperatures (Fig.~\ref{fig:Br}). 

The out-of-plane displacements represent deviations from the equilibrium crystal structure and can be largely understood as a dynamic effect, because the magnitude of the displacement increases with temperature (Fig.~\ref{fig:Br}). These displacements involve a small change in the total energy of the system (Fig.~\ref{fig:Br-energy}), hence they are likely driven by soft phonon modes, although a comprehensive study of the phonon spectrum would be necessary to verify this conjecture. The splitting of the Br position is an intrinsic feature of (CuBr)LaNb$_2$O$_7$ driven by the small energy difference between two possible arrangements of the [CuBr] ribbons in the adjacent layers (Fig.~\ref{fig:models}). While the coordination preference of Cu atoms ensures similar arrangement of all ribbons in the $ab$ plane, the disorder between the adjacent planes has little influence on the structure and only slightly changes the energy. Therefore, diffraction data show the short-range order of Br evidenced by the diffuse scattering that reflects the preferential ABAB-type arrangement of the [CuBr] ribbons (Sec.~\ref{sec:short-range}).

The effect of split halogen position in the averaged crystal structure and the ensuing short-range structural order may be relevant for all compounds of the (CuCl)LaNb$_2$O$_7$ family. For example, single crystals of (CuCl)LaNb$_2$O$_7$ also revealed two positions of Cl atoms, although one of them was preferentially occupied (about 90~\% and 10~\% of Cl atoms, respectively).\cite{hernandez2011} The increase in the $c$ lattice parameter should further reduce the energy difference between the AAAA and ABAB configurations, thus randomizing the arrangement of the [CuX] (X = Cl, Br) zigzag ribbons in adjacent planes. This effectively eliminates the respective superstructure reflections and complicates experimental structural studies. For example, recent neutron diffraction experiments\cite{yusuf2011} did not show any signatures of the superstructure formation in (CuBr)Sr$_2$Nb$_3$O$_{10}$, although the magnetism of this compound, including the peculiar $\frac13$-magnetization plateau,\cite{tsujimoto2007,*tsujimoto2008} can be hardly understood in terms of the square-lattice magnetic model derived from the parent, tetragonal crystal structure. A plausible explanation is the local ordering of Cu and Br atoms that will not produce sharp superstructure reflections, which would be observable in an x-ray or neutron diffraction experiment. 

The effects of short-range structural order may also be relevant for solid solutions obtained by the Cl/Br substitution. For example, Cs$_2$CuCl$_4$ and Cs$_2$CuBr$_4$ are orthorhombic compounds with fully ordered crystal structures, but the Cs$_2$CuCl$_{4-x}$Br$_x$ solid solutions have the tetragonal symmetry and presumably contain split positions of Cl and Br at $1<x<2$ (Ref.~\onlinecite{krueger2010}). This tendency to the structural disorder should not be overlooked in the ongoing studies of quantum magnets, where the Cl/Br substitution is used as a handy tool for changing the parameter regime and introducing bond randomness. The Br atoms not only occupy a crystallographic position different from Cl, thus introducing random exchange couplings, but also deteriorate the long-range structural order in the compound. Particularly, the (CuCl$_{1-x}$Br$_x$)LaNb$_2$O$_7$ solid solutions,\cite{tsujimoto2010,*uemura2009} showing the transition from the gapped magnetic phase at low $x$ toward the long-range AFM order at high $x$, can not be considered as a simple experimental example of a quantum phase transition driven by a change in microscopic parameters. Structural changes are an equally important ingredient of the physics and require a further elaborate investigation.

Despite the large spatial separation between the magnetic [CuBr] units, (CuBr)LaNb$_2$O$_7$ does not show any experimental signatures of low-dimensional magnetic behavior. Further, numerous couplings in the $ab$ plane do not induce any notable frustration. Spin systems with reduced dimensionality, as well as frustrated spin systems, feature strong quantum fluctuations that impede long-range magnetic order and trigger short-range order above $T_N$. In antiferromagnetic systems, a typical signature of the short-range order is the broad and symmetric susceptibility maximum, which is, for example, present in (CuCl)LaNb$_2$O$_7$ and (CuCl)LaTa$_2$O$_7$ (Ref.~\onlinecite{kitada2009}) yet absent in (CuBr)LaNb$_2$O$_7$ (see Fig.~\ref{fig:fits}). Upon cooling, the susceptibility of (CuBr)LaNb$_2$O$_7$ drops down right below the magnetic transition at $T_N\simeq 32$~K, thus rendering the susceptibility maximum narrow and asymmetric. 

The short-range magnetic order evidenced by the susceptibility maxima should be also observable in neutron-diffraction experiments above $T_N$ as a broad feature (diffuse scattering) preceding the formation of magnetic reflections. The lack of diffuse magnetic scattering at 40~K (see Fig.~\ref{fig:neutron}) is another argument disfavoring the low-dimensional and/or frustrated nature of (CuBr)LaNb$_2$O$_7$. Finally, the relatively high ordered magnetic moment $\mu\simeq 0.72$~$\mu_B$ is also a strong evidence of suppressed quantum fluctuations. In Sec.~\ref{sec:simul}, we have shown that the staggered magnetization $m_s\simeq 0.88$~$\mu_B$ (i.e., the ordered magnetic moment corrected for the hybridization with ligand orbitals and for the spin-orbit coupling) is similar to $m_s\simeq 0.83$~$\mu_B$ for the cubic spin lattice, the archetype 3D spin system with weak quantum fluctuations.

In (CuBr)LaNb$_2$O$_7$, weak quantum fluctuations should be ascribed to the large number of non-frustrated exchange couplings per magnetic site. Each Cu atom is involved in as many as ten couplings, compared to, e.g., six and four couplings on the cubic lattice and square lattice, respectively. Although quantum fluctuations recede because of the enhanced connectivity, they are not removed completely, as the reduction in the ordered magnetic moment is partially related to the AFM exchange and the quantum behavior of Cu$^{+2}$ (spin-$\frac12$). For example, the staggered magnetization is reduced for 12~\% compared to the classical value of 1~$\mu_B$. Using classical Monte Carlo simulations for the same spin lattice and same parameter regime, we arrive at $T_N/J_4\simeq 0.8$ and $T_N\simeq 38$~K compared to the experimental values of $T_N/J_4\simeq 0.67$ and $T_N\simeq 32$~K. Therefore, weak quantum effects are still present in the spin system of (CuBr)LaNb$_2$O$_7$. This compound may be interesting as a system lacking both low dimensionality and strong frustration, so that weak quantum effects in a spin-$\frac12$ magnet can be observed. In this respect, (CuBr)LaNb$_2$O$_7$ is rather unique, because Cu$^{+2}$ compounds are prone to the formation of low-dimensional spin systems. In rare cases of geometrically 3D spin lattices, as in the green dioptase Cu$_6$Si$_6$O$_{18}\cdot 6$H$_2$O, the low coordination number (weak connectivity) still triggers strong quantum fluctuations that are clearly observed experimentally.\cite{janson2010} The spin lattice of (CuBr)LaNb$_2$O$_7$ not only reveals the 3D geometry (the sizable coupling $J_{\perp}$ along the $c$ direction), but also features weak quantum fluctuations.

(CuBr)LaNb$_2$O$_7$ and (CuCl)LaNb$_2$O$_7$ feature similar spin lattices, but strongly differ in the relevant parameter regimes. The main ingredient of the spin lattice is the strong fourth-neighbor AFM coupling $J_4$. This coupling runs via the long-range Cu--X--X--Cu superexchange pathway (X = Cl, Br) and strongly depends on the deviation of this pathway from the straight line. Therefore, the second fourth-neighbor coupling $J_4'$ is much weaker than $J_4$, and spin dimers rather than bond-alternating spin chains are formed (see the right panel of Fig.~\ref{fig:lattice}). Although fits to the experimental data retain a certain ambiguity with respect to smaller individual exchange couplings, they yield a robust estimate of the effective interdimer coupling given by a linear combination of all interdimer couplings taken with appropriate coordination numbers: $J_{\eff}=2|J_1|+J_1'+J_1''+2J_2+J_4'+2J_{\perp}$. In (CuCl)LaNb$_2$O$_7$, $J_{\eff}/J_4\simeq 0.5$ so that the system is close to the limit of isolated spin dimers.\cite{tsirlin2010b} Therefore, (CuCl)LaNb$_2$O$_7$ reveals a sizable spin gap\cite{kageyama2005a,*kageyama2005b} and undergoes Bose-Einstein condensation of magnons in high magnetic fields.\cite{kitada2007} On the contrary, (CuBr)LaNb$_2$O$_7$ features $J_{\eff}/J_4\simeq 3$ and belongs to the opposite limit of strong interdimer couplings. The interdimer couplings close the spin gap and establish the long-range AFM order with a relatively high $T_N/J_4\simeq 0.67$. 

The strongly enhanced interdimer couplings are the crucial microscopic difference between (CuBr)LaNb$_2$O$_7$ and its Cl-containing counterpart. Although DFT results are less accurate than fits to the experimental data, computational estimates show a similar trend of increased interdimer couplings in the Br compound and, particularly, reveal sizable nearest-neighbor interactions $J_1'$ and $J_1''$ that are missing in (CuCl)LaNb$_2$O$_7$ (compare Table~\ref{tab:exchange} to Table~I in Ref.~\onlinecite{tsirlin2010b}). The strong interdimer exchange can be ascribed to the larger spatial extent of Br $4p$ orbitals compared to Cl $3p$ orbitals. The Br atoms enable stronger interactions via short superexchange pathways $J_1'$, $J_1''$, and $J_2$ that lack direct connections between the CuO$_2$X$_2$ magnetic plaquettes (see the left panel of Fig.~\ref{fig:lattice}). Moreover, the larger size of Br reduces the Cu--X--Cu angle within the zigzag ribbon (compare $101.8^{\circ}$ to $109.0^{\circ}$ for Br and Cl, respectively), thus increasing the FM coupling $J_1$.

In summary, we studied the crystal structure, electronic structure, and magnetism of (CuBr)LaNb$_2$O$_7$. This compound is very similar to its Cl counterpart in general, yet different in several important aspects. First, the Br atoms are detrimental for the structural ordering and develop a short-range order, only. Second, the Br atoms are prone to out-of-plane displacements, especially at elevated temperatures. Third, the introduction of Br does not change the nature of the dimer-based spin lattice, although the interdimer couplings are increased dramatically, thus triggering the transition from a gapped ground state in (CuCl)LaNb$_2$O$_7$ toward the long-range stripe AFM order in (CuBr)LaNb$_2$O$_7$. 

\acknowledgments
We are grateful to ESRF, ILL, and HMI for granting the measurement time. Experimental assistance of Andy Fitch, Carolina Curfs, Adrian Hill, and Monika Gam\.za at the ID31 beamline of ESRF is acknowledged. We would also like to thank Yurii Prots and Horst Borrmann for laboratory XRD measurements, Stefan Hoffmann for thermal analysis, as well as Daria Mikhailova and Stefan Hoffmann for careful reading of the manuscript and fruitful suggestions. A.T. was funded by Alexander von Humboldt Foundation.


\begin{thebibliography}{66}%
\makeatletter
\providecommand \@ifxundefined [1]{%
 \@ifx{#1\undefined}
}%
\providecommand \@ifnum [1]{%
 \ifnum #1\expandafter \@firstoftwo
 \else \expandafter \@secondoftwo
 \fi
}%
\providecommand \@ifx [1]{%
 \ifx #1\expandafter \@firstoftwo
 \else \expandafter \@secondoftwo
 \fi
}%
\providecommand \natexlab [1]{#1}%
\providecommand \enquote  [1]{``#1''}%
\providecommand \bibnamefont  [1]{#1}%
\providecommand \bibfnamefont [1]{#1}%
\providecommand \citenamefont [1]{#1}%
\providecommand \href@noop [0]{\@secondoftwo}%
\providecommand \href [0]{\begingroup \@sanitize@url \@href}%
\providecommand \@href[1]{\@@startlink{#1}\@@href}%
\providecommand \@@href[1]{\endgroup#1\@@endlink}%
\providecommand \@sanitize@url [0]{\catcode `\\12\catcode `\$12\catcode
  `\&12\catcode `\#12\catcode `\^12\catcode `\_12\catcode `\%12\relax}%
\providecommand \@@startlink[1]{}%
\providecommand \@@endlink[0]{}%
\providecommand \url  [0]{\begingroup\@sanitize@url \@url }%
\providecommand \@url [1]{\endgroup\@href {#1}{\urlprefix }}%
\providecommand \urlprefix  [0]{URL }%
\providecommand \Eprint [0]{\href }%
\providecommand \doibase [0]{http://dx.doi.org/}%
\providecommand \selectlanguage [0]{\@gobble}%
\providecommand \bibinfo  [0]{\@secondoftwo}%
\providecommand \bibfield  [0]{\@secondoftwo}%
\providecommand \translation [1]{[#1]}%
\providecommand \BibitemOpen [0]{}%
\providecommand \bibitemStop [0]{}%
\providecommand \bibitemNoStop [0]{.\EOS\space}%
\providecommand \EOS [0]{\spacefactor3000\relax}%
\providecommand \BibitemShut  [1]{\csname bibitem#1\endcsname}%
\let\auto@bib@innerbib\@empty
\bibitem [{\citenamefont {Manaka}\ \emph {et~al.}(1997)\citenamefont {Manaka},
  \citenamefont {Yamada},\ and\ \citenamefont {Yamaguchi}}]{manaka1997a}%
  \BibitemOpen
  \bibfield  {author} {\bibinfo {author} {\bibfnamefont {H.}~\bibnamefont
  {Manaka}}, \bibinfo {author} {\bibfnamefont {I.}~\bibnamefont {Yamada}}, \
  and\ \bibinfo {author} {\bibfnamefont {K.}~\bibnamefont {Yamaguchi}},\
  }\href@noop {} {\bibfield  {journal} {\bibinfo  {journal} {J. Phys. Soc.
  Jpn.}\ }\textbf {\bibinfo {volume} {66}},\ \bibinfo {pages} {564} (\bibinfo
  {year} {1997})}\BibitemShut {NoStop}%
\bibitem [{\citenamefont {Manaka}\ and\ \citenamefont
  {Yamada}(1997)}]{manaka1997b}%
  \BibitemOpen
  \bibfield  {author} {\bibinfo {author} {\bibfnamefont {H.}~\bibnamefont
  {Manaka}}\ and\ \bibinfo {author} {\bibfnamefont {I.}~\bibnamefont
  {Yamada}},\ }\href@noop {} {\bibfield  {journal} {\bibinfo  {journal} {J.
  Phys. Soc. Jpn.}\ }\textbf {\bibinfo {volume} {66}},\ \bibinfo {pages} {1908}
  (\bibinfo {year} {1997})}\BibitemShut {NoStop}%
\bibitem [{\citenamefont {Valent{\'\i}}\ \emph {et~al.}(2003)\citenamefont
  {Valent{\'\i}}, \citenamefont {Saha-Dasgupta}, \citenamefont {Gros},\ and\
  \citenamefont {Rosner}}]{valenti2003}%
  \BibitemOpen
  \bibfield  {author} {\bibinfo {author} {\bibfnamefont {R.}~\bibnamefont
  {Valent{\'\i}}}, \bibinfo {author} {\bibfnamefont {T.}~\bibnamefont
  {Saha-Dasgupta}}, \bibinfo {author} {\bibfnamefont {C.}~\bibnamefont {Gros}},
  \ and\ \bibinfo {author} {\bibfnamefont {H.}~\bibnamefont {Rosner}},\
  }\href@noop {} {\bibfield  {journal} {\bibinfo  {journal} {Phys. Rev. B}\
  }\textbf {\bibinfo {volume} {67}},\ \bibinfo {pages} {245110} (\bibinfo
  {year} {2003}), cond-mat/0301119}\BibitemShut {NoStop}%
\bibitem [{\citenamefont {Zaharko}\ \emph {et~al.}(2006)\citenamefont
  {Zaharko}, \citenamefont {R{\o}nnow}, \citenamefont {Mesot}, \citenamefont
  {Crowe}, \citenamefont {Paul}, \citenamefont {Brown}, \citenamefont
  {Daoud-Aladine}, \citenamefont {Meents}, \citenamefont {Wagner},
  \citenamefont {Prester},\ and\ \citenamefont {Berger}}]{zaharko2006}%
  \BibitemOpen
  \bibfield  {author} {\bibinfo {author} {\bibfnamefont {O.}~\bibnamefont
  {Zaharko}}, \bibinfo {author} {\bibfnamefont {H.}~\bibnamefont {R{\o}nnow}},
  \bibinfo {author} {\bibfnamefont {J.}~\bibnamefont {Mesot}}, \bibinfo
  {author} {\bibfnamefont {S.~J.}\ \bibnamefont {Crowe}}, \bibinfo {author}
  {\bibfnamefont {D.~M.}\ \bibnamefont {Paul}}, \bibinfo {author}
  {\bibfnamefont {P.~J.}\ \bibnamefont {Brown}}, \bibinfo {author}
  {\bibfnamefont {A.}~\bibnamefont {Daoud-Aladine}}, \bibinfo {author}
  {\bibfnamefont {A.}~\bibnamefont {Meents}}, \bibinfo {author} {\bibfnamefont
  {A.}~\bibnamefont {Wagner}}, \bibinfo {author} {\bibfnamefont
  {M.}~\bibnamefont {Prester}}, \ and\ \bibinfo {author} {\bibfnamefont
  {H.}~\bibnamefont {Berger}},\ }\href@noop {} {\bibfield  {journal} {\bibinfo
  {journal} {Phys. Rev. B}\ }\textbf {\bibinfo {volume} {73}},\ \bibinfo
  {pages} {064422} (\bibinfo {year} {2006}), cond-mat/0512617}\BibitemShut {NoStop}%
\bibitem [{\citenamefont {Ono}\ \emph {et~al.}(2005)\citenamefont {Ono},
  \citenamefont {Tanaka}, \citenamefont {Nakagomi}, \citenamefont {Kolomiyets},
  \citenamefont {Mitamura}, \citenamefont {Ishikawa}, \citenamefont {Goto},
  \citenamefont {Nakajima}, \citenamefont {Oosawa}, \citenamefont {Koike},
  \citenamefont {Kakurai}, \citenamefont {Klenke}, \citenamefont {Smeibidle},
  \citenamefont {Mei{\ss}ner},\ and\ \citenamefont {Katori}}]{ono2005}%
  \BibitemOpen
  \bibfield  {author} {\bibinfo {author} {\bibfnamefont {T.}~\bibnamefont
  {Ono}}, \bibinfo {author} {\bibfnamefont {H.}~\bibnamefont {Tanaka}},
  \bibinfo {author} {\bibfnamefont {T.}~\bibnamefont {Nakagomi}}, \bibinfo
  {author} {\bibfnamefont {O.}~\bibnamefont {Kolomiyets}}, \bibinfo {author}
  {\bibfnamefont {H.}~\bibnamefont {Mitamura}}, \bibinfo {author}
  {\bibfnamefont {F.}~\bibnamefont {Ishikawa}}, \bibinfo {author}
  {\bibfnamefont {T.}~\bibnamefont {Goto}}, \bibinfo {author} {\bibfnamefont
  {K.}~\bibnamefont {Nakajima}}, \bibinfo {author} {\bibfnamefont
  {A.}~\bibnamefont {Oosawa}}, \bibinfo {author} {\bibfnamefont
  {Y.}~\bibnamefont {Koike}}, \bibinfo {author} {\bibfnamefont
  {K.}~\bibnamefont {Kakurai}}, \bibinfo {author} {\bibfnamefont
  {J.}~\bibnamefont {Klenke}}, \bibinfo {author} {\bibfnamefont
  {P.}~\bibnamefont {Smeibidle}}, \bibinfo {author} {\bibfnamefont
  {M.}~\bibnamefont {Mei{\ss}ner}}, \ and\ \bibinfo {author} {\bibfnamefont
  {H.~A.}\ \bibnamefont {Katori}},\ }\href@noop {} {\bibfield  {journal}
  {\bibinfo  {journal} {J. Phys. Soc. Jpn. Suppl.}\ }\textbf {\bibinfo {volume}
  {74}},\ \bibinfo {pages} {135} (\bibinfo {year} {2005})}\BibitemShut
  {NoStop}%
\bibitem [{\citenamefont {Foyevtsova}\ \emph {et~al.}(2011)\citenamefont
  {Foyevtsova}, \citenamefont {Opahle}, \citenamefont {Zhang}, \citenamefont
  {Jeschke},\ and\ \citenamefont {Valent{\'\i}}}]{foyevtsova2011}%
  \BibitemOpen
  \bibfield  {author} {\bibinfo {author} {\bibfnamefont {K.}~\bibnamefont
  {Foyevtsova}}, \bibinfo {author} {\bibfnamefont {I.}~\bibnamefont {Opahle}},
  \bibinfo {author} {\bibfnamefont {Y.-Z.}\ \bibnamefont {Zhang}}, \bibinfo
  {author} {\bibfnamefont {H.~O.}\ \bibnamefont {Jeschke}}, \ and\ \bibinfo
  {author} {\bibfnamefont {R.}~\bibnamefont {Valent{\'\i}}},\ }\href@noop {}
  {\bibfield  {journal} {\bibinfo  {journal} {Phys. Rev. B}\ }\textbf {\bibinfo
  {volume} {83}},\ \bibinfo {pages} {125126} (\bibinfo {year}
  {2011}), arXiv:1009.0697}\BibitemShut {NoStop}%
\bibitem [{\citenamefont {Kr\"uger}\ \emph {et~al.}(2010)\citenamefont
  {Kr\"uger}, \citenamefont {Belz}, \citenamefont {Schossau}, \citenamefont
  {Haghighirad}, \citenamefont {Cong}, \citenamefont {Wolf}, \citenamefont
  {Gottlieb-Schoenmeyer}, \citenamefont {Ritter},\ and\ \citenamefont
  {Assmus}}]{krueger2010}%
  \BibitemOpen
  \bibfield  {author} {\bibinfo {author} {\bibfnamefont {N.}~\bibnamefont
  {Kr\"uger}}, \bibinfo {author} {\bibfnamefont {S.}~\bibnamefont {Belz}},
  \bibinfo {author} {\bibfnamefont {F.}~\bibnamefont {Schossau}}, \bibinfo
  {author} {\bibfnamefont {A.~A.}\ \bibnamefont {Haghighirad}}, \bibinfo
  {author} {\bibfnamefont {P.~T.}\ \bibnamefont {Cong}}, \bibinfo {author}
  {\bibfnamefont {B.}~\bibnamefont {Wolf}}, \bibinfo {author} {\bibfnamefont
  {S.}~\bibnamefont {Gottlieb-Schoenmeyer}}, \bibinfo {author} {\bibfnamefont
  {F.}~\bibnamefont {Ritter}}, \ and\ \bibinfo {author} {\bibfnamefont
  {W.}~\bibnamefont {Assmus}},\ }\href@noop {} {\bibfield  {journal} {\bibinfo
  {journal} {Cryst. Growth Design}\ }\textbf {\bibinfo {volume} {10}},\
  \bibinfo {pages} {4456} (\bibinfo {year} {2010}), arXiv:1103.0139}\BibitemShut {NoStop}%
\bibitem [{\citenamefont {Manaka}\ \emph {et~al.}(2001)\citenamefont {Manaka},
  \citenamefont {Yamada},\ and\ \citenamefont {Katori}}]{manaka2001}%
  \BibitemOpen
  \bibfield  {author} {\bibinfo {author} {\bibfnamefont {H.}~\bibnamefont
  {Manaka}}, \bibinfo {author} {\bibfnamefont {I.}~\bibnamefont {Yamada}}, \
  and\ \bibinfo {author} {\bibfnamefont {H.~A.}\ \bibnamefont {Katori}},\
  }\href@noop {} {\bibfield  {journal} {\bibinfo  {journal} {Phys. Rev. B}\
  }\textbf {\bibinfo {volume} {63}},\ \bibinfo {pages} {104408} (\bibinfo
  {year} {2001})}\BibitemShut {NoStop}%
\bibitem [{\citenamefont {Manaka}\ \emph {et~al.}(2002)\citenamefont {Manaka},
  \citenamefont {Yamada}, \citenamefont {Mitamura},\ and\ \citenamefont
  {Goto}}]{manaka2002}%
  \BibitemOpen
  \bibfield  {author} {\bibinfo {author} {\bibfnamefont {H.}~\bibnamefont
  {Manaka}}, \bibinfo {author} {\bibfnamefont {I.}~\bibnamefont {Yamada}},
  \bibinfo {author} {\bibfnamefont {H.}~\bibnamefont {Mitamura}}, \ and\
  \bibinfo {author} {\bibfnamefont {T.}~\bibnamefont {Goto}},\ }\href@noop {}
  {\bibfield  {journal} {\bibinfo  {journal} {Phys. Rev. B}\ }\textbf {\bibinfo
  {volume} {66}},\ \bibinfo {pages} {064402} (\bibinfo {year}
  {2002})}\BibitemShut {NoStop}%
\bibitem [{\citenamefont {Goto}\ \emph {et~al.}(2008)\citenamefont {Goto},
  \citenamefont {Suzuki}, \citenamefont {Kanada}, \citenamefont {Saito},
  \citenamefont {Oosawa}, \citenamefont {Watanabe},\ and\ \citenamefont
  {Manaka}}]{goto2008}%
  \BibitemOpen
  \bibfield  {author} {\bibinfo {author} {\bibfnamefont {T.}~\bibnamefont
  {Goto}}, \bibinfo {author} {\bibfnamefont {T.}~\bibnamefont {Suzuki}},
  \bibinfo {author} {\bibfnamefont {K.}~\bibnamefont {Kanada}}, \bibinfo
  {author} {\bibfnamefont {T.}~\bibnamefont {Saito}}, \bibinfo {author}
  {\bibfnamefont {A.}~\bibnamefont {Oosawa}}, \bibinfo {author} {\bibfnamefont
  {I.}~\bibnamefont {Watanabe}}, \ and\ \bibinfo {author} {\bibfnamefont
  {H.}~\bibnamefont {Manaka}},\ }\href@noop {} {\bibfield  {journal} {\bibinfo
  {journal} {Phys. Rev. B}\ }\textbf {\bibinfo {volume} {78}},\ \bibinfo
  {pages} {054422} (\bibinfo {year} {2008}), arXiv:0807.3380}\BibitemShut {NoStop}%
\bibitem [{\citenamefont {Oba}\ \emph {et~al.}(2006)\citenamefont {Oba},
  \citenamefont {Kageyama}, \citenamefont {Kitano}, \citenamefont {Yasuda},
  \citenamefont {Baba}, \citenamefont {Nishi}, \citenamefont {Hirota},
  \citenamefont {Narumi}, \citenamefont {Hagiwara}, \citenamefont {Kindo},
  \citenamefont {Saito}, \citenamefont {Ajiro},\ and\ \citenamefont
  {Yoshimura}}]{oba2006}%
  \BibitemOpen
  \bibfield  {author} {\bibinfo {author} {\bibfnamefont {N.}~\bibnamefont
  {Oba}}, \bibinfo {author} {\bibfnamefont {H.}~\bibnamefont {Kageyama}},
  \bibinfo {author} {\bibfnamefont {T.}~\bibnamefont {Kitano}}, \bibinfo
  {author} {\bibfnamefont {J.}~\bibnamefont {Yasuda}}, \bibinfo {author}
  {\bibfnamefont {Y.}~\bibnamefont {Baba}}, \bibinfo {author} {\bibfnamefont
  {M.}~\bibnamefont {Nishi}}, \bibinfo {author} {\bibfnamefont
  {K.}~\bibnamefont {Hirota}}, \bibinfo {author} {\bibfnamefont
  {Y.}~\bibnamefont {Narumi}}, \bibinfo {author} {\bibfnamefont
  {M.}~\bibnamefont {Hagiwara}}, \bibinfo {author} {\bibfnamefont
  {K.}~\bibnamefont {Kindo}}, \bibinfo {author} {\bibfnamefont
  {T.}~\bibnamefont {Saito}}, \bibinfo {author} {\bibfnamefont
  {Y.}~\bibnamefont {Ajiro}}, \ and\ \bibinfo {author} {\bibfnamefont
  {K.}~\bibnamefont {Yoshimura}},\ }\href@noop {} {\bibfield  {journal}
  {\bibinfo  {journal} {J. Phys. Soc. Jpn}\ }\textbf {\bibinfo {volume} {75}},\
  \bibinfo {pages} {113601} (\bibinfo {year} {2006})}\BibitemShut {NoStop}%
\bibitem [{\citenamefont {Kageyama}\ \emph
  {et~al.}(2005{\natexlab{a}})\citenamefont {Kageyama}, \citenamefont {Kitano},
  \citenamefont {Oba}, \citenamefont {Nishi}, \citenamefont {Nagai},
  \citenamefont {Hirota}, \citenamefont {Viciu}, \citenamefont {Wiley},
  \citenamefont {Yasuda}, \citenamefont {Baba}, \citenamefont {Ajiro},\ and\
  \citenamefont {Yoshimura}}]{kageyama2005a}%
  \BibitemOpen
  \bibfield  {author} {\bibinfo {author} {\bibfnamefont {H.}~\bibnamefont
  {Kageyama}}, \bibinfo {author} {\bibfnamefont {T.}~\bibnamefont {Kitano}},
  \bibinfo {author} {\bibfnamefont {N.}~\bibnamefont {Oba}}, \bibinfo {author}
  {\bibfnamefont {M.}~\bibnamefont {Nishi}}, \bibinfo {author} {\bibfnamefont
  {S.}~\bibnamefont {Nagai}}, \bibinfo {author} {\bibfnamefont
  {K.}~\bibnamefont {Hirota}}, \bibinfo {author} {\bibfnamefont
  {L.}~\bibnamefont {Viciu}}, \bibinfo {author} {\bibfnamefont {J.~B.}\
  \bibnamefont {Wiley}}, \bibinfo {author} {\bibfnamefont {J.}~\bibnamefont
  {Yasuda}}, \bibinfo {author} {\bibfnamefont {Y.}~\bibnamefont {Baba}},
  \bibinfo {author} {\bibfnamefont {Y.}~\bibnamefont {Ajiro}}, \ and\ \bibinfo
  {author} {\bibfnamefont {K.}~\bibnamefont {Yoshimura}},\ }\href@noop {}
  {\bibfield  {journal} {\bibinfo  {journal} {J. Phys. Soc. Jpn.}\ }\textbf
  {\bibinfo {volume} {74}},\ \bibinfo {pages} {1702} (\bibinfo {year}
  {2005}{\natexlab{a}})}\BibitemShut {NoStop}%
\bibitem [{\citenamefont {Kageyama}\ \emph
  {et~al.}(2005{\natexlab{b}})\citenamefont {Kageyama}, \citenamefont {Yasuda},
  \citenamefont {Kitano}, \citenamefont {Totsuka}, \citenamefont {Narumi},
  \citenamefont {Hagiwara}, \citenamefont {Kindo}, \citenamefont {Baba},
  \citenamefont {Oba}, \citenamefont {Ajiro},\ and\ \citenamefont
  {Yoshimura}}]{kageyama2005b}%
  \BibitemOpen
  \bibfield  {author} {\bibinfo {author} {\bibfnamefont {H.}~\bibnamefont
  {Kageyama}}, \bibinfo {author} {\bibfnamefont {J.}~\bibnamefont {Yasuda}},
  \bibinfo {author} {\bibfnamefont {T.}~\bibnamefont {Kitano}}, \bibinfo
  {author} {\bibfnamefont {K.}~\bibnamefont {Totsuka}}, \bibinfo {author}
  {\bibfnamefont {Y.}~\bibnamefont {Narumi}}, \bibinfo {author} {\bibfnamefont
  {M.}~\bibnamefont {Hagiwara}}, \bibinfo {author} {\bibfnamefont
  {K.}~\bibnamefont {Kindo}}, \bibinfo {author} {\bibfnamefont
  {Y.}~\bibnamefont {Baba}}, \bibinfo {author} {\bibfnamefont {N.}~\bibnamefont
  {Oba}}, \bibinfo {author} {\bibfnamefont {Y.}~\bibnamefont {Ajiro}}, \ and\
  \bibinfo {author} {\bibfnamefont {K.}~\bibnamefont {Yoshimura}},\ }\href@noop
  {} {\bibfield  {journal} {\bibinfo  {journal} {J. Phys. Soc. Jpn.}\ }\textbf
  {\bibinfo {volume} {74}},\ \bibinfo {pages} {3155} (\bibinfo {year}
  {2005}{\natexlab{b}})}\BibitemShut {NoStop}%
\bibitem [{\citenamefont {Kodenkandath}\ \emph {et~al.}(1999)\citenamefont
  {Kodenkandath}, \citenamefont {Lalena}, \citenamefont {Zhou}, \citenamefont
  {Carpenter}, \citenamefont {Sangregorio}, \citenamefont {Falster},
  \citenamefont {Simmons}, \citenamefont {O'Connor},\ and\ \citenamefont
  {Wiley}}]{koden1999}%
  \BibitemOpen
  \bibfield  {author} {\bibinfo {author} {\bibfnamefont {T.~A.}\ \bibnamefont
  {Kodenkandath}}, \bibinfo {author} {\bibfnamefont {J.~N.}\ \bibnamefont
  {Lalena}}, \bibinfo {author} {\bibfnamefont {W.~L.}\ \bibnamefont {Zhou}},
  \bibinfo {author} {\bibfnamefont {E.~E.}\ \bibnamefont {Carpenter}}, \bibinfo
  {author} {\bibfnamefont {C.}~\bibnamefont {Sangregorio}}, \bibinfo {author}
  {\bibfnamefont {A.~U.}\ \bibnamefont {Falster}}, \bibinfo {author}
  {\bibfnamefont {W.~B.~J.}\ \bibnamefont {Simmons}}, \bibinfo {author}
  {\bibfnamefont {C.~J.}\ \bibnamefont {O'Connor}}, \ and\ \bibinfo {author}
  {\bibfnamefont {J.~B.}\ \bibnamefont {Wiley}},\ }\href@noop {} {\bibfield
  {journal} {\bibinfo  {journal} {J. Amer. Chem. Soc.}\ }\textbf {\bibinfo
  {volume} {121}},\ \bibinfo {pages} {10743} (\bibinfo {year}
  {1999})}\BibitemShut {NoStop}%
\bibitem [{\citenamefont {Tsirlin}\ \emph
  {et~al.}(2010{\natexlab{a}})\citenamefont {Tsirlin}, \citenamefont
  {Abakumov}, \citenamefont {{Van Tendeloo}},\ and\ \citenamefont
  {Rosner}}]{tsirlin2010a}%
  \BibitemOpen
  \bibfield  {author} {\bibinfo {author} {\bibfnamefont {A.~A.}\ \bibnamefont
  {Tsirlin}}, \bibinfo {author} {\bibfnamefont {A.~M.}\ \bibnamefont
  {Abakumov}}, \bibinfo {author} {\bibfnamefont {G.}~\bibnamefont {{Van
  Tendeloo}}}, \ and\ \bibinfo {author} {\bibfnamefont {H.}~\bibnamefont
  {Rosner}},\ }\href@noop {} {\bibfield  {journal} {\bibinfo  {journal} {Phys.
  Rev. B}\ }\textbf {\bibinfo {volume} {82}},\ \bibinfo {pages} {054107}
  (\bibinfo {year} {2010}{\natexlab{a}}), arXiv:1005.4898}\BibitemShut {NoStop}%
\bibitem [{\citenamefont {Tassel}\ \emph {et~al.}(2010)\citenamefont {Tassel},
  \citenamefont {Kang}, \citenamefont {Lee}, \citenamefont {Hernandez},
  \citenamefont {Qiu}, \citenamefont {Paulus}, \citenamefont {Collet},
  \citenamefont {Lake}, \citenamefont {Guidi}, \citenamefont {Whangbo},
  \citenamefont {Ritter}, \citenamefont {Kageyama},\ and\ \citenamefont
  {Lee}}]{tassel2010}%
  \BibitemOpen
  \bibfield  {author} {\bibinfo {author} {\bibfnamefont {C.}~\bibnamefont
  {Tassel}}, \bibinfo {author} {\bibfnamefont {J.}~\bibnamefont {Kang}},
  \bibinfo {author} {\bibfnamefont {C.}~\bibnamefont {Lee}}, \bibinfo {author}
  {\bibfnamefont {O.}~\bibnamefont {Hernandez}}, \bibinfo {author}
  {\bibfnamefont {Y.}~\bibnamefont {Qiu}}, \bibinfo {author} {\bibfnamefont
  {W.}~\bibnamefont {Paulus}}, \bibinfo {author} {\bibfnamefont
  {E.}~\bibnamefont {Collet}}, \bibinfo {author} {\bibfnamefont
  {B.}~\bibnamefont {Lake}}, \bibinfo {author} {\bibfnamefont {T.}~\bibnamefont
  {Guidi}}, \bibinfo {author} {\bibfnamefont {M.-H.}\ \bibnamefont {Whangbo}},
  \bibinfo {author} {\bibfnamefont {C.}~\bibnamefont {Ritter}}, \bibinfo
  {author} {\bibfnamefont {H.}~\bibnamefont {Kageyama}}, \ and\ \bibinfo
  {author} {\bibfnamefont {S.-H.}\ \bibnamefont {Lee}},\ }\href@noop {}
  {\bibfield  {journal} {\bibinfo  {journal} {Phys. Rev. Lett.}\ }\textbf
  {\bibinfo {volume} {105}},\ \bibinfo {pages} {167205} (\bibinfo {year}
  {2010}), arXiv:1006.0755}\BibitemShut {NoStop}%
\bibitem [{\citenamefont {Hernandez}\ \emph {et~al.}(2011)\citenamefont
  {Hernandez}, \citenamefont {Tassel}, \citenamefont {Nakano}, \citenamefont
  {Paulus}, \citenamefont {Ritter}, \citenamefont {Collet}, \citenamefont
  {Kitada}, \citenamefont {Yoshimura},\ and\ \citenamefont
  {Kageyama}}]{hernandez2011}%
  \BibitemOpen
  \bibfield  {author} {\bibinfo {author} {\bibfnamefont {O.~J.}\ \bibnamefont
  {Hernandez}}, \bibinfo {author} {\bibfnamefont {C.}~\bibnamefont {Tassel}},
  \bibinfo {author} {\bibfnamefont {K.}~\bibnamefont {Nakano}}, \bibinfo
  {author} {\bibfnamefont {W.}~\bibnamefont {Paulus}}, \bibinfo {author}
  {\bibfnamefont {C.}~\bibnamefont {Ritter}}, \bibinfo {author} {\bibfnamefont
  {E.}~\bibnamefont {Collet}}, \bibinfo {author} {\bibfnamefont
  {A.}~\bibnamefont {Kitada}}, \bibinfo {author} {\bibfnamefont
  {K.}~\bibnamefont {Yoshimura}}, \ and\ \bibinfo {author} {\bibfnamefont
  {H.}~\bibnamefont {Kageyama}},\ }\href@noop {} {\bibfield  {journal}
  {\bibinfo  {journal} {Dalton Trans.}\ }\textbf {\bibinfo {volume} {40}},\
  \bibinfo {pages} {4605} (\bibinfo {year} {2011})}\BibitemShut {NoStop}%
\bibitem [{\citenamefont {Tsirlin}\ and\ \citenamefont
  {Rosner}(2009)}]{tsirlin2009}%
  \BibitemOpen
  \bibfield  {author} {\bibinfo {author} {\bibfnamefont {A.~A.}\ \bibnamefont
  {Tsirlin}}\ and\ \bibinfo {author} {\bibfnamefont {H.}~\bibnamefont
  {Rosner}},\ }\href@noop {} {\bibfield  {journal} {\bibinfo  {journal} {Phys.
  Rev. B}\ }\textbf {\bibinfo {volume} {79}},\ \bibinfo {pages} {214416}
  (\bibinfo {year} {2009})}\BibitemShut {NoStop}%
\bibitem [{\citenamefont {Ren}\ and\ \citenamefont {Cheng}(2010)}]{ren2010}%
  \BibitemOpen
  \bibfield  {author} {\bibinfo {author} {\bibfnamefont {C.-Y.}\ \bibnamefont
  {Ren}}\ and\ \bibinfo {author} {\bibfnamefont {C.}~\bibnamefont {Cheng}},\
  }\href@noop {} {\bibfield  {journal} {\bibinfo  {journal} {Phys. Rev. B}\
  }\textbf {\bibinfo {volume} {82}},\ \bibinfo {pages} {024404} (\bibinfo
  {year} {2010}), arXiv:0901.0154}\BibitemShut {NoStop}%
\bibitem [{\citenamefont {Tsirlin}\ and\ \citenamefont
  {Rosner}(2010{\natexlab{a}})}]{tsirlin2010b}%
  \BibitemOpen
  \bibfield  {author} {\bibinfo {author} {\bibfnamefont {A.~A.}\ \bibnamefont
  {Tsirlin}}\ and\ \bibinfo {author} {\bibfnamefont {H.}~\bibnamefont
  {Rosner}},\ }\href@noop {} {\bibfield  {journal} {\bibinfo  {journal} {Phys.
  Rev. B}\ }\textbf {\bibinfo {volume} {82}},\ \bibinfo {pages} {060409(R)}
  (\bibinfo {year} {2010}{\natexlab{a}}), arXiv:1007.3883}\BibitemShut {NoStop}%
\bibitem [{Note1()}]{Note1}%
  \BibitemOpen
  \bibinfo {note} {More precise estimates of the magnetic couplings can be
  found in Sec.~\ref {sec:simul}}\BibitemShut {NoStop}%
\bibitem [{\citenamefont {Shannon}\ \emph {et~al.}(2004)\citenamefont
  {Shannon}, \citenamefont {Schmidt}, \citenamefont {Penc},\ and\ \citenamefont
  {Thalmeier}}]{shannon2004}%
  \BibitemOpen
  \bibfield  {author} {\bibinfo {author} {\bibfnamefont {N.}~\bibnamefont
  {Shannon}}, \bibinfo {author} {\bibfnamefont {B.}~\bibnamefont {Schmidt}},
  \bibinfo {author} {\bibfnamefont {K.}~\bibnamefont {Penc}}, \ and\ \bibinfo
  {author} {\bibfnamefont {P.}~\bibnamefont {Thalmeier}},\ }\href@noop {}
  {\bibfield  {journal} {\bibinfo  {journal} {Eur. Phys. J. B}\ }\textbf
  {\bibinfo {volume} {38}},\ \bibinfo {pages} {599} (\bibinfo {year}
  {2004}), cond-mat/0312160}\BibitemShut {NoStop}%
\bibitem [{\citenamefont {Yoshida}\ \emph {et~al.}(2008)\citenamefont
  {Yoshida}, \citenamefont {Ogata}, \citenamefont {Takigawa}, \citenamefont
  {Kitano}, \citenamefont {Kageyama}, \citenamefont {Ajiro},\ and\
  \citenamefont {Yoshimura}}]{yoshida2008}%
  \BibitemOpen
  \bibfield  {author} {\bibinfo {author} {\bibfnamefont {M.}~\bibnamefont
  {Yoshida}}, \bibinfo {author} {\bibfnamefont {N.}~\bibnamefont {Ogata}},
  \bibinfo {author} {\bibfnamefont {M.}~\bibnamefont {Takigawa}}, \bibinfo
  {author} {\bibfnamefont {T.}~\bibnamefont {Kitano}}, \bibinfo {author}
  {\bibfnamefont {H.}~\bibnamefont {Kageyama}}, \bibinfo {author}
  {\bibfnamefont {Y.}~\bibnamefont {Ajiro}}, \ and\ \bibinfo {author}
  {\bibfnamefont {K.}~\bibnamefont {Yoshimura}},\ }\href@noop {} {\bibfield
  {journal} {\bibinfo  {journal} {J. Phys. Soc. Jpn.}\ }\textbf {\bibinfo
  {volume} {77}},\ \bibinfo {pages} {104705} (\bibinfo {year}
  {2008}), arXiv:0805.2218}\BibitemShut {NoStop}%
\bibitem [{\citenamefont {Yoshida}\ \emph {et~al.}(2007)\citenamefont
  {Yoshida}, \citenamefont {Ogata}, \citenamefont {Takigawa}, \citenamefont
  {Yamaura}, \citenamefont {Ichihara}, \citenamefont {Kitano}, \citenamefont
  {Kageyama}, \citenamefont {Ajiro},\ and\ \citenamefont
  {Yoshimura}}]{yoshida2007}%
  \BibitemOpen
  \bibfield  {author} {\bibinfo {author} {\bibfnamefont {M.}~\bibnamefont
  {Yoshida}}, \bibinfo {author} {\bibfnamefont {N.}~\bibnamefont {Ogata}},
  \bibinfo {author} {\bibfnamefont {M.}~\bibnamefont {Takigawa}}, \bibinfo
  {author} {\bibfnamefont {J.}~\bibnamefont {Yamaura}}, \bibinfo {author}
  {\bibfnamefont {M.}~\bibnamefont {Ichihara}}, \bibinfo {author}
  {\bibfnamefont {T.}~\bibnamefont {Kitano}}, \bibinfo {author} {\bibfnamefont
  {H.}~\bibnamefont {Kageyama}}, \bibinfo {author} {\bibfnamefont
  {Y.}~\bibnamefont {Ajiro}}, \ and\ \bibinfo {author} {\bibfnamefont
  {K.}~\bibnamefont {Yoshimura}},\ }\href@noop {} {\bibfield  {journal}
  {\bibinfo  {journal} {J. Phys. Soc. Jpn.}\ }\textbf {\bibinfo {volume}
  {76}},\ \bibinfo {pages} {104703} (\bibinfo {year} {2007}), arXiv:0706.3559}\BibitemShut
  {NoStop}%
\bibitem [{\citenamefont {Pet{\v r}i{\v c}ek}\ \emph
  {et~al.}(2006)\citenamefont {Pet{\v r}i{\v c}ek}, \citenamefont {Du{\v
  s}ek},\ and\ \citenamefont {Palatinus}}]{jana2006}%
  \BibitemOpen
  \bibfield  {author} {\bibinfo {author} {\bibfnamefont {V.}~\bibnamefont
  {Pet{\v r}i{\v c}ek}}, \bibinfo {author} {\bibfnamefont {M.}~\bibnamefont
  {Du{\v s}ek}}, \ and\ \bibinfo {author} {\bibfnamefont {L.}~\bibnamefont
  {Palatinus}},\ }\href@noop {} {\enquote {\bibinfo {title} {Jana2006. {T}he
  crystallographic computing system},}\ } (\bibinfo {year} {2006}),\ \bibinfo
  {note} {{I}nstitute of Physics, Praha, Czech Republic}\BibitemShut {NoStop}%
\bibitem [{\citenamefont {Rodr{\'\i}guez-Carvajal}(1993)}]{fullprof}%
  \BibitemOpen
  \bibfield  {author} {\bibinfo {author} {\bibfnamefont {J.}~\bibnamefont
  {Rodr{\'\i}guez-Carvajal}},\ }\href@noop {} {\bibfield  {journal} {\bibinfo
  {journal} {Physica B}\ }\textbf {\bibinfo {volume} {192}},\ \bibinfo {pages}
  {55} (\bibinfo {year} {1993})}\BibitemShut {NoStop}%
\bibitem [{\citenamefont {Koepernik}\ and\ \citenamefont
  {Eschrig}(1999)}]{fplo}%
  \BibitemOpen
  \bibfield  {author} {\bibinfo {author} {\bibfnamefont {K.}~\bibnamefont
  {Koepernik}}\ and\ \bibinfo {author} {\bibfnamefont {H.}~\bibnamefont
  {Eschrig}},\ }\href@noop {} {\bibfield  {journal} {\bibinfo  {journal} {Phys.
  Rev. B}\ }\textbf {\bibinfo {volume} {59}},\ \bibinfo {pages} {1743}
  (\bibinfo {year} {1999})}\BibitemShut {NoStop}%
\bibitem [{\citenamefont {Perdew}\ and\ \citenamefont {Wang}(1992)}]{pw92}%
  \BibitemOpen
  \bibfield  {author} {\bibinfo {author} {\bibfnamefont {J.~P.}\ \bibnamefont
  {Perdew}}\ and\ \bibinfo {author} {\bibfnamefont {Y.}~\bibnamefont {Wang}},\
  }\href@noop {} {\bibfield  {journal} {\bibinfo  {journal} {Phys. Rev. B}\
  }\textbf {\bibinfo {volume} {45}},\ \bibinfo {pages} {13244} (\bibinfo {year}
  {1992})}\BibitemShut {NoStop}%
\bibitem [{\citenamefont {Perdew}\ \emph {et~al.}(1996)\citenamefont {Perdew},
  \citenamefont {Burke},\ and\ \citenamefont {Ernzerhof}}]{pbe}%
  \BibitemOpen
  \bibfield  {author} {\bibinfo {author} {\bibfnamefont {J.~P.}\ \bibnamefont
  {Perdew}}, \bibinfo {author} {\bibfnamefont {K.}~\bibnamefont {Burke}}, \
  and\ \bibinfo {author} {\bibfnamefont {M.}~\bibnamefont {Ernzerhof}},\
  }\href@noop {} {\bibfield  {journal} {\bibinfo  {journal} {Phys. Rev. Lett.}\
  }\textbf {\bibinfo {volume} {77}},\ \bibinfo {pages} {3865} (\bibinfo {year}
  {1996})}\BibitemShut {NoStop}%
\bibitem [{\citenamefont {Kresse}\ and\ \citenamefont
  {Furthm\"uller}(1996{\natexlab{a}})}]{vasp1}%
  \BibitemOpen
  \bibfield  {author} {\bibinfo {author} {\bibfnamefont {G.}~\bibnamefont
  {Kresse}}\ and\ \bibinfo {author} {\bibfnamefont {J.}~\bibnamefont
  {Furthm\"uller}},\ }\href@noop {} {\bibfield  {journal} {\bibinfo  {journal}
  {Comput. Mater. Sci.}\ }\textbf {\bibinfo {volume} {6}},\ \bibinfo {pages}
  {15} (\bibinfo {year} {1996}{\natexlab{a}})}\BibitemShut {NoStop}%
\bibitem [{\citenamefont {Kresse}\ and\ \citenamefont
  {Furthm\"uller}(1996{\natexlab{b}})}]{vasp2}%
  \BibitemOpen
  \bibfield  {author} {\bibinfo {author} {\bibfnamefont {G.}~\bibnamefont
  {Kresse}}\ and\ \bibinfo {author} {\bibfnamefont {J.}~\bibnamefont
  {Furthm\"uller}},\ }\href@noop {} {\bibfield  {journal} {\bibinfo  {journal}
  {Phys. Rev. B}\ }\textbf {\bibinfo {volume} {54}},\ \bibinfo {pages} {11169}
  (\bibinfo {year} {1996}{\natexlab{b}})}\BibitemShut {NoStop}%
\bibitem [{\citenamefont {Bl\"ochl}(1994)}]{paw1}%
  \BibitemOpen
  \bibfield  {author} {\bibinfo {author} {\bibfnamefont {P.~E.}\ \bibnamefont
  {Bl\"ochl}},\ }\href@noop {} {\bibfield  {journal} {\bibinfo  {journal}
  {Phys. Rev. B}\ }\textbf {\bibinfo {volume} {50}},\ \bibinfo {pages} {17953}
  (\bibinfo {year} {1994})}\BibitemShut {NoStop}%
\bibitem [{\citenamefont {Kresse}\ and\ \citenamefont {Joubert}(1999)}]{paw2}%
  \BibitemOpen
  \bibfield  {author} {\bibinfo {author} {\bibfnamefont {G.}~\bibnamefont
  {Kresse}}\ and\ \bibinfo {author} {\bibfnamefont {D.}~\bibnamefont
  {Joubert}},\ }\href@noop {} {\bibfield  {journal} {\bibinfo  {journal} {Phys.
  Rev. B}\ }\textbf {\bibinfo {volume} {59}},\ \bibinfo {pages} {1758}
  (\bibinfo {year} {1999})}\BibitemShut {NoStop}%
\bibitem [{\citenamefont {Todo}\ and\ \citenamefont {Kato}(2001)}]{loop}%
  \BibitemOpen
  \bibfield  {author} {\bibinfo {author} {\bibfnamefont {S.}~\bibnamefont
  {Todo}}\ and\ \bibinfo {author} {\bibfnamefont {K.}~\bibnamefont {Kato}},\
  }\href@noop {} {\bibfield  {journal} {\bibinfo  {journal} {Phys. Rev. Lett.}\
  }\textbf {\bibinfo {volume} {87}},\ \bibinfo {pages} {047203} (\bibinfo
  {year} {2001}), cond-mat/9911047}\BibitemShut {NoStop}%
\bibitem [{\citenamefont {Alet}\ \emph {et~al.}(2005)\citenamefont {Alet},
  \citenamefont {Wessel},\ and\ \citenamefont {Troyer}}]{dirloop}%
  \BibitemOpen
  \bibfield  {author} {\bibinfo {author} {\bibfnamefont {F.}~\bibnamefont
  {Alet}}, \bibinfo {author} {\bibfnamefont {S.}~\bibnamefont {Wessel}}, \ and\
  \bibinfo {author} {\bibfnamefont {M.}~\bibnamefont {Troyer}},\ }\href@noop {}
  {\bibfield  {journal} {\bibinfo  {journal} {Phys. Rev. E}\ }\textbf {\bibinfo
  {volume} {71}},\ \bibinfo {pages} {036706} (\bibinfo {year} {2005}), cond-mat/0308495},\
  \bibinfo {note} {and references therein}\BibitemShut {NoStop}%
\bibitem [{\citenamefont {Albuquerque}\ \emph {et~al.}(2007)\citenamefont
  {Albuquerque}, \citenamefont {Alet}, \citenamefont {Corboz}, \citenamefont
  {Dayal}, \citenamefont {Feiguin}, \citenamefont {Fuchs}, \citenamefont
  {Gamper}, \citenamefont {Gull}, \citenamefont {G\"urtler}, \citenamefont
  {Honecker}, \citenamefont {Igarashi}, \citenamefont {K\"orner}, \citenamefont
  {Kozhevnikov}, \citenamefont {L\"auchli}, \citenamefont {Manmana},
  \citenamefont {Matsumoto}, \citenamefont {McCulloch}, \citenamefont {Michel},
  \citenamefont {Noack}, \citenamefont {Paw{\l}owski}, \citenamefont {Pollet},
  \citenamefont {Pruschke}, \citenamefont {Schollw\"ock}, \citenamefont {Todo},
  \citenamefont {Trebst}, \citenamefont {Troyer}, \citenamefont {Werner},\ and\
  \citenamefont {Wessel}}]{alps}%
  \BibitemOpen
  \bibfield  {author} {\bibinfo {author} {\bibfnamefont {A.}~\bibnamefont
  {Albuquerque}}, \bibinfo {author} {\bibfnamefont {F.}~\bibnamefont {Alet}},
  \bibinfo {author} {\bibfnamefont {P.}~\bibnamefont {Corboz}}, \bibinfo
  {author} {\bibfnamefont {P.}~\bibnamefont {Dayal}}, \bibinfo {author}
  {\bibfnamefont {A.}~\bibnamefont {Feiguin}}, \bibinfo {author} {\bibfnamefont
  {S.}~\bibnamefont {Fuchs}}, \bibinfo {author} {\bibfnamefont
  {L.}~\bibnamefont {Gamper}}, \bibinfo {author} {\bibfnamefont
  {E.}~\bibnamefont {Gull}}, \bibinfo {author} {\bibfnamefont {S.}~\bibnamefont
  {G\"urtler}}, \bibinfo {author} {\bibfnamefont {A.}~\bibnamefont {Honecker}},
  \bibinfo {author} {\bibfnamefont {R.}~\bibnamefont {Igarashi}}, \bibinfo
  {author} {\bibfnamefont {M.}~\bibnamefont {K\"orner}}, \bibinfo {author}
  {\bibfnamefont {A.}~\bibnamefont {Kozhevnikov}}, \bibinfo {author}
  {\bibfnamefont {A.}~\bibnamefont {L\"auchli}}, \bibinfo {author}
  {\bibfnamefont {S.}~\bibnamefont {Manmana}}, \bibinfo {author} {\bibfnamefont
  {M.}~\bibnamefont {Matsumoto}}, \bibinfo {author} {\bibfnamefont
  {I.}~\bibnamefont {McCulloch}}, \bibinfo {author} {\bibfnamefont
  {F.}~\bibnamefont {Michel}}, \bibinfo {author} {\bibfnamefont
  {R.}~\bibnamefont {Noack}}, \bibinfo {author} {\bibfnamefont
  {G.}~\bibnamefont {Paw{\l}owski}}, \bibinfo {author} {\bibfnamefont
  {L.}~\bibnamefont {Pollet}}, \bibinfo {author} {\bibfnamefont
  {T.}~\bibnamefont {Pruschke}}, \bibinfo {author} {\bibfnamefont
  {U.}~\bibnamefont {Schollw\"ock}}, \bibinfo {author} {\bibfnamefont
  {S.}~\bibnamefont {Todo}}, \bibinfo {author} {\bibfnamefont {S.}~\bibnamefont
  {Trebst}}, \bibinfo {author} {\bibfnamefont {M.}~\bibnamefont {Troyer}},
  \bibinfo {author} {\bibfnamefont {P.}~\bibnamefont {Werner}}, \ and\ \bibinfo
  {author} {\bibfnamefont {S.}~\bibnamefont {Wessel}},\ }\href@noop {}
  {\bibfield  {journal} {\bibinfo  {journal} {J. Magn. Magn. Mater.}\ }\textbf
  {\bibinfo {volume} {310}},\ \bibinfo {pages} {1187} (\bibinfo {year}
  {2007})}\BibitemShut {NoStop}%
\bibitem [{Note2()}]{Note2}%
  \BibitemOpen
  \bibinfo {note} {Although the electron diffraction study was done at RT, its
  results are relevant to the low-temperature structure, because
  (CuBr)LaNb$_2$O$_7$ does not demonstrate phase transitions upon cooling below
  room temperature (Sec.~\ref {sec:evolution}).}\BibitemShut {Stop}%
\bibitem [{Note3()}]{Note3}%
  \BibitemOpen
  \bibinfo {note} {We refer to the conventional Glazer's notation, where $a$,
  $b$, and $c$ show different rotations around the respective crystallographic
  directions, $+/-$ denote the in-phase/out-of-phase tilt, and 0 indicates the
  absence of rotations.}\BibitemShut {Stop}%
\bibitem [{sup()}]{supplement}%
  \BibitemOpen
  \href@noop {} {}\bibinfo {note} {See Supplementary information for details of
  structure refinement, individual x-ray and neutron diffraction patterns, and
  the TGA data.}\BibitemShut {Stop}%
\bibitem [{foo()}]{foot1}%
  \BibitemOpen
  \href@noop {} {}\bibinfo {note} {At 20~K, the lower ADPs of La and Nb
  compared to that of other atoms might be related to systematic errors of the
  powder refinement. Single-crystal experiments would be necessary for a
  reliable evaluation of ADPs.}\BibitemShut {Stop}%
\bibitem [{Note4()}]{Note4}%
  \BibitemOpen
  \bibinfo {note} {Although the very weak diffuse scattering at $hkl$ with odd
  $h+k$ and integer $l$ is still present above $T_2$ (e.g., the broad feature
  at $2\theta \simeq 10.5$~deg in Fig.~\ref {fig:xrd}), it can not be taken
  into account in the conventional Rietveld refinement and should be considered
  in future neutron and/or single-crystal studies.}\BibitemShut {Stop}%
\bibitem [{Note5()}]{Note5}%
  \BibitemOpen
  \bibinfo {note} {The refinement of the XRD data presented in Ref.~\protect
  \rev@citealpnum {tsirlin2010a} results in $U_{11}=0.048(6)$~\r A$^2$,
  $U_{22}=0.024(5)$~\r A$^2$, and $U_{33}=0.03(1)$~\r A$^2$ for Cl atoms at
  660~K.}\BibitemShut {Stop}%
\bibitem [{Note6()}]{Note6}%
  \BibitemOpen
  \bibinfo {note} {Following Ref.~\protect \rev@citealpnum {tsirlin2010a}, we
  performed structure optimization using GGA+$U$. Although the LSDA+$U$
  functional delivers similar structural models, it underestimates the Cu--Br
  distances and leads to less accurate structural data. Relaxed structures are
  nearly independent of the $U_d$ and $J_d$ parameters as well as the
  double-counting-correction scheme.}\BibitemShut {Stop}%
\bibitem [{Note7()}]{Note7}%
  \BibitemOpen
  \bibinfo {note} {In the $P2$ space group, we also fixed the $z$ coordinate
  for one of the Cu sites. Otherwise, the origin can be arbitrarily shifted
  along the $c$ direction.}\BibitemShut {Stop}%
\bibitem [{\citenamefont {Eschrig}\ and\ \citenamefont
  {Koepernik}(2009)}]{wannier}%
  \BibitemOpen
  \bibfield  {author} {\bibinfo {author} {\bibfnamefont {H.}~\bibnamefont
  {Eschrig}}\ and\ \bibinfo {author} {\bibfnamefont {K.}~\bibnamefont
  {Koepernik}},\ }\href@noop {} {\bibfield  {journal} {\bibinfo  {journal}
  {Phys. Rev. B}\ }\textbf {\bibinfo {volume} {80}},\ \bibinfo {pages} {104503}
  (\bibinfo {year} {2009}), arXiv:0905.4844}\BibitemShut {NoStop}%
\bibitem [{\citenamefont {Schmitt}\ \emph {et~al.}(2009)\citenamefont
  {Schmitt}, \citenamefont {Janson}, \citenamefont {Schmidt}, \citenamefont
  {Hoffmann}, \citenamefont {Schnelle}, \citenamefont {Drechsler},\ and\
  \citenamefont {Rosner}}]{schmitt2009}%
  \BibitemOpen
  \bibfield  {author} {\bibinfo {author} {\bibfnamefont {M.}~\bibnamefont
  {Schmitt}}, \bibinfo {author} {\bibfnamefont {O.}~\bibnamefont {Janson}},
  \bibinfo {author} {\bibfnamefont {M.}~\bibnamefont {Schmidt}}, \bibinfo
  {author} {\bibfnamefont {S.}~\bibnamefont {Hoffmann}}, \bibinfo {author}
  {\bibfnamefont {W.}~\bibnamefont {Schnelle}}, \bibinfo {author}
  {\bibfnamefont {S.-L.}\ \bibnamefont {Drechsler}}, \ and\ \bibinfo {author}
  {\bibfnamefont {H.}~\bibnamefont {Rosner}},\ }\href@noop {} {\bibfield
  {journal} {\bibinfo  {journal} {Phys. Rev. B}\ }\textbf {\bibinfo {volume}
  {79}},\ \bibinfo {pages} {245119} (\bibinfo {year} {2009}), arXiv:0905.4038}\BibitemShut
  {NoStop}%
\bibitem [{\citenamefont {Tsirlin}\ \emph
  {et~al.}(2010{\natexlab{b}})\citenamefont {Tsirlin}, \citenamefont {Janson},\
  and\ \citenamefont {Rosner}}]{tsirlin2010c}%
  \BibitemOpen
  \bibfield  {author} {\bibinfo {author} {\bibfnamefont {A.~A.}\ \bibnamefont
  {Tsirlin}}, \bibinfo {author} {\bibfnamefont {O.}~\bibnamefont {Janson}}, \
  and\ \bibinfo {author} {\bibfnamefont {H.}~\bibnamefont {Rosner}},\
  }\href@noop {} {\bibfield  {journal} {\bibinfo  {journal} {Phys. Rev. B}\
  }\textbf {\bibinfo {volume} {82}},\ \bibinfo {pages} {144416} (\bibinfo
  {year} {2010}{\natexlab{b}}), arXiv:1007.1646}\BibitemShut {NoStop}%
\bibitem [{\citenamefont {Tsirlin}\ and\ \citenamefont
  {Rosner}(2010{\natexlab{b}})}]{tsirlin2010d}%
  \BibitemOpen
  \bibfield  {author} {\bibinfo {author} {\bibfnamefont {A.~A.}\ \bibnamefont
  {Tsirlin}}\ and\ \bibinfo {author} {\bibfnamefont {H.}~\bibnamefont
  {Rosner}},\ }\href@noop {} {\bibfield  {journal} {\bibinfo  {journal} {Phys.
  Rev. B}\ }\textbf {\bibinfo {volume} {81}},\ \bibinfo {pages} {024424}
  (\bibinfo {year} {2010}{\natexlab{b}}), arXiv:0910.2056}\BibitemShut {NoStop}%
\bibitem [{\citenamefont {Imai}\ \emph {et~al.}(2008)\citenamefont {Imai},
  \citenamefont {Nytko}, \citenamefont {Bartlett}, \citenamefont {Shores},\
  and\ \citenamefont {Nocera}}]{imai2008}%
  \BibitemOpen
  \bibfield  {author} {\bibinfo {author} {\bibfnamefont {T.}~\bibnamefont
  {Imai}}, \bibinfo {author} {\bibfnamefont {E.~A.}\ \bibnamefont {Nytko}},
  \bibinfo {author} {\bibfnamefont {B.~M.}\ \bibnamefont {Bartlett}}, \bibinfo
  {author} {\bibfnamefont {M.~P.}\ \bibnamefont {Shores}}, \ and\ \bibinfo
  {author} {\bibfnamefont {D.~G.}\ \bibnamefont {Nocera}},\ }\href@noop {}
  {\bibfield  {journal} {\bibinfo  {journal} {Phys. Rev. Lett.}\ }\textbf
  {\bibinfo {volume} {100}},\ \bibinfo {pages} {077203} (\bibinfo {year}
  {2008}), cond-mat/0703141}\BibitemShut {NoStop}%
\bibitem [{Note8()}]{Note8}%
  \BibitemOpen
  \bibinfo {note} {Impurity spins typically arise from local defects and/or
  surface effects that are inevitable in powder samples, especially those
  prepared at low temperatures. The contribution of impurity spins is usually
  approximated by the Curie-Weiss law with a small Weiss constant $\theta
  _{\protect \text {imp}}$, as shown, e.g., for the impurity spins in
  herbertsmithite (Ref.~\protect \rev@citealpnum {imai2008}).}\BibitemShut
  {Stop}%
\bibitem [{Note9()}]{Note9}%
  \BibitemOpen
  \bibinfo {note} {We performed simulations for the finite lattice with
  $12\times 12\times 6$ sites and periodic boundary conditions. This lattice
  size is sufficient to avoid finite-size effects in the relevant temperature
  range. The magnetization curve was simulated at $T/J_4=0.03$ corresponding to
  1.45~K vs. the experimental temperature of 1.3~K (Ref.~\protect
  \rev@citealpnum {oba2006}).}\BibitemShut {Stop}%
\bibitem [{\citenamefont {Banks}\ \emph {et~al.}(2009)\citenamefont {Banks},
  \citenamefont {Kremer}, \citenamefont {Hoch}, \citenamefont {Simon},
  \citenamefont {Ouladdiaf}, \citenamefont {Broto}, \citenamefont {Rakoto},
  \citenamefont {Lee},\ and\ \citenamefont {Whangbo}}]{banks2009}%
  \BibitemOpen
  \bibfield  {author} {\bibinfo {author} {\bibfnamefont {M.~G.}\ \bibnamefont
  {Banks}}, \bibinfo {author} {\bibfnamefont {R.~K.}\ \bibnamefont {Kremer}},
  \bibinfo {author} {\bibfnamefont {C.}~\bibnamefont {Hoch}}, \bibinfo {author}
  {\bibfnamefont {A.}~\bibnamefont {Simon}}, \bibinfo {author} {\bibfnamefont
  {B.}~\bibnamefont {Ouladdiaf}}, \bibinfo {author} {\bibfnamefont {J.-M.}\
  \bibnamefont {Broto}}, \bibinfo {author} {\bibfnamefont {H.}~\bibnamefont
  {Rakoto}}, \bibinfo {author} {\bibfnamefont {C.}~\bibnamefont {Lee}}, \ and\
  \bibinfo {author} {\bibfnamefont {M.-H.}\ \bibnamefont {Whangbo}},\
  }\href@noop {} {\bibfield  {journal} {\bibinfo  {journal} {Phys. Rev. B}\
  }\textbf {\bibinfo {volume} {80}},\ \bibinfo {pages} {024404} (\bibinfo
  {year} {2009}), arXiv:0904.2929}\BibitemShut {NoStop}%
\bibitem [{\citenamefont {Janson}\ \emph {et~al.}(2011)\citenamefont {Janson},
  \citenamefont {Tsirlin}, \citenamefont {Sichelschmidt}, \citenamefont
  {Skourski}, \citenamefont {Weickert},\ and\ \citenamefont
  {Rosner}}]{janson2011}%
  \BibitemOpen
  \bibfield  {author} {\bibinfo {author} {\bibfnamefont {O.}~\bibnamefont
  {Janson}}, \bibinfo {author} {\bibfnamefont {A.~A.}\ \bibnamefont {Tsirlin}},
  \bibinfo {author} {\bibfnamefont {J.}~\bibnamefont {Sichelschmidt}}, \bibinfo
  {author} {\bibfnamefont {Y.}~\bibnamefont {Skourski}}, \bibinfo {author}
  {\bibfnamefont {F.}~\bibnamefont {Weickert}}, \ and\ \bibinfo {author}
  {\bibfnamefont {H.}~\bibnamefont {Rosner}},\ }\href@noop {} {\bibfield
  {journal} {\bibinfo  {journal} {Phys. Rev. B}\ }\textbf {\bibinfo {volume}
  {83}},\ \bibinfo {pages} {094435} (\bibinfo {year} {2011}), arXiv:1011.5393}\BibitemShut
  {NoStop}%
\bibitem [{\citenamefont {Povarov}\ \emph {et~al.}(2011)\citenamefont
  {Povarov}, \citenamefont {Smirnov}, \citenamefont {Starykh}, \citenamefont
  {Petrov},\ and\ \citenamefont {Shapiro}}]{povarov2011}%
  \BibitemOpen
  \bibfield  {author} {\bibinfo {author} {\bibfnamefont {K.~Y.}\ \bibnamefont
  {Povarov}}, \bibinfo {author} {\bibfnamefont {A.~I.}\ \bibnamefont
  {Smirnov}}, \bibinfo {author} {\bibfnamefont {O.~A.}\ \bibnamefont
  {Starykh}}, \bibinfo {author} {\bibfnamefont {S.~V.}\ \bibnamefont {Petrov}},
  \ and\ \bibinfo {author} {\bibfnamefont {A.~Y.}\ \bibnamefont {Shapiro}},\
  }\href@noop {} {\bibfield  {journal} {\bibinfo  {journal} {Phys. Rev. Lett.}\
  }\textbf {\bibinfo {volume} {107}},\ \bibinfo {pages} {037204} (\bibinfo
  {year} {2011}), arXiv:1101.5275}\BibitemShut {NoStop}%
\bibitem [{\citenamefont {Sandvik}(1997)}]{sandvik1997}%
  \BibitemOpen
  \bibfield  {author} {\bibinfo {author} {\bibfnamefont {A.~W.}\ \bibnamefont
  {Sandvik}},\ }\href@noop {} {\bibfield  {journal} {\bibinfo  {journal} {Phys.
  Rev. B}\ }\textbf {\bibinfo {volume} {56}},\ \bibinfo {pages} {11678}
  (\bibinfo {year} {1997})}\BibitemShut {NoStop}%
\bibitem [{Note10()}]{Note10}%
  \BibitemOpen
  \bibinfo {note} {The inverse temperature was set to $\beta =(T/J_4)^{-1}=4L$
  for the $2L\times 2L\times L$ finite lattices.}\BibitemShut {Stop}%
\bibitem [{\citenamefont {Schmidt}\ \emph {et~al.}(2002)\citenamefont
  {Schmidt}, \citenamefont {Schulenburg}, \citenamefont {Richter},\ and\
  \citenamefont {Betts}}]{schmidt2002}%
  \BibitemOpen
  \bibfield  {author} {\bibinfo {author} {\bibfnamefont {R.}~\bibnamefont
  {Schmidt}}, \bibinfo {author} {\bibfnamefont {J.}~\bibnamefont
  {Schulenburg}}, \bibinfo {author} {\bibfnamefont {J.}~\bibnamefont
  {Richter}}, \ and\ \bibinfo {author} {\bibfnamefont {D.~D.}\ \bibnamefont
  {Betts}},\ }\href@noop {} {\bibfield  {journal} {\bibinfo  {journal} {Phys.
  Rev. B}\ }\textbf {\bibinfo {volume} {66}},\ \bibinfo {pages} {224406}
  (\bibinfo {year} {2002}), cond-mat/0203085}\BibitemShut {NoStop}%
\bibitem [{Note11()}]{Note11}%
  \BibitemOpen
  \bibinfo {note} {More specifically, the magnetic moment on the Cu site is
  0.64~$\mu _B$ in the AMF calculation with $U_d=5$~eV and 0.80~$\mu _B$ in the
  FLL calculation with $U_d=12$~eV, compared to 1~$\mu _B$ in a system without
  hybridization. Note that these numbers are obtained from DFT+$U$ calculations
  that miss both quantum effects and spin-orbit coupling. Therefore, neither
  the deviation of $g$ from the free-electron value, nor quantum fluctuations
  are taken into account.}\BibitemShut {Stop}%
\bibitem [{\citenamefont {Yusuf}\ \emph {et~al.}(2011)\citenamefont {Yusuf},
  \citenamefont {Bera}, \citenamefont {Ritter}, \citenamefont {Tsujimoto},
  \citenamefont {Ajiro}, \citenamefont {Kageyama},\ and\ \citenamefont
  {Attfield}}]{yusuf2011}%
  \BibitemOpen
  \bibfield  {author} {\bibinfo {author} {\bibfnamefont {S.~M.}\ \bibnamefont
  {Yusuf}}, \bibinfo {author} {\bibfnamefont {A.~K.}\ \bibnamefont {Bera}},
  \bibinfo {author} {\bibfnamefont {C.}~\bibnamefont {Ritter}}, \bibinfo
  {author} {\bibfnamefont {Y.}~\bibnamefont {Tsujimoto}}, \bibinfo {author}
  {\bibfnamefont {Y.}~\bibnamefont {Ajiro}}, \bibinfo {author} {\bibfnamefont
  {H.}~\bibnamefont {Kageyama}}, \ and\ \bibinfo {author} {\bibfnamefont
  {J.~P.}\ \bibnamefont {Attfield}},\ }\href@noop {} {\bibfield  {journal}
  {\bibinfo  {journal} {Phys. Rev. B}\ }\textbf {\bibinfo {volume} {84}},\
  \bibinfo {pages} {064407} (\bibinfo {year} {2011})}\BibitemShut {NoStop}%
\bibitem [{\citenamefont {Tsujimoto}\ \emph {et~al.}(2007)\citenamefont
  {Tsujimoto}, \citenamefont {Baba}, \citenamefont {Oba}, \citenamefont
  {Kageyama}, \citenamefont {Fukui}, \citenamefont {Narumi}, \citenamefont
  {Kindo}, \citenamefont {Saito}, \citenamefont {Takano}, \citenamefont
  {Ajiro},\ and\ \citenamefont {Yoshimura}}]{tsujimoto2007}%
  \BibitemOpen
  \bibfield  {author} {\bibinfo {author} {\bibfnamefont {Y.}~\bibnamefont
  {Tsujimoto}}, \bibinfo {author} {\bibfnamefont {Y.}~\bibnamefont {Baba}},
  \bibinfo {author} {\bibfnamefont {N.}~\bibnamefont {Oba}}, \bibinfo {author}
  {\bibfnamefont {H.}~\bibnamefont {Kageyama}}, \bibinfo {author}
  {\bibfnamefont {T.}~\bibnamefont {Fukui}}, \bibinfo {author} {\bibfnamefont
  {Y.}~\bibnamefont {Narumi}}, \bibinfo {author} {\bibfnamefont
  {K.}~\bibnamefont {Kindo}}, \bibinfo {author} {\bibfnamefont
  {T.}~\bibnamefont {Saito}}, \bibinfo {author} {\bibfnamefont
  {M.}~\bibnamefont {Takano}}, \bibinfo {author} {\bibfnamefont
  {Y.}~\bibnamefont {Ajiro}}, \ and\ \bibinfo {author} {\bibfnamefont
  {K.}~\bibnamefont {Yoshimura}},\ }\href@noop {} {\bibfield  {journal}
  {\bibinfo  {journal} {J. Phys. Soc. Jpn.}\ }\textbf {\bibinfo {volume}
  {76}},\ \bibinfo {pages} {063711} (\bibinfo {year} {2007})}\BibitemShut
  {NoStop}%
\bibitem [{\citenamefont {Tsujimoto}\ \emph {et~al.}(2008)\citenamefont
  {Tsujimoto}, \citenamefont {Kageyama}, \citenamefont {Baba}, \citenamefont
  {Kitada}, \citenamefont {Yamamoto}, \citenamefont {Narumi}, \citenamefont
  {Kindo}, \citenamefont {Nishi}, \citenamefont {Carlo}, \citenamefont {Aczel},
  \citenamefont {Williams}, \citenamefont {Goko}, \citenamefont {Luke},
  \citenamefont {Uemura}, \citenamefont {Ueda}, \citenamefont {Ajiro},\ and\
  \citenamefont {Yoshimura}}]{tsujimoto2008}%
  \BibitemOpen
  \bibfield  {author} {\bibinfo {author} {\bibfnamefont {Y.}~\bibnamefont
  {Tsujimoto}}, \bibinfo {author} {\bibfnamefont {H.}~\bibnamefont {Kageyama}},
  \bibinfo {author} {\bibfnamefont {Y.}~\bibnamefont {Baba}}, \bibinfo {author}
  {\bibfnamefont {A.}~\bibnamefont {Kitada}}, \bibinfo {author} {\bibfnamefont
  {T.}~\bibnamefont {Yamamoto}}, \bibinfo {author} {\bibfnamefont
  {Y.}~\bibnamefont {Narumi}}, \bibinfo {author} {\bibfnamefont
  {K.}~\bibnamefont {Kindo}}, \bibinfo {author} {\bibfnamefont
  {M.}~\bibnamefont {Nishi}}, \bibinfo {author} {\bibfnamefont {J.~P.}\
  \bibnamefont {Carlo}}, \bibinfo {author} {\bibfnamefont {A.~A.}\ \bibnamefont
  {Aczel}}, \bibinfo {author} {\bibfnamefont {T.~J.}\ \bibnamefont {Williams}},
  \bibinfo {author} {\bibfnamefont {T.}~\bibnamefont {Goko}}, \bibinfo {author}
  {\bibfnamefont {G.~M.}\ \bibnamefont {Luke}}, \bibinfo {author}
  {\bibfnamefont {Y.~J.}\ \bibnamefont {Uemura}}, \bibinfo {author}
  {\bibfnamefont {Y.}~\bibnamefont {Ueda}}, \bibinfo {author} {\bibfnamefont
  {Y.}~\bibnamefont {Ajiro}}, \ and\ \bibinfo {author} {\bibfnamefont
  {K.}~\bibnamefont {Yoshimura}},\ }\href@noop {} {\bibfield  {journal}
  {\bibinfo  {journal} {Phys. Rev. B}\ }\textbf {\bibinfo {volume} {78}},\
  \bibinfo {pages} {214410} (\bibinfo {year} {2008})}\BibitemShut {NoStop}%
\bibitem [{\citenamefont {Tsujimoto}\ \emph {et~al.}(2010)\citenamefont
  {Tsujimoto}, \citenamefont {Kitada}, \citenamefont {Kageyama}, \citenamefont
  {Nishi}, \citenamefont {Narumi}, \citenamefont {Kindo}, \citenamefont
  {Kiuchi}, \citenamefont {Ueda}, \citenamefont {Uemura}, \citenamefont
  {Ajiro},\ and\ \citenamefont {Yoshimura}}]{tsujimoto2010}%
  \BibitemOpen
  \bibfield  {author} {\bibinfo {author} {\bibfnamefont {Y.}~\bibnamefont
  {Tsujimoto}}, \bibinfo {author} {\bibfnamefont {A.}~\bibnamefont {Kitada}},
  \bibinfo {author} {\bibfnamefont {H.}~\bibnamefont {Kageyama}}, \bibinfo
  {author} {\bibfnamefont {M.}~\bibnamefont {Nishi}}, \bibinfo {author}
  {\bibfnamefont {Y.}~\bibnamefont {Narumi}}, \bibinfo {author} {\bibfnamefont
  {K.}~\bibnamefont {Kindo}}, \bibinfo {author} {\bibfnamefont
  {Y.}~\bibnamefont {Kiuchi}}, \bibinfo {author} {\bibfnamefont
  {Y.}~\bibnamefont {Ueda}}, \bibinfo {author} {\bibfnamefont {Y.~J.}\
  \bibnamefont {Uemura}}, \bibinfo {author} {\bibfnamefont {Y.}~\bibnamefont
  {Ajiro}}, \ and\ \bibinfo {author} {\bibfnamefont {K.}~\bibnamefont
  {Yoshimura}},\ }\href@noop {} {\bibfield  {journal} {\bibinfo  {journal} {J.
  Phys. Soc. Jpn.}\ }\textbf {\bibinfo {volume} {79}},\ \bibinfo {pages}
  {014709} (\bibinfo {year} {2010}), arXiv:0907.5103}\BibitemShut {NoStop}%
\bibitem [{\citenamefont {Uemura}\ \emph {et~al.}(2009)\citenamefont {Uemura},
  \citenamefont {Aczel}, \citenamefont {Ajiro}, \citenamefont {Carlo},
  \citenamefont {Goko}, \citenamefont {Goldfeld}, \citenamefont {Kitada},
  \citenamefont {Luke}, \citenamefont {MacDougall}, \citenamefont {Mihailescu},
  \citenamefont {Rodriguez}, \citenamefont {Russo}, \citenamefont {Tsujimoto},
  \citenamefont {Wiebe}, \citenamefont {Williams}, \citenamefont {Yamamoto},
  \citenamefont {Yoshimura},\ and\ \citenamefont {Kageyama}}]{uemura2009}%
  \BibitemOpen
  \bibfield  {author} {\bibinfo {author} {\bibfnamefont {Y.~J.}\ \bibnamefont
  {Uemura}}, \bibinfo {author} {\bibfnamefont {A.~A.}\ \bibnamefont {Aczel}},
  \bibinfo {author} {\bibfnamefont {Y.}~\bibnamefont {Ajiro}}, \bibinfo
  {author} {\bibfnamefont {J.~P.}\ \bibnamefont {Carlo}}, \bibinfo {author}
  {\bibfnamefont {T.}~\bibnamefont {Goko}}, \bibinfo {author} {\bibfnamefont
  {D.~A.}\ \bibnamefont {Goldfeld}}, \bibinfo {author} {\bibfnamefont
  {A.}~\bibnamefont {Kitada}}, \bibinfo {author} {\bibfnamefont {G.~M.}\
  \bibnamefont {Luke}}, \bibinfo {author} {\bibfnamefont {G.~J.}\ \bibnamefont
  {MacDougall}}, \bibinfo {author} {\bibfnamefont {I.~G.}\ \bibnamefont
  {Mihailescu}}, \bibinfo {author} {\bibfnamefont {J.~A.}\ \bibnamefont
  {Rodriguez}}, \bibinfo {author} {\bibfnamefont {P.~L.}\ \bibnamefont
  {Russo}}, \bibinfo {author} {\bibfnamefont {Y.}~\bibnamefont {Tsujimoto}},
  \bibinfo {author} {\bibfnamefont {C.~R.}\ \bibnamefont {Wiebe}}, \bibinfo
  {author} {\bibfnamefont {T.~J.}\ \bibnamefont {Williams}}, \bibinfo {author}
  {\bibfnamefont {T.}~\bibnamefont {Yamamoto}}, \bibinfo {author}
  {\bibfnamefont {K.}~\bibnamefont {Yoshimura}}, \ and\ \bibinfo {author}
  {\bibfnamefont {H.}~\bibnamefont {Kageyama}},\ }\href@noop {} {\bibfield
  {journal} {\bibinfo  {journal} {Phys. Rev. B}\ }\textbf {\bibinfo {volume}
  {80}},\ \bibinfo {pages} {174408} (\bibinfo {year} {2009}), arXiv:0806.2021}\BibitemShut
  {NoStop}%
\bibitem [{\citenamefont {Kitada}\ \emph {et~al.}(2009)\citenamefont {Kitada},
  \citenamefont {Tsujimoto}, \citenamefont {Kageyama}, \citenamefont {Ajiro},
  \citenamefont {Nishi}, \citenamefont {Narumi}, \citenamefont {Kindo},
  \citenamefont {Ichihara}, \citenamefont {Ueda}, \citenamefont {Uemura},\ and\
  \citenamefont {Yoshimura}}]{kitada2009}%
  \BibitemOpen
  \bibfield  {author} {\bibinfo {author} {\bibfnamefont {A.}~\bibnamefont
  {Kitada}}, \bibinfo {author} {\bibfnamefont {Y.}~\bibnamefont {Tsujimoto}},
  \bibinfo {author} {\bibfnamefont {H.}~\bibnamefont {Kageyama}}, \bibinfo
  {author} {\bibfnamefont {Y.}~\bibnamefont {Ajiro}}, \bibinfo {author}
  {\bibfnamefont {M.}~\bibnamefont {Nishi}}, \bibinfo {author} {\bibfnamefont
  {Y.}~\bibnamefont {Narumi}}, \bibinfo {author} {\bibfnamefont
  {K.}~\bibnamefont {Kindo}}, \bibinfo {author} {\bibfnamefont
  {M.}~\bibnamefont {Ichihara}}, \bibinfo {author} {\bibfnamefont
  {Y.}~\bibnamefont {Ueda}}, \bibinfo {author} {\bibfnamefont {Y.~J.}\
  \bibnamefont {Uemura}}, \ and\ \bibinfo {author} {\bibfnamefont
  {K.}~\bibnamefont {Yoshimura}},\ }\href@noop {} {\bibfield  {journal}
  {\bibinfo  {journal} {Phys. Rev. B}\ }\textbf {\bibinfo {volume} {80}},\
  \bibinfo {pages} {174409} (\bibinfo {year} {2009})}\BibitemShut {NoStop}%
\bibitem [{\citenamefont {Janson}\ \emph {et~al.}(2010)\citenamefont {Janson},
  \citenamefont {Tsirlin}, \citenamefont {Schmitt},\ and\ \citenamefont
  {Rosner}}]{janson2010}%
  \BibitemOpen
  \bibfield  {author} {\bibinfo {author} {\bibfnamefont {O.}~\bibnamefont
  {Janson}}, \bibinfo {author} {\bibfnamefont {A.~A.}\ \bibnamefont {Tsirlin}},
  \bibinfo {author} {\bibfnamefont {M.}~\bibnamefont {Schmitt}}, \ and\
  \bibinfo {author} {\bibfnamefont {H.}~\bibnamefont {Rosner}},\ }\href@noop {}
  {\bibfield  {journal} {\bibinfo  {journal} {Phys. Rev. B}\ }\textbf {\bibinfo
  {volume} {82}},\ \bibinfo {pages} {014424} (\bibinfo {year}
  {2010}), arXiv:1004.3765}\BibitemShut {NoStop}%
\bibitem [{\citenamefont {Kitada}\ \emph {et~al.}(2007)\citenamefont {Kitada},
  \citenamefont {Hiroi}, \citenamefont {Tsujimoto}, \citenamefont {Kitano},
  \citenamefont {Kageyama}, \citenamefont {Ajiro},\ and\ \citenamefont
  {Yoshimura}}]{kitada2007}%
  \BibitemOpen
  \bibfield  {author} {\bibinfo {author} {\bibfnamefont {A.}~\bibnamefont
  {Kitada}}, \bibinfo {author} {\bibfnamefont {Z.}~\bibnamefont {Hiroi}},
  \bibinfo {author} {\bibfnamefont {Y.}~\bibnamefont {Tsujimoto}}, \bibinfo
  {author} {\bibfnamefont {T.}~\bibnamefont {Kitano}}, \bibinfo {author}
  {\bibfnamefont {H.}~\bibnamefont {Kageyama}}, \bibinfo {author}
  {\bibfnamefont {Y.}~\bibnamefont {Ajiro}}, \ and\ \bibinfo {author}
  {\bibfnamefont {K.}~\bibnamefont {Yoshimura}},\ }\href@noop {} {\bibfield
  {journal} {\bibinfo  {journal} {J. Phys. Soc. Jpn.}\ }\textbf {\bibinfo
  {volume} {76}},\ \bibinfo {pages} {093706} (\bibinfo {year}
  {2007})}\BibitemShut {NoStop}%
\end{thebibliography}
%

\pagebreak
\renewcommand{\thefigure}{S\arabic{figure}}
\setcounter{figure}{0}
\setcounter{table}{1}
\renewcommand{\thetable}{S\arabic{table}}

\begin{widetext}
\begin{center}
{\large
Supplementary information for 
\smallskip

\textbf{``Short-range order of Br and three-dimensional magnetism of (CuBr)LaNb$_2$O$_7$''}}
\medskip

A. A. Tsirlin, A. M. Abakumov, C. Ritter, P. Henry, O. Janson, and H. Rosner
\end{center}
\medskip

\begin{table}[!ht]
\begin{minipage}{14cm}
\caption{Details of structure refinements for (CuBr)LaNb$_2$O$_7$.}
\begin{ruledtabular}
\begin{tabular}{cccccccc}
   $T$        & $a$       & $b$       & $c$        & Space group & Radiation   & Source & $R_p/R_{wp}$ \\\hline
   10         & 7.7856(2) & 7.7983(2) & 11.6938(2) & $Cmmm$      &  neutron    &   D2B  & 0.029/0.037  \\
              &           &           &            &             &  neutron    &   D20  & 0.025/0.033  \\
              &           &           &            &             & synchrotron &  ID31  & 0.054/0.079  \\\hline
   300        & 7.7926(2) & 7.8009(2) & 11.7017(2) & $Cmmm$      &  neutron    &  E9    & 0.026/0.033  \\
              &           &           &            &             & synchrotron &  ID31  & 0.102/0.124  \\\hline
   720        & 3.9077(1) & 3.9077(1) & 11.7192(1) & $P4/mmm$    & synchrotron &  ID31  & 0.064/0.082  \\
\end{tabular}
\end{ruledtabular}
\end{minipage}
\end{table}

\begin{figure}[!h]
\includegraphics[width=11cm]{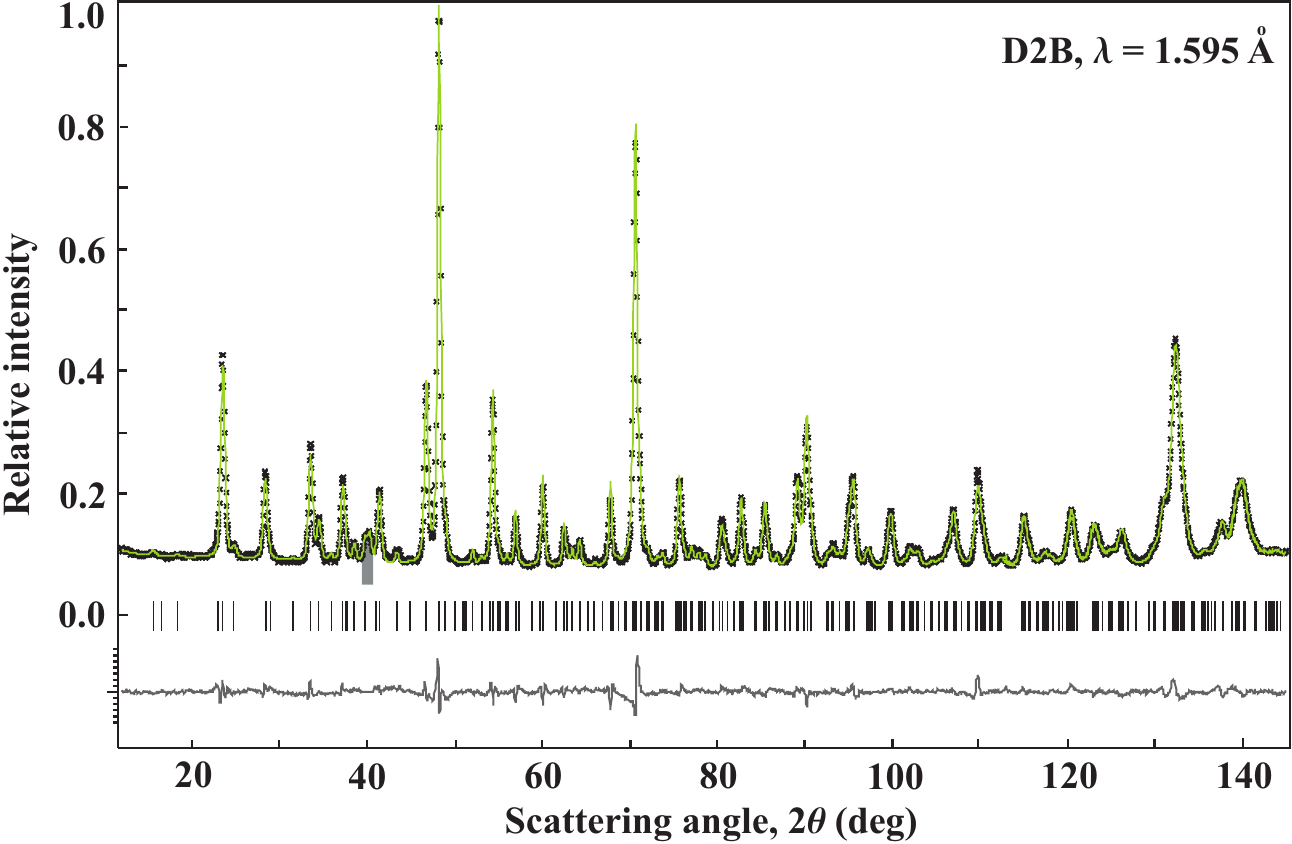}
\begin{minipage}{14cm}
\caption{\label{fig:s1}\normalsize
Rietveld refinement of the low-temperature D2B neutron data. The excluded region around $2\theta=40$~deg is due to the cryostat window.
}
\end{minipage}
\end{figure}
\bigskip

\begin{figure}[!h]
\includegraphics[width=11cm]{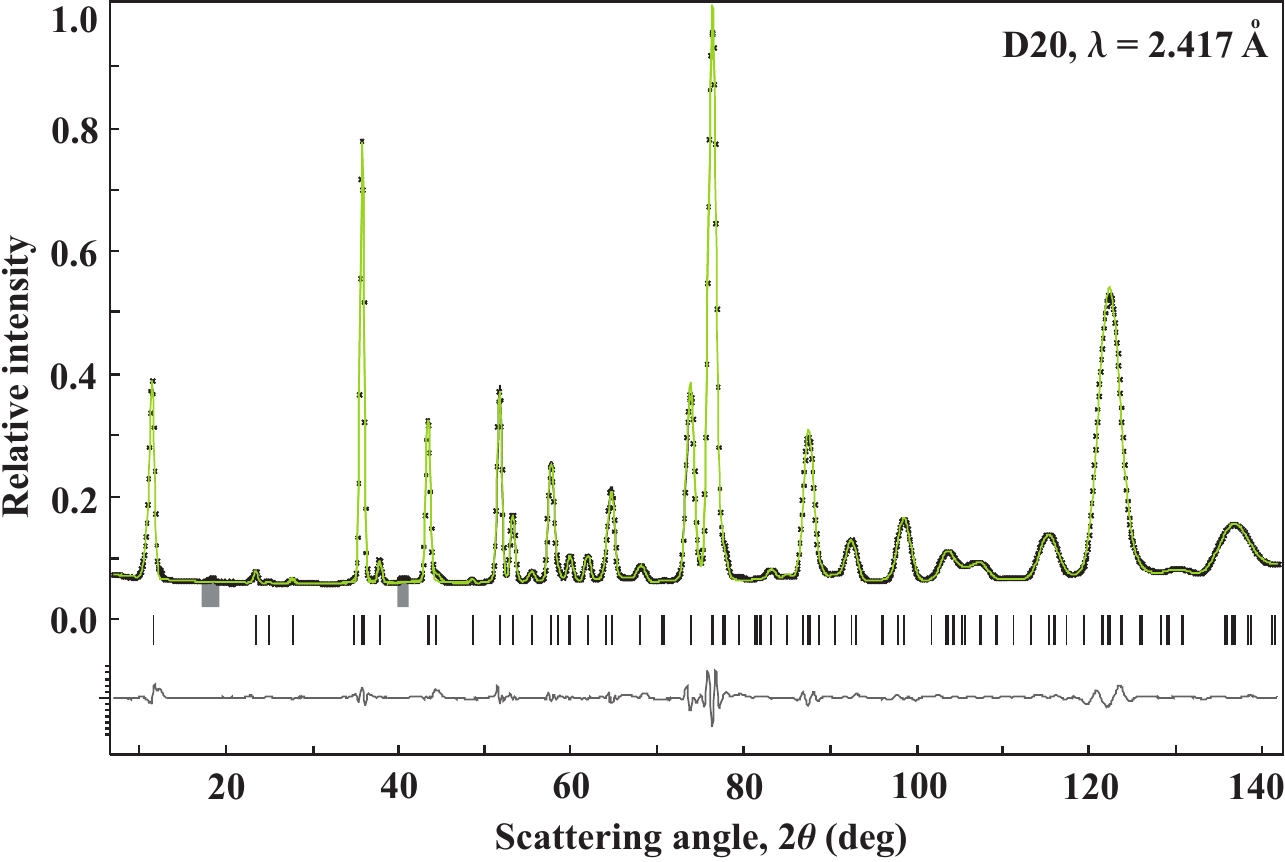}
\begin{minipage}{14cm}
\caption{\label{fig:s2}\normalsize
Rietveld refinement of the low-temperature D20 neutron data. The excluded regions around $2\theta=18$~deg and 40~deg are due to the magnetic scattering and diffuse scattering, respectively (see Fig.~6 of the manuscript).
}
\end{minipage}
\end{figure}
\bigskip

\begin{figure}[!h]
\includegraphics[width=11cm]{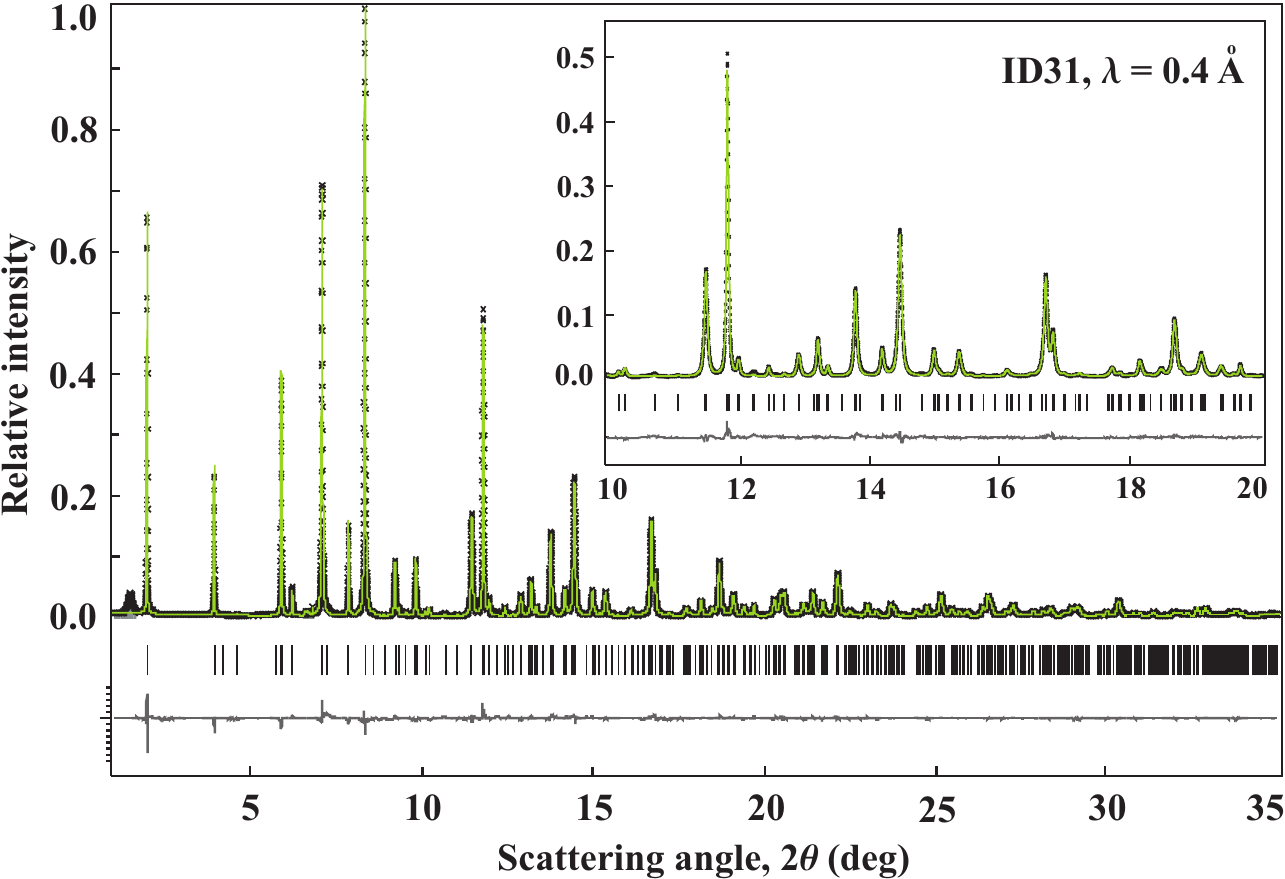}
\begin{minipage}{14cm}
\caption{\label{fig:s3}\normalsize
Rietveld refinement of the low-temperature synchrotron XRD data. The broad feature around $2\theta=1.5$~deg is due to the cryostat window.
}
\end{minipage}
\end{figure}
\bigskip

\begin{figure}[!h]
\includegraphics[width=11cm]{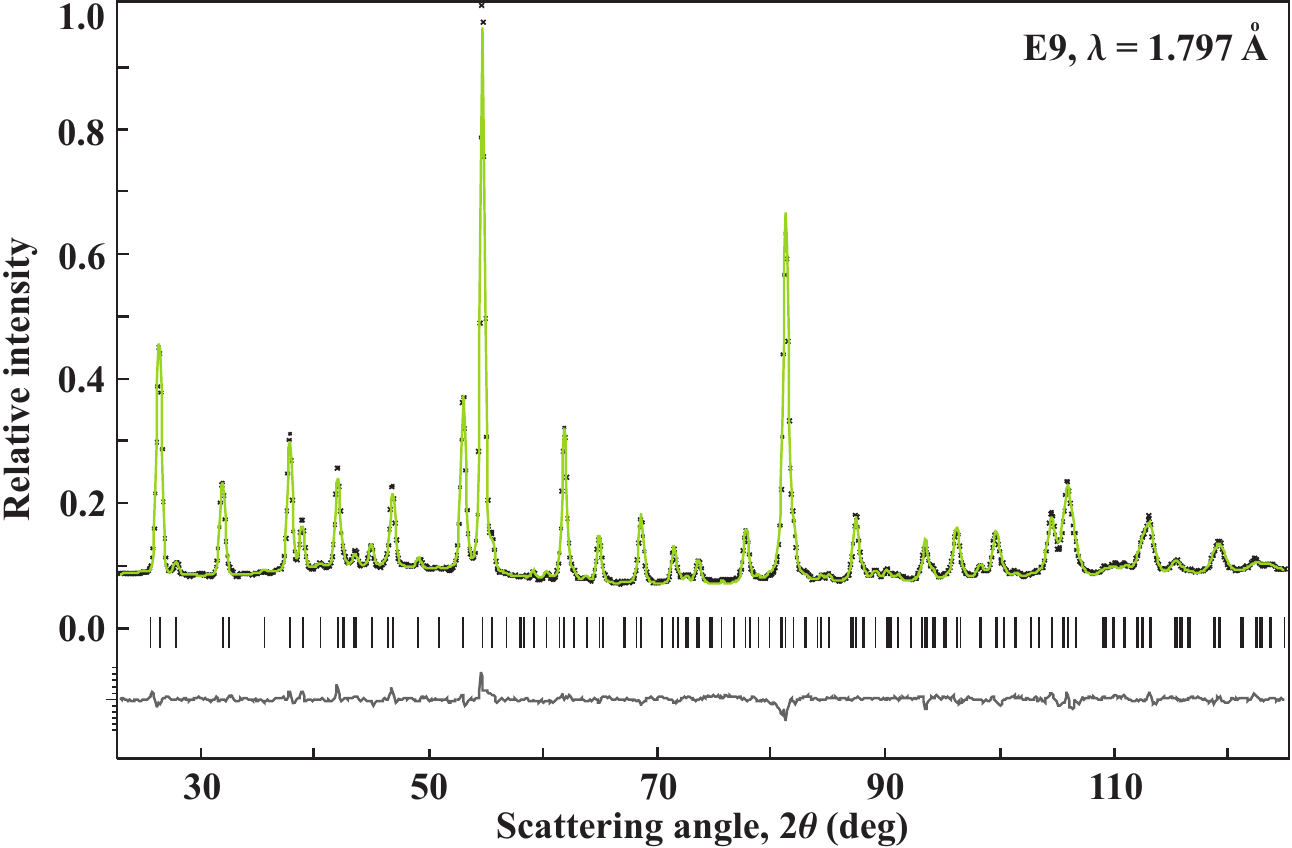}
\begin{minipage}{14cm}
\caption{\label{fig:s4}\normalsize
Rietveld refinement of the room-temperature E9 neutron data.
}
\end{minipage}
\end{figure}
\bigskip

\begin{figure}[!h]
\includegraphics[width=11cm]{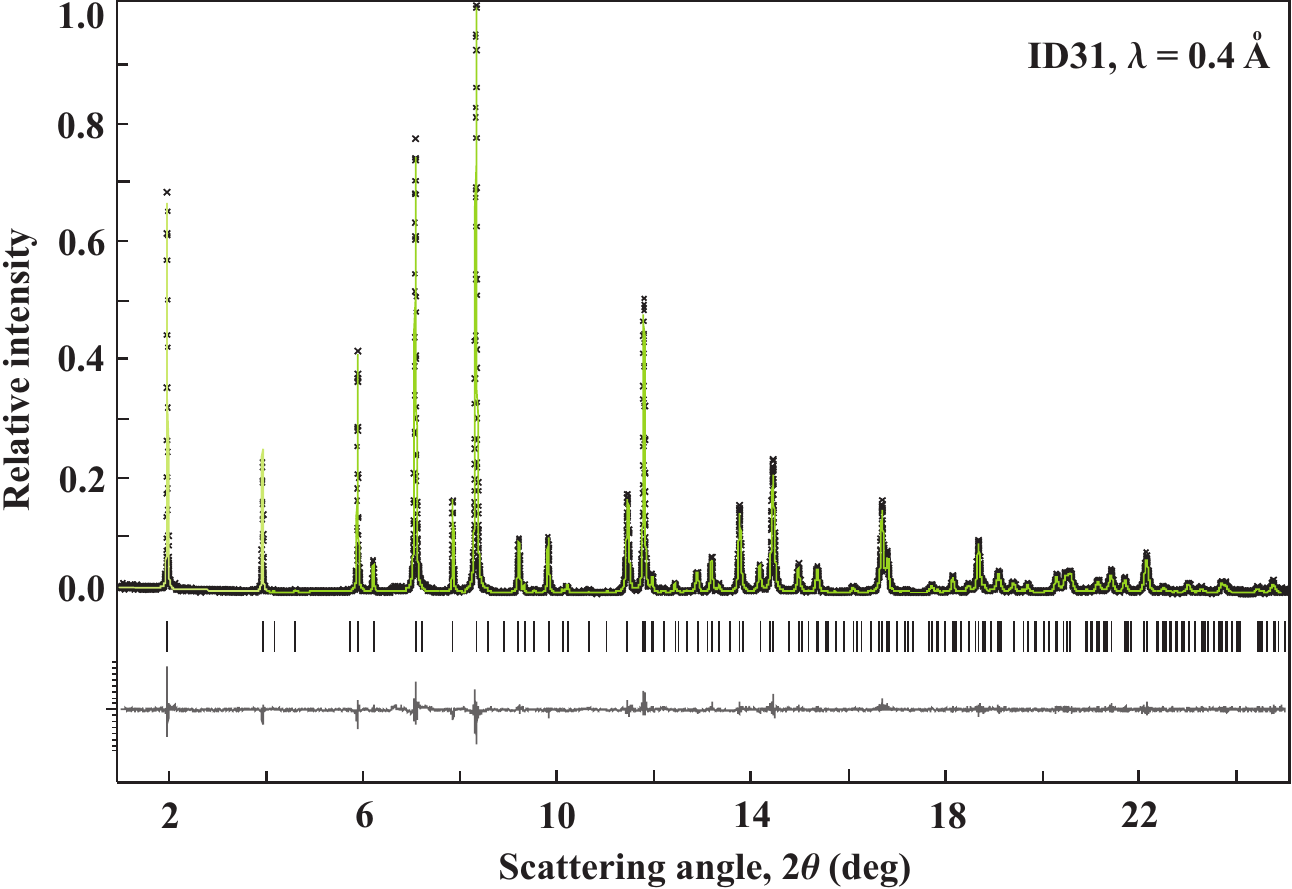}
\begin{minipage}{14cm}
\caption{\label{fig:s5}\normalsize
Rietveld refinement of the room-temperature synchrotron XRD data. 
}
\end{minipage}
\end{figure}
\bigskip

\begin{figure}[!h]
\includegraphics[width=11cm]{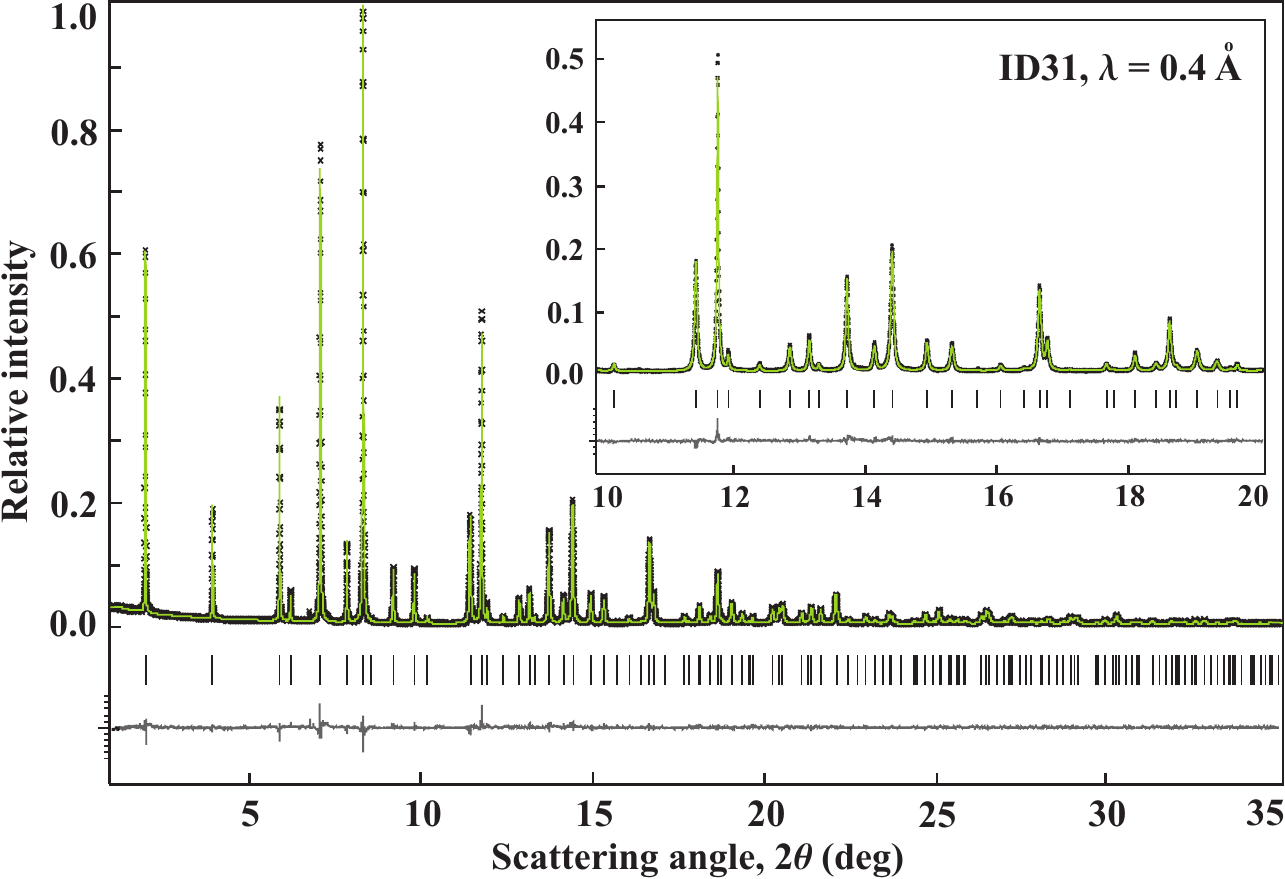}
\begin{minipage}{14cm}
\caption{\label{fig:s6}\normalsize
Rietveld refinement of the synchrotron XRD data collected at 720~K. 
}
\end{minipage}
\end{figure}
\bigskip

\begin{figure}[!h]
\includegraphics[width=11cm]{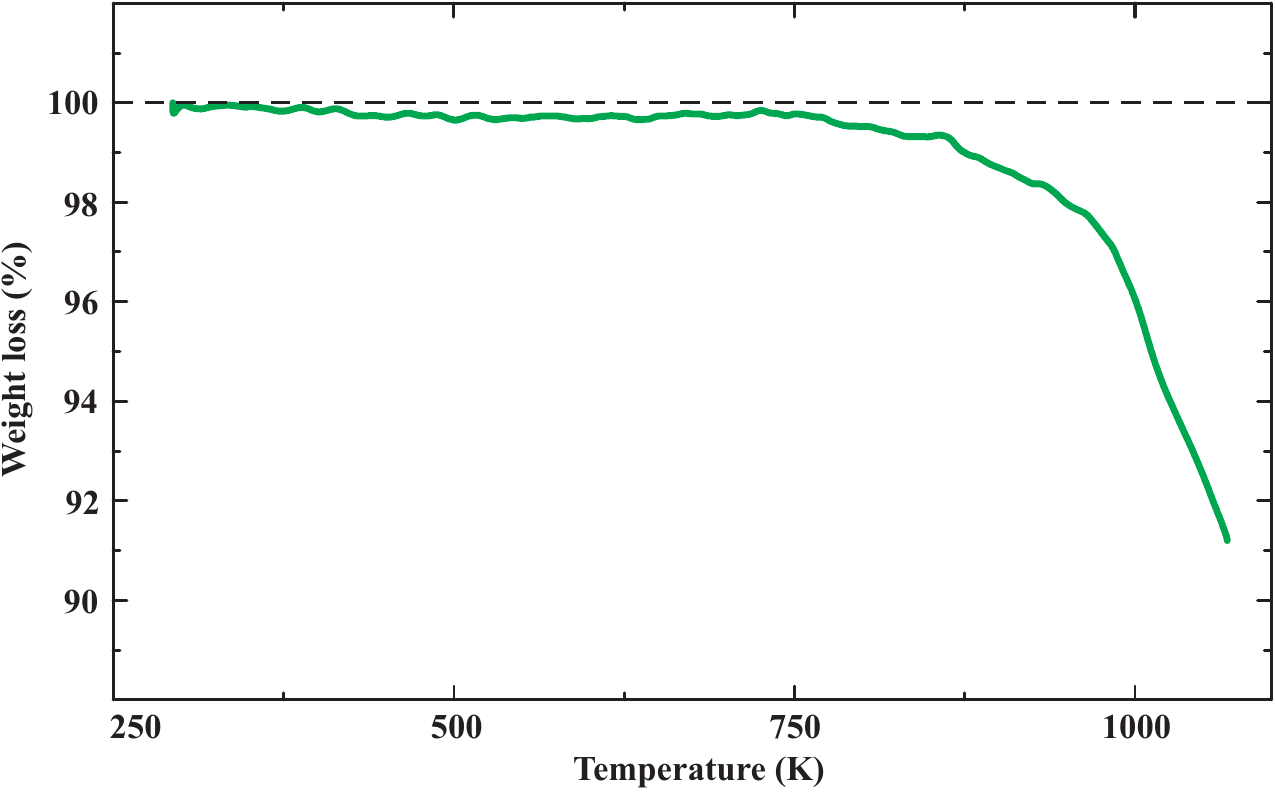}
\begin{minipage}{14cm}
\caption{\label{fig:s7}\normalsize
Thermogravimetric data for (CuBr)LaNb$_2$O$_7$. Note the onset of the weight loss around 750~K.
}
\end{minipage}
\end{figure}

\end{widetext}

\end{document}